\documentclass[english,aps,prd,showpacs,nofootinbib,titlepage,subeqn]{revtex4}
\usepackage{amssymb}
\usepackage{amsmath}
\usepackage{mathtools}
\usepackage{eucal}
\usepackage[dvips]{graphics}
\usepackage{epsfig}
\usepackage{bm}
\usepackage[T1]{fontenc}
\usepackage{emerald}
\usepackage{babel}
\usepackage{slantsc}
\usepackage{array}
\usepackage[sort&compress]{natbib}

\def\st{\scriptstyle}
\def\sst{\scriptscriptstyle}
\def\be{\begin{equation}}
\def\ee{\end{equation}}
\def\ba{\begin{eqnarray}}
\def\ea{\end{eqnarray}}
\def\bsu{\begin{subequations}}
\def\esu{\end{subequations}}

\def\gg{\mathfrak{g}}

\def\a{\alpha}
\def\b{\beta}
\def\g{\gamma}     \def\G{\Gamma}
\def\d{\delta}

\def\e{\epsilon}

\def\m{\mu}
\def\n{\nu}
   
\def\p{\Phi}         \def\vp{\phi}

\def\vt{\vartheta}
\def\r{\rho}
\def\s{\sigma}
\def\t{\tau}
\def\up{\upsilon}

\def\H{{\textit H}}
\def\la{\label}
\def\pd{\partial}
\def\le{\left}
\def\ri{\right}
\def\lag{{\cal L}}                            % Lagrangian density

\def\mm{{\mathtt f}}

\begin{document}

\title{Post-Newtonian Celestial Dynamics in Cosmology: Field Equations}
\author{Sergei M. Kopeikin}
\affiliation{Department of Physics \& Astronomy, University of Missouri, 322 Physics Bldg., Columbia, MO 65211, USA}
\email{kopeikins@missouri.edu}
\author{Alexander N. Petrov}
\affiliation{Sternberg Astronomical Institute, \\Moscow M.~V. Lomonosov State University, \\Universitetskii Prospect 13, Moscow 119992, Russia}
\email{alex.petrov55@gmail.com}
\date{\today}
\begin{abstract}
 Post-Newtonian celestial dynamics is a relativistic theory of motion of massive bodies and test particles under the influence of relatively weak gravitational forces. Standard approach for development of this theory relies upon the key concept of the isolated astronomical system supplemented by  the assumption that the background space-time is flat. The standard post-Newtonian theory of motion was instrumental in explanation of the existing experimental data on binary pulsars, satellite and lunar laser ranging, and in building precise ephemerides of planets in the solar system. Recent studies of the formation of large-scale structure in our universe indicate that the standard post-Newtonian mechanics fails to describe more subtle dynamical effects in motion of the bodies comprising the astronomical systems of larger size  - galaxies and clusters of galaxies - where the Riemann curvature of the expanding FLRW universe interacts with the local gravitational field of the astronomical system and, as such, can not be ignored.

 The present paper outlines theoretical principles of the post-Newtonian mechanics in the expanding universe. It is based upon the gauge-invariant theory of the Lagrangian perturbations of cosmological manifold caused by an isolated astronomical N-body system  (the solar system, a binary star, a galaxy, a cluster of galaxies). We postulate that the geometric properties of the background manifold are described by a homogeneous and isotropic Friedman-Lema\^itre-Robertson-Walker (FLRW) metric governed by two primary components - the dark matter and the dark energy. The dark matter is treated as an ideal fluid with the Lagrangian taken in the form of pressure along with the scalar Clebsch potential as a dynamic variable. The dark energy is associated with a single scalar field with a potential which is hold unspecified as long as the theory permits. Both the Lagrangians of the dark matter and the scalar field are formulated in terms of the field variables which play a role of generalized coordinates in the Lagrangian formalism. It allows us to implement the powerful methods of variational calculus to derive the gauge-invariant field equations of the post-Newtonian celestial mechanics of an isolated astronomical system in an expanding universe. These equations generalize the field equations of the post-Newtonian theory in asymptotically-flat spacetime by taking into account the cosmological effects explicitly and in a self-consistent manner without assuming the principle of liner superposition of the fields or a vacuole model of the isolated system, etc. The field equations for matter dynamic variables and gravitational field perturbations are coupled in the most general case of arbitrary equation of state of matter of the background universe. 
 
 We introduce a new cosmological gauge which generalizes the de Donder (harmonic) gauge of the post-Newtonian theory in asymptotically flat spacetime. This gauge significantly simplifies the gravitational field equations and allows to find out the approximations where the field equations can be fully decoupled and solved analytically. The residual gauge freedom is explored and the residula gauge transformations are formulated in the form of the wave equations for the gauge functions. We demonstrate how the cosmological effects interfere with the local system and affect the local distribution of matter of the isolated system and its orbital dynamics. Finally, we worked out the precise mathematical definition of the Newtonian limit for an isolated system residing on the cosmological manifold. The results of the present paper can be useful in the solar system for calculating more precise ephemerides of the solar system bodies on extremely long time intervals, in galactic astronomy to study the dynamics of clusters of galaxies, and in gravitational wave astronomy for discussing the impact of cosmology on generation and propagation of gravitational waves emitted by coalescing binaries and/or merging galactic nuclei. 
\end{abstract}
\maketitle
%\tableofcontents
\section{Notations}\la{notat}
This section summarizes notations used in the present paper.
We use $G$ to denote the universal gravitational constant and $c$ for the ultimate speed in Minkowski spacetime. Every time, when there is no confusion about the system of units, we shall choose a geometrized system of units such that $G=c=1$.
We put a bar over any function that belongs to the background manifold of the FLRW cosmological model. Any function without such a bar belongs to the perturbed manifold.

The notations used in the present paper are as follows:
\begin{itemize}
\item Greek indices $\a,\b,\g,\ldots$ run through values $0,1,2,3$, and Roman indices $i,j,k,\ldots$ take values $1,2,3$,
\item Einstein summation rule is applied for repeated (dummy) indices, for example,  $P^\a Q_\a\equiv P^0 Q_0+P^1 Q_1+P^2 Q_2 + P^3 Q_3$, and $P^i Q_i\equiv P^1 Q_1+P^2 Q_2 + P^3 Q_3$,
\item $g_{\a\b}$ is a full metric on the cosmological spacetime manifold,
\item $\bar g_{\a\b}$ is the FLRW metric on the background spacetime manifold,
\item $\mm_{\a\b}$ is the metric on the conformal spacetime manifold,
\item $\eta_{\a\b}={\rm diag}\{-1,+1,+1,+1\}$ is the Minkowski metric,
\item $T$ and $X^i=\{X,Y,Z\}$ are the coordinate time and isotropic spatial coordinates on the background manifold,
\item $X^\a=\{X^0,X^i\}=\{c\eta,X^i\}$ are the conformal coordinates with $\eta$ being a conformal time,
\item $x^\a=\{x^0,x^i\}=\{ct,x^i\}$ is an arbitrary coordinate chart on the background manifold,
\item a bar, $\bar F$ above a geometric object $F$, denotes the unperturbed value of $F$ on the background manifold,
\item a prime $F'=dF/d\eta$ denotes a total derivative with respect to the conformal time $\eta$,
\item a dot $\dot F=dF/d\eta$ denotes a total derivative with respect to the cosmic time $T$,
\item $\pd_\a=\pd/\pd x^\a$ is a partial derivative with respect to the coordinate $x^\a$,
\item a comma with a following index $F_{,\a}=\pd_\a F$ is another designation of a partial derivative with respect to a coordinate $x^\a$,
\item a vertical bar, $F_{|\a}$ denotes a covariant derivative of a geometric object $F$ (a scalar, a vector, a tensor) with respect to the background metric $\bar g_{\a\b}$,
\item a semicolon, $F_{;\a}$ denotes a covariant derivative of a geometric object $F$ (a scalar, a vector, a tensor) with respect to the conformal metric $\mm_{\a\b}$,
\item the tensor indices of geometric objects on the background manifold are raised and lowered with the background metric $\bar g_{\a\b}$,
\item the tensor indices of geometric objects on the conformal spacetime are raised and lowered with the conformal metric $\mm_{\a\b}$,
\item the scale factor of the FLRW metric is denoted as $R=R(T)$, or as $a=a(\eta)=R[T(\eta)]$,
\item the Hubble parameter, $\H=\dot R/R$, and the conformal Hubble parameter, ${\cal H}= a'/a$.
\end{itemize}
Other notations will be introduced and explained in the main text of the paper.

\section{Introduction}\la{intda}

Post-Newtonian celestial mechanics is a branch of fundamental gravitational physics  \citep{1989racm.book.....S, 1991ercm.book.....B,2011rcms.book.....K} that deals with the theoretical concepts and experimental methods of measuring gravitational fields and testing general theory of relativity both in the solar system and beyond \citep{2006LRR.....9....3W,2010AIPC.1256....3T}. In particular, the relativistic celestial mechanics of binary pulsars (see \citep{2005hpa..book.....L}, and references therein) was instrumental in providing conclusive evidence for the existence of gravitational radiation as predicted by Einstein's theory \citep{2010IAUS..261..218S,2010ApJ...722.1030W}.

Over the last few decades, various groups within the International Astronomical Union (IAU) have been active in exploring the application of the general theory of relativity to the modeling and interpretation of high-accuracy astrometric observations in the solar system and beyond. A Working Group on Relativity in Celestial Mechanics and Astrometry was formed in 1994 to define and implement a relativistic theory of reference frames and time scales. This task was successfully completed with the adoption of a series of resolutions on astronomical reference systems, time scales, and Earth rotation models by 24th General Assembly of the IAU, held in Manchester, UK, in 2000. The IAU resolutions are based on the first post-Newtonian approximation of general relativity which is a conceptual basis of the fundamental astronomy in the solar system \citep{2003AJ....126.2687S}.

The mathematical formalism of the post-Newtonian approximations is getting progressively complicated as one goes from the Newtonian to higher orders \citep{1987thyg.book..128D,2011mmgr.book..167S}. For this reason the theory has been primarily developed under a basic assumption that the background spacetime is asymptotically flat. Mathematically, it means that the full spacetime metric, $g_{\a\b}$, is decomposed around the background Minkowskian metric, $\eta_{\a\b}={\rm diag}(-1,1,1,1)$, into a linear combination
\be\la{i1}
g_{\a\b}=\eta_{\a\b}+h_{\a\b}\;,
\ee
where the perturbation,
\be\la{zz1}
h_{\a\b}=c^{-2}h^{[2]}_{\a\b}+c^{-3}h^{[3]}_{\a\b}+c^{-4}h^{[4]}_{\a\b}+\ldots\;,
\ee
is the post-Newtonian series with respect to the powers of $1/c$, where $c$ is the ultimate speed in general relativity (equal to the speed of light). Post-Newtonian approximations is the method to determine $h_{\a\b}$ by solving Einstein's field equations with the tensor of energy-momentum  of matter of a localized astronomical system, $\mathfrak{T}_{\a\b}(\Phi,h_{\a\b})$, taken as a source of the gravitational field $h_{\a\b}$, by iterations starting from $h_{\a\b}=0$ in the expression for $\mathfrak{T}_{\a\b}$. The solution of the field equations and the equations of motion of the astronomical bodies are derived in some coordinates $r^\a=\{ct,{\bm r}\}$ where $t$ is the coordinate time, and ${\bm r}=\{x,y,z\}$ are spatial coordinates. The post-Newtonian theory in asymptotically-flat spacetime has a well-defined Newtonian limit determined by:
\begin{enumerate}
\item[1)] equation for the Newtonian potential, $U=h^{[2]}_{00}/2$,
\be\la{z1}
U(t,{\bm r})=\int_{\cal V}\frac{\rho(t,{\bm r}')d^3r'}{|{\bm r}-{\bm r}'|}\;,
\ee
where $\rho=c^{-2}\mathfrak{T}_{00}$, is the density of matter  producing the gravitational field,
\item[2)] equation of motion for massive particles
\be\la{z2}
\ddot{\bm r}={\bm\nabla}U\;,
\ee
where ${\bm\nabla}=\{\pd_x,\pd_y,\pd_z\}$ is the operator of gradient, ${\bm r}={\bm r}(t)$ is time-dependent position of a particle (worldline of the particle), and the dot denotes a total derivative with respect to time $t$,
\item[3)] equations of motion for light (massless particles)
\be\la{z3}
\ddot{\bm r}=0\;.
\ee
\end{enumerate}
These equations are considered as fundamentals for creation of astronomical ephemerides of celestial bodies in the solar system \citep{2011rcms.book.....K} and in any other localized system of self-gravitating bodies like a binary pulsar \citep{2005hpa..book.....L}. In all practical cases they have to be extended to take into account the post-Newtonian corrections sometimes up to the 3-d post-Newtonian order of magnitude \citep{2011PNAS..108.5938W}. It is important to notice that in the Newtonian limit the coordinate time $t$ of the gravitational equations of motion (\ref{z2}), (\ref{z3}) coincides with the proper time of observer $\t$ that is practically measured with an atomic clock. The formalism of the present paper has been employed in \citep{2012PhRvD..86f4004K} to check the theoretical consistency of (\ref{z1})--(\ref{z3}) and to analyse the outcome of some experiments like the anomalous Doppler effect discovered by J. Anderson et al \citep{2002PhRvD..65h2004A,2010IAUS..261..189A} in the orbital motion of Pioneer 10 and 11 space probes. 

So far, the post-Newtonian theory was mathematically successful (''unreasonably effective'' as Clifford Will states \citep{2011PNAS..108.5938W}) and passed through numerous experimental tests with a flying color. Nevertheless, it hides several pitfalls. The first one is the problem of convergence of the post-Newtonian series and regularization of divergent integrals that appear in the post-Newtonian calculations at higher post-Newtonian orders \citep{2011mmgr.book..167S}.
The second problem is that the background manifold is not asymptotically-flat Minkowskian spacetime but the FLRW metric, $\bar g_{\a\b}$. We live in the expanding universe which rate of expansion is determined by the Hubble constant $\H_0$. Therefore, the right thing would be to replace the post-Newtonian decomposition (\ref{i1}) with a more adequate post-Friedmannian series \citep{2002PhRvD..66j3507T}
\be\la{i2}
g_{\a\b}=\bar g_{\a\b}+\varkappa_{\a\b}\;,
\ee
where
\be\la{zz2}
\varkappa_{\a\b}=\varkappa^{\{0\}}_{\a\b}+\H \varkappa^{\{1\}}_{\a\b}+\H^2 \varkappa^{\{2\}}_{\a\b}+\ldots\;,
\ee
is the metric perturbation around the cosmological background represented as a series with respect to the Hubble parameter, $\H$. Each term of the series has its own expansion into post-Newtonian series like (\ref{zz1}).
Generalization of the theory of post-Newtonian approximations from the Minkowski spacetime to that of the expanding universe is important for extending the applicability of the post-Newtonian celestial mechanics to testing cosmological effects and for more deep understanding of the process of formation of the large-scale structure in the universe and gravitational interaction between galaxies and clusters of galaxies.

Whether cosmological expansion affects gravitational dynamics of bodies inside a localized astronomical system was a matter of considerable efforts of many researchers \citep{1945RvMP...17..120E,1946RvMP...18..148E,1954ZPhy..137..595S,2000CQGra..17.2739B,1933MNRAS..93..325M,2004CeMDA..90..267K,2007CQGra..24.5031M}. A recent article \citep{2010RvMP...82..169C} summarizes the previous results and provides the reader with a number of other valuable resources. Most of the previous works on celestial mechanics in cosmology were based on assumption of spherical symmetry of gravitational field and matching two (for example, Schwarzschild and Friedmann) exact solutions of Einstein's equations. The matching was achieved in many different ways. McVittie's solution \citep{1933MNRAS..93..325M} is perhaps the most successful mathematically but yet lacks a clear physical interpretation \citep{2010RvMP...82..169C}. Moreover, its practical application is doubtful since it is valid only for spherically-symmetric case.

We need a precise mathematical formulation of the post-Newtonian theory for a self-gravitating localized astronomical system not limited by the assumption of the spherical symmetry, embedded to the expanding universe and coupled through the gravitational interaction with the time-dependent background geometry. Theoretical description of the post-Newtonian theory for a localized astronomical system in expanding universe should correspond in the limit of vanishing $\H$ to the post-Newtonian theory obtained in the asymptotically-flat spacetime. Such a description will allow us to directly compare the equations of the standard post-Newtonian celestial mechanics with its cosmological counterpart. Therefore, the task is to derive a set of the post-Newtonian equations in cosmology in some coordinates introduced on the background manifold, and to map them onto the set of the Newtonian equations (\ref{z1})--(\ref{z3}) in asymptotically-flat spacetime. Such a theory of the post-Newtonian celestial mechanics would be of a paramount importance for extending the tools of experimental gravitational physics to the field of cosmology, for example, to properly formulate the cosmological extension of the PPN formalism \citep{1993tegp.book.....W}. The present article discusses the main ideas and principal results of such a theoretical approach in the linearized approximation with respect to the gravitational perturbations of the cosmological background caused by the presence of a localized astronomical system.

The paper is organized as follows. In the following section we describe a brief history of the development of the theory of cosmological perturbations. Section \ref{act} describes the Lagrangian of gravitational field, the matter of the background cosmological model, and an isolated astronomical system which perturbs the background cosmological manifold. Section \ref{bma} describes the geometric structure of the background spacetime manifold of the cosmological model and the corresponding equations of motion of the matter and field variables. Section \ref{lpsm} introduces the reader to the theory of the Lagrangian perturbations of the cosmological manifold and the dynamic variables. Section \ref{gife3} makes use of the preceding sections in order to derive the field equations in the gauge-invariant form. Beginning from section \ref{pnfe} we focus on the spatially-flat universe in order to derive the post-Newtonian field equations that generalize the post-Newtonian equations in the asymptotically-flat spacetime. These equations are coupled in the scalar sector of the proposed theory. Therefore, we consider in section \ref{dspn} a few particular cases when the equations can be fully decoupled one from another, and solved in terms of the retarded potentials. Appendix provides a proof of the Lorentz-invariance of the retarded potentials for the wave equations describing propagation of weak gravitational and sound waves on the background cosmological manifold.

\section{Brief History of Cosmological Perturbation Theory}\la{bhcper}

In order to solve the problem of the interaction of the gravitational field of an isolated astronomical system with expanding universe one has to resort to the theory of cosmological perturbations. The immediate goal of cosmological perturbation theory is to relate the physics of the early universe to CMB anisotropy and large-scale structure and to provide the initial conditions for numerical simulations of structure formations. The ultimate goal of this theory is to establish a mathematical link between the fundamental physical laws at the Planck epoch and the output of the gravitational wave detectors which are the only experimental devices being able to map parameters of the universe at that time \citep{2008RvMA...20..140K}.

Originally, two basic approximation schemes for calculation of cosmological perturbations have been proposed by Lifshitz with his collaborators \citep{lif,1963AdPhy..12..185L} and, later on, by Bardeen \citep{1980PhRvD..22.1882B}. Lifshitz \citep{lif} worked out a coordinate-dependent theory of cosmological perturbations while Bardeen \citep{1980PhRvD..22.1882B} concentrated on finding the gauge-invariant combinations for perturbed quantities and derivation of a perturbation technique based on gauge-invariant field equations. At the same time, \citet{1980JETPL..31..596L} had suggested an original approach for deriving the gauge-invariant scalar equations based on the thermodynamic theory of the Clebsch potential also known in cosmology as the scalar velocity potential \citep{1959flme.book.....L,1970PhRvD...2.2762S} or the Taub potential \citep{1967rta1.book..170T,1970PhRvD...2.2762S}. It turns out that the variational principle with a Lagrangian of cosmological matter formulated in terms of the Clebsch potential, is the most useful mathematical device for developing the theory of relativistic celestial mechanics of localized astronomical systems embedded in expanding cosmological manifold. It is for this reason, we use the Clebsch potential in the present paper.

A few words of clarification regarding a comparison between Lifshitz's and Bardeen's  approaches should be relayed to the reader. The approach established in the papers by Lifshitz \citep{lif,1963AdPhy..12..185L}, is fully correct. Lifshitz decided to work in synchronous gauge and realized that this fixing of coordinates allows for a residual gauge ambiguity, which can also be fixed by picking a synchronous, comoving coordinate system. Bardeen \citep{1980PhRvD..22.1882B} wrote his paper, because there was some confusion in the 1970s about that issue (which could have beed avoided if people would have studied Lifshitz's papers carefully). He demonstrated, that any coordinate could be chosen and that there exist quantities which are independent of that choice, which he identified with the physical degrees of freedom. However, this is not where the story of the cosmological perturbation theory ends. 
Closer inspection shows that what is really relevant is not the choice of coordinates (which do not have a physical meaning), rather the choice of spacetime foliation is relevant, for example, it makes a physical difference if one defines the Harrison-Zel'dovich spectrum \citep{Harrison_1970PhRvD,Zeldovich_1972MNRAS} of the primordial perturbations on a synchronous, comoving hypersurface or a shear free hypersurface. Pitfalls in understanding this issue are subtle and, sometimes, may be not easily recognized (see \citep{grish94,Der_Mukh_1995PhRvD,1998PhRvD..57.3302M} and \citep{Gourgoulhon_2012LNP} for a detailed discussion of the role of foliations in cosmology and in general relativity).

In the years that followed, the gauge-invariant formalism was refined and improved by Durrer and Straumann \citep{durstr,1989A&A...208....1D}, Ellis et al. \citep{ellis1,ellis2,ellis3} and, especially, by Mukhanov et al. \citep{mukh,2005pfc..book.....M}. Irrespectively of the approach a specific gauge must be fixed in order to solve equations for cosmological perturbations. Any gauge is allowed and its particular choice is simply a matter of convenience. Imposing a gauge condition eliminates four gauge degrees of freedom in the cosmological pertrubations and brings the differential equations for them to a solvable form. Nonetheless, the residual gauge freedom originated from the tensor nature of the gravitational field remains. This residual gauge freedom leads to appearance of unphysical perturbations which must be disentangled from the physical modes. Lifshitz theory of cosmological perturbations \citep{lif,1963AdPhy..12..185L} is worked out in a synchronous gauge and contains the spurious modes but they are easily isolated from the physical perturbations \citep{1974JETP...40..409G}. Other gauges used in cosmology are described in Bardeen's paper \citep{1980PhRvD..22.1882B} and used in cosmological perturbation theory. Among them, the longitudinal  (conformal or Newtonian) gauge is one of the most common. This gauge is advocated by Mukhanov \citep{2005pfc..book.....M} because it removes spurious coordinate degrees of freedom out of scalar perturbations.
Detailed comparison of the cosmological perturbation theory in the synchronous and conformal gauges was given by Ma and Bertschinger \citep{maber}.

Unfortunately, none of the previously known cosmological gauges can be applied for analysis of the cosmological perturbations caused by localized matter distributions like an isolated astronomical system which can be a single star, a planetary system, a galaxy, or even a cluster of galaxies. The reason is that the synchronous gauge has no the Newtonian limit and is applicable only for freely falling test particles while the longitudinal gauge separates the scalar, vector and tensor modes present in the metric tensor perturbation in the way that is incompatible with the technique of the approximation schemes having been worked out in asymptotically flat space-time \citep{2011rcms.book.....K}. We also notice that standard cosmological perturbation technique often operates with harmonic (Fourier) decomposition of both the metric tensor and matter perturbations when one is interested in statistical statements based on the cosmological principle. This technique is unsuitable and must be avoided in sub-horizon approximation for working out the post-Newtonian celestial mechanics of self-gravitating isolated systems. Current paradigm is that the cosmological generalization of the Newtonian field equations of an isolated gravitating system like the solar system or a galaxy or a cluster of galaxies can be easily obtained by just making use of the linear principle of superposition with a simple algebraic addition of the local system to the tensor of energy momentum of the background matter. It is assumed that the superposition procedure is equivalent to operating with the Newtonian equations of motion derived in asymptotically-flat spacetime and adding to them (''by hands'') the tidal force due to the presence of the external universe (see, for example, \citep{2007CQGra..24.5031M}). Though such a procedure may look pretty obvious it lacks a rigorous mathematical analysis of the perturbations induced on the background cosmological manifold by the local system. This analysis should be done in the way that embeds cosmological variables to the field equations of standard post-Newtonian approximations not by ``hands'' but by precise mathematical technique which is the goal of the present paper. The variational calculus on manifolds is the most convenient for joining the standard theory of cosmological perturbations with the post-Newtonian approximations in asymptotically-flat spacetime. It allows us to track down the rich interplay between the perturbations of the background manifold with the dynamic variables of the local system which cause these perturbations. The output is the system of the post-Newtonian field equations with the cosmological effects incorporated to them in a physically-transparent and mathematically-rigorous way. This system can be used to solve a variety of physical problems starting from celestial mechanics of localised systems in cosmology to gravitational wave astronomy in expanding universe that can be useful for deeper exploration on scientific capability of such missions as LISA and Big Bang Observer (BBO) \citep{BBO_2005PhRvD} 

In fact, the problem of whether the cosmological expansion affects the long-term evolution of an isolated N-body system (galaxy, solar system, binary system, etc.) has a long controversial history. The reason is that there was no an adequate mathematical formalism for describing cosmological perturbations caused by the isolated system so that different authors have arrived to opposite opinions. It seems that McVittie \citep{1933MNRAS..93..325M} was first who had considered the influence of the expansion of the universe on the dynamics of test particles orbiting around a massive point-like body immersed to the cosmological background. He found an exact solution of the Einstein equations in his model which assumed that the  mass of the central body is not constant but decreases as the universe expands. Einstein and Straus \citep{1945RvMP...17..120E,1946RvMP...18..148E} suggested a different approach to discuss motion of particles in gravitationally self-interacting systems residing on the expanding background. They showed that a Schwarzschild solution could be smoothly matched to the Friedman universe on a spherical surface separating the two solutions. Inside the surace ("vacuole") the motion of the test particles is totally unaffected by the expansion. Thus, Einstein and Straus \citep{1945RvMP...17..120E,1946RvMP...18..148E} concluded that the cosmic expansion is irrelevant for the Solar system. Bonnor \citep{2000CQGra..17.2739B} generalized the Einstein-Straus vacuole and matched the Schwarzschild region to the inhomogeneous Lema\^itre-Tolman-Bondi model thus, making the average energy density inside the vacuole be independent of the exterior energy density while in the Einstein-Straus model they must be equal. Bonnor \citep{2000CQGra..17.2739B} concluded that the local systems expand but at a rate which is negligible compared with the general cosmic expansion. Similar conclusion was reached by Mashhoon et al. \citep{2007CQGra..24.5031M} who analysed the tidal dynamics of test particles in the Fermi coordinates.

The vacuole solutions are not appropriate for adequate physical solution of the N-body problem in the expanding universe. There are several reasons for it. First, the vacuole is spherically-symmetric while majority of real astronomical systems are not. Second, the vacuole solution imposes physically unrealistic boundary conditions on the matching surface that relates the central mass to the size of the surface and to the cosmic energy density. Third, the vacuole is unstable against small perturbations. In order to overcome these difficulties a realistic approach based on the approximate analytic solution of the Einstein equations for the N-body problem immersed to the cosmological background, is required.
In the case of a flat space-time there are two the most advanced techniques for finding approximate solution of the Einstein equations describing gravitational field of an isolated astronomical system. The first is called the post-Newtonian approximations and we have briefly discussed this technique in the introduction. The second technique is called post-Minkowskian approximations \citep{1987thyg.book..128D}. The post-Newtonian approximations is applicable to the systems with weak gravitational field and slow motion of matter. The post-Minkowskian approximations also assume that the field is weak but does not imply any limitation on the speed of matter. The post-Newtonian iterations are based on solving the elliptic-type Poisson equations while the post-Minkowskain approach operates with the hyperbolic-type Dalambert equations. The post-Minkowskian approximations naturally includes description of the gravitational radiation emitted by the isolated system while the post-Newtonian scheme has to use additional mathematical methods to describe generation of the gravitational waves \citep{1970ApJ...160..153C}. In the present paper we concentrate on the development of a generic scheme for calculation of cosmological perturbations caused by a localized distribution of matter which preserves many advantages of the
post-Minkowskian approximation scheme. The cosmological post-Newtonian approximations are derived from the general perturbation scheme by making use of the slow-motion expansion with respect to a small parameter $v/c$ where $v$ is the characteristic velocity of matter in the N-body system and $c$ is the fundamental speed.

There were several attempts to work out a physically-adequate and mathematically-rigorous approximation schemes in general relativity in order to construct and to adequately describe small-scale self-gravitating structures in the universe. The most notable works in this direction have been done by Kurskov and Ozernoi \citep{1976SvA....19..569K}, Futamase et al. \citep{1988PhRvL..61.2175F,1989MNRAS.237..187F,1991GReGr..23.1251B}, Buchert and Ehlers \citep{1993MNRAS.264..375B,1997GReGr..29..733E}, Mukhanov et al. \citep{mukh,2005pfc..book.....M,mukh2,mukh3}, Zalaletdinov \citep{2008IJMPA..23.1173Z}. These approximation schemes have been designed to track the temporal evolution of the cosmological perturbations from a very large down to a small scale up to the epoch when the perturbation becomes isolated from the expanding cosmological background. These approaches looked hardly connected between each other until recent works by Clarkson et al \citep{2011CQGra..28v5002C,2012CQGra..29g9601C}, Li \& Schwarz \citep{2007PhRvD..76h3011L,2008PhRvD..78h3531L}, R{\"a}s{\"a}nen \citep{2010PhRvD..81j3512R}, Buchert \& R{\"a}s{\"a}nen \citep{2012ARNPS..62...57B} and Wiegand \& Schwarz \citep{2012A&A...538A.147W}. In particular, Wiegand \& Schwarz \citep{2012A&A...538A.147W} have shown that the idea of cosmic variance (that isa standard way of thinking) is closely related to the cosmic averages defined by Buchert and Ehlers \citep{1993MNRAS.264..375B,1997GReGr..29..733E}. All researchers agree that the second and higher-order non-linear approximations are important to understand the back-reaction of the cosmological perturbations propagating on the cosmological background used in the linearized theory (see, for example, \citep{1988PhRvL..61.2175F,mukh,mukh3,2008IJMPA..23.1173Z,1968PhRv..166.1263I,1968PhRv..166.1272I}).

Papers \citep{1999PhRvD..60d4007D,2001PhLA..292..173K,2002PhLB..532....1R} attempted to construct an approximation scheme being compatible with both the post-Newtonian and post-Minkowskian approximations in asymptotically-flat space-time and the gauge-invariant theory of cosmological perturbations caused by a localized astronomical system. We have succeeded in solving this problem in the work \citep{2001PhLA..292..173K} which assumes the dust-dominated background universe with spatial curvature $k=0$. We remind that in standard Bardeen's approach \citep{1980PhRvD..22.1882B} the metric tensor perturbations $h_{\alpha\beta}$ are decomposed in irreducible scalar, vector, and tensor parts which are combined with themselves and with matter perturbations in order to obtain some gauge-invariant quantities that do not contain spurious modes invoked by the freedom in doing coordinate transformations on cosmological background manifold. Bardeen \citep{1980PhRvD..22.1882B} reformulates Einstein's field equations in terms of these gauge-invariant variables which are further decomposed in Fourier harmonics. The field equations become then of the Helmholtz type and are solved by constructing Green's function. This specific procedure of Bardeen's approach is incompatible with the post-Newtonian or post-Minkowskian approximations which do not decompose the metric tensor in scalar, vector, and tensor harmonics and do not expand them to the Fourier series. Therefore, we have used a different procedure based on introduction of auxiliary scalar, $\phi$, and vector, $\zeta$, fields which are used along with the metric tensor perturbation $h_{\alpha\beta}$ as basic elements for decomposing the perturbed stress-energy tensor of matter $\delta T_{\alpha\beta}$ and selecting from this decomposition that part of $\delta T_{\alpha\beta}$ which has the same transformation property as the perturbed Einstein tensor $\delta G_{\alpha\beta}$. This process makes the rest of the perturbation of $\delta T_{\alpha\beta}$ gauge-invariant so that it can be identified with the {\it bare} (external) perturbation imposed on the cosmological background by the presence of a material system like a single star, solar system, galaxy, etc. The auxiliary scalar, $\phi$, and vector, $\zeta$, fields are also determined by this procedure. For example, it is proved out \citep{2001PhLA..292..173K} that the vector field $\zeta$ is identically equal to zero while the scalar field $\phi$ is found from the equation following from the Bianchi identity of the perturbed Einstein equations. The entire approach is gauge-invariant but the equations for the scalar and metric perturbations are strongly coupled in general case. We have shown that there is a special cosmological gauge generalizing the harmonic gauge of general relativity in such a way that the reduced field equations are completely decoupled and significantly simplified. More specifically, the linearized Einstein equations for the metric tensor perturbations $h_{0i}$ and $h_{ij}$ are decoupled both from each other and from $h_{00}$ component which couples only with the auxiliary scalar field $\phi$. However, it turns out that our gauge \citep{2001PhLA..292..173K} admits a simple linear combination, $\chi$, of $h_{00}$, $h_{kk}$, and $\phi$ such that the equation for $\chi$ decouples from any other perturbation. The equations for $\chi$, $h_{0i}$ and $h_{ij}$ are of a wave-type and have the {\it bare} stress energy-tensor of matter as a source in their right-hand sides. These equations have simple Green functions given in terms of the retarded integrals with the Minkowskian null cone defined by the conformally-flat part of the FLRW metric.

In the work \citep{2002PhLB..532....1R} this linearized approach
has been extended to the background cosmological models governed by a perfect fluid with the barotropic equation of state $\bar p=q\bar\epsilon$, where $\bar p$ and $\bar\epsilon$ are pressure and energy density of the background matter respectively, and $q$ is a constant parameter taking value in the range from -1 to +1. We have shown \citep{2002PhLB..532....1R} that the overall perturbative scheme for calculation of the cosmological perturbations in such model can be significantly streamlined and simplified if one formally replaces the stress-energy tensor of the perfect fluid with one of a classic scalar field minimally coupled with metric. A specific (exponential) form of the potential $V(\Phi)$ of the scalar field $\Phi$ is fixed by two conditions:
\begin{enumerate}
\item it reproduces all functional relationships of the background FLRW cosmological model;
\item it maintains the background equation of state $\bar p=q\bar\epsilon$.
\end{enumerate}
Although a minimally coupled scalar field can be viewed to a certain extent as a perfect fluid one should keep in mind that its barotropic equation of state does not hold generally in the perturbed universe (see \citep{1992CQGra...9..921B} and discussion in \citep{grish94,1998PhRvD..57.3302M}). This may impose some technical difficulties in handling the mathematical analysis of cosmological perturbations caused by localized distributions of matter. We explain how to get around these difficulties by making use of the Lagrangian-based variational technique of the gauge-invariant perturbations of curved manifolds. This makes our perturbative approach \citep{2001PhLA..292..173K,2002PhLB..532....1R} more efficient in developing the post-Newtonian celestial mechanics in cosmology as compared with the standard technique \citep{mukh,2005pfc..book.....M,lif,1963AdPhy..12..185L,1980PhRvD..22.1882B,durstr,1989A&A...208....1D,ellis1,ellis2,ellis3,maber}.

Development of observational cosmology and gravitational wave astronomy demands to
extend the linearized theory of cosmological perturbations to second and higher orders of approximation. A fair number of works have been devoted to solving this problem. Non-linear perturbations of the metric tensor and matter affect evolution of the universe and this back-reaction of the perturbations should be taken into account. This requires derivation of the effective stress-energy tensor for cosmological perturbations like freely-propagating gravitational waves and scalar field \citep{mukh,2005pfc..book.....M,mukh2,mukh3}. The laws of conservation for the effective stress-energy tensor are important for better understanding of physics of the expanding universe \citep{2000PhRvD..61b4038B,2009SSRv..148..315G}.

In the present paper we construct a non-linear theory of cosmological perturbations for isolated systems which generalizes the post-Minkowskian approximation scheme for calculation perturbations of gravitational field in asymptotically flat space-time. We rely upon the basic results of the linearized gauge-invariant theory from our previous works \citep{2001PhLA..292..173K,2002PhLB..532....1R} in order to derive a decoupled system of equations for
quadratic cosmological perturbations of a spatially flat FLRW universe. We implement the Lagrangian-based theory of dynamical perturbations of gravitational field on a curved background manifold which has been worked out in \citep{1984CMaPh..94..379G,1988IJMPA...3.2651P} (see also \citep{2000PhRvD..61b4038B}). This theory has a number of specific advantages over other perturbation methods among which the most important are:
\begin{itemize}
\item Lagrangian-based approach is covariant and can be implemented for any curved
background spacetime that is an exact solution of the Einstein gravity field equations;
\item the system of the partial differential equations describing dynamics of the perturbations is determined by a dynamic Lagrangian $\lag^D$ which is derived from the total Lagrangian $\lag$ by making use of its Taylor expansion with respect to the perturbations and accounting for the background field equations. The dynamic Lagrangian $\lag^D$ defines the conserved quantities for the perturbations (energy, angular momentum, etc.) that depend on the symmetries of the background manifold;
\item the dynamic Lagrangian $\lag^D$ and the corresponding field equations for the perturbations are gauge-invariant in any order of the perturbation theory. Gauge transformations map the background manifold onto itself and are associated with arbitrary (analytic) coordinate transformations on the background space-time;
\item the entire perturbation theory is self-reproductive and is extended to the next perturbative order out of a previous iteration so that the linearized approximation is the basic starting point.
\end{itemize}

\section{Lagrangian and Field Variables}\la{act}

We accept the Einstein's theory of general relativity and consider a universe filled up with matter consisting of three components. The first two components are: (1) an ideal fluid composed of particles of one type with transmutations excluded; (2) a scalar field; and (3) a matter of the localized astronomical system. The ideal fluid consists of baryons and cold dark matter, while the scalar field describes dark energy \citep{2010deto.book.....A}.
We assume that these two components do not interact with each other directly, and are the source of the {\it Friedmann-Lem\^itre-Robertson-Walker} (FLRW) geometry. There is no dissipation in the ideal fluid and in the scalar field so that they can only interact through the gravitational field. It means that the equations of motion for the fluid and the scalar field are decoupled, and we can calculate their evolution separately. In other words, the Lagrangian of the ideal fluid and that of the scalar field depend only on their own field variables. 

The tensor of energy-momentum of matter of the localized astronomical system is not specified in agreement with the approach adopted in the post-Newtonian approximation scheme developed in the asymptotically-flat spacetime \citep{1987thyg.book..128D,2004PhR...400..209K}. This allows us to generate all possible types of cosmological perturbations: scalar, vector and tensor modes. We are the most interested in developing our formalism for application to the astronomical system of massive bodies bound together by intrinsic gravitational forces like the solar system, galaxy, or a cluster of galaxies. It means that our approach admits a large density contrast between the background matter and the matter of the localized system. The localized system perturbs the background matter and gravitational field of FLRW universe locally but it is not included to the matter source of the background geometry, at least, in the approximation being linearized with respect to the metric tensor perturbation. Our goal is to study how the perturbations of the background matter and gravitational field are incorporated to the gravitational field perturbations of the standard post-Newtonian theory of relativistic celestial mechanics.   

Let us now consider the action functional and the Lagrangian of each component.

\subsection{The Action Functional}\la{acfu}

We shall consider a theory with the action functional 
\be\la{qa1}
S=\int_{\cal M} \lag d^4x\;,
\ee
where the integration is performed over the entire spacetime manifold ${\cal M}$.
The Lagrangian $\lag$ is comprised of four terms
\be\la{qa2}
\lag=\lag^{\rm g}+\lag^{\rm m} +\lag^{\rm q}+\lag^{\rm p}\;,
\ee
where $\lag^{\rm g}$, $\lag^{\rm m}$, $\lag^{\rm q}$ are the Lagrangians of gravitational field, the dark matter, the scalar field that governs the accelerated expansion of the universe \citep{2009AIPC.1166...53G}, and $\lag^{\rm p}$ is the Lagrangian describing the source of the cosmological perturbations. Gravitational field Lagrangian is
\be\la{hl1}
\lag^{\rm g}=-\frac{1}{16\pi}\sqrt{-g}R\;,
\ee
where $R$ is the Ricci scalar built of the metric $g_{\a\b}$ and its first and second derivatives \citep{mtw}. Other Lagrangians depend on the metric and 
the matter variables. Correct choice of the matter variables is a key element in the development of the Lagrangian theory of the post-Newtonian perturbations of the cosmological manifold caused by a localized astronomical system.

\subsection{The Lagrangian of the Ideal Fluid}

The ideal fluid is characterized by the following thermodynamic parameters: the rest-mass density ${\r_{\rm m}}$, the specific internal energy $\Pi_{\rm m}$ (per unit of mass), pressure $p_{\rm m}$, and entropy $s_{\rm m}$ where the sub-index 'm' stands for 'matter'. We shall assume that the entropy of the ideal fluid remains constant, that excludes it from further consideration. The standard approach to the theory of cosmological perturbations preassumes that the constant entropy excludes rotational (vector) perturbations of the fluid component from the start, and only scalar (adiabatic) perturbations are generated \citep{1972gcpa.book.....W,weinberg_2008,2005pfc..book.....M,2010deto.book.....A}. However, the present paper deals with the cosmological perturbations that are generated by a localized astronomical system which is described by its own Lagrangian (see section \ref{laloas}) which is left as general as possible. This leads to the tensor of energy-momentum of the matter of the localized system that incorporates the rotational motion of matter which is the source of the rotational perturbations of the background ideal fluid. This extrapolates the concept of the gravitomagnetic field of the post-Newtonian dynamics of localized systems in the asymptotically-flat spacetime \citep{Ciufolini_1995,1991ercm.book.....B,2011rcms.book.....K} to cosmology. Further details regarding the vector perturbations are given in section \ref{feveper} of the present paper.  

The total energy density of the fluid
\be\la{gug1}
\e_{\rm m}={\r_{\rm m}}(1+\Pi_{\rm m})\;.
\ee
One more thermodynamic parameter is the specific enthalpy of the fluid defined as
\be\la{pf2}
\mu_{\rm m}=\frac{\e_{\rm m}+p_{\rm m}}{{\r_{\rm m}}}=1+\Pi_{\rm m}+\frac{p_{\rm m}}{{\r_{\rm m}}}\;.
\ee
In the most general case, the thermodynamic equation of state of the fluid is given by equation $p_{\rm m}=p_{\rm m}({\r_{\rm m}},\Pi_{\rm m})$, where the specific internal energy $\Pi_{\rm m}$ is related to pressure by the first law of thermodynamics.

Since the entropy has been assumed to be constant, the first law of thermodynamics reads
\be\la{pf1}
d\Pi_{\rm m}+p_{\rm m}d\le(\frac{1}{{\r_{\rm m}}}\ri)=0\;.
\ee
It can be used to derive the following thermodynamic relationships
\ba\la{pf3} dp_{\rm m}&=&{\r_{\rm m}} d\m_{\rm m}\;,\\\la{pf3a} d\e_{\rm m}&=&\mu_{\rm m} d{\r_{\rm m}}\;,
\ea
which means that all thermodynamic quantities are solely functions of the specific enthalpy $\m_{\rm m}$, for example, ${\r_{\rm m}}={\r_{\rm m}}(\m_{\rm m})$, $\Pi_{\rm m}=\Pi_{\rm m}(\m_{\rm m})$, etc. The equation of state is also a function of the variable $\m_{\rm m}$, that is
\be\la{pf4}
p_{\rm m}=p_{\rm m}(\m_{\rm m})\;.
\ee

Derivatives of the thermodynamic quantities with respect to $\m_{\rm m}$ can be calculated by making use of equations (\ref{pf3}) and (\ref{pf3a}), and the definition of the (adiabatic) speed of sound $c_{\rm s}$ of the fluid
\be\la{pf4a}
\frac{\pd p_{\rm m}}{\pd\e_{\rm m}}=\frac{c^2_{\rm s}}{c^2}\;,
\ee
where the partial derivative is taken under a condition that the entropy, $s_{\rm m}$, of the fluid does not change.
Then, the derivatives of the thermodynamic quantities take on the following form
\be\la{pf5}
\frac{\pd p_{\rm m}}{\pd\m_{\rm m}}=\r_{\rm m}\;,\qquad \frac{\pd\e_{\rm m}}{\pd\m_{\rm m}}=\frac{c^2}{c^2_{\rm s}}{\r_{\rm m}}\;,\qquad \frac{\pd{\r_{\rm m}}}{\pd\m_{\rm m}}=\frac{c^2}{c^2_{\rm s}}\frac{{\r_{\rm m}}}{\m_{\rm m}}\;,
\ee
where all partial derivatives are performed under the same condition of constant entropy.

The Lagrangian of the ideal fluid is usually taken in the form of the total energy density, $L^{\rm m}=\sqrt{-g}\e_{\rm m}$ \citep{mtw}. However, this form is less convenient for applying the variational calculus on manifolds. The above thermodynamic relationships and the integration by parts of the action (\ref{qa1}) allows us to recast the Lagrangian $L^{\rm m}=\sqrt{-g}\e_{\rm m}$ to the form of pressure, $L^{\rm m}=-\sqrt{-g}p_{\rm m}$, so that the Lagrangian density becomes (see \citep[pp. 334-335 ]{2011rcms.book.....K} for more detail)
\be\la{pfl1}
\lag^{\rm m}=-\sqrt{-g}p_{\rm m}=\sqrt{-g}\le(\e_{\rm m}-{\r_{\rm m}}\m_{\rm m}\ri)\;.
\ee

Theoretical description of the ideal fluid as a dynamic system on space-time manifold is given the most conveniently in terms of the Clebsch potential, $\p$ which is also called the velocity potential \citep{1970PhRvD...2.2762S}. In the case of a single-component ideal fluid the Clebsch potential is introduced by the following relationship
\be\la{pf6}
\mu_{\rm m} w_\a=-\p_{,\a}\;.
\ee
In fact, equation (\ref{pf6}) is a solution of relativistic equations of motion of the ideal fluid \citep{1959flme.book.....L}.

The Clebsch potential is a primary field variable in the Lagrangian description of the isentropic ideal fluid. The four-velocity is normalized to $w^\a w_\a=g_{\a\b}w^\a w^\b=-1$, so that the specific enthalpy can be expressed in the following form
\be
\la{pf7}
\mu_{\rm m}=\sqrt{-g^{\a\b}\p_{,\a}\p_{,\b}}\;.
\ee
One may also notice that
\be\la{pf7a}
\m_{\rm m}=w^\a\p_{,\a}\;.
\ee
It is important to notice that the Clebsch potential $\p$ has no direct physical meaning as it can be changed to another value $\p\rightarrow\p'=\p+\tilde\p$ such that the gauge function, $\tilde\p$, is constant along the worldlines of the fluid: $w^\a\tilde\p_{,\a}=0$.

In terms of the Clebsch potential the Lagrangian (\ref{pfl1}) of the ideal fluid is
\be\la{pzq1}
\lag^{\rm m}=\sqrt{-g}\le(\e_{\rm m}-{\r_{\rm m}}\sqrt{-g^{\a\b}\p_{,\a}\p_{,\b}}\ri)\;.
\ee
Metrical tensor of energy-momentum of the ideal fluid is obtained by taking a variational derivative of the Lagrangian (\ref{pzq1}) with respect to the metric tensor,
\be\la{pf2a}
T^{\rm m}_{\a\b}=\frac{2}{\sqrt{-g}}\frac{\d\lag^{\rm m}}{\d g^{\a\b}}\;.
\ee
Calculation yields
\be\la{pf2b}
T^{\rm m}_{\a\b}=\le(\e_{\rm m}+p_{\rm m}\ri)w_\a w_\b+p_{\rm m}g_{\a\b}\;,
\ee
where $w^\a=dx^\a/d\t$ is the four-velocity of the fluid, and $\t$ is the proper time of the fluid element taken along its worldline. This is a standard form of the tensor of energy-momentum of the ideal fluid \citep{mtw}. Because the Lagrangian (\ref{pzq1}) is expressed in terms of the dynamical variable $\p$, the Noether approach based on taking the variational derivative of the Lagrangian with respect to the field variable, can be applied to derive the canonical tensor of the energy-momentum of the ideal fluid. This calculation has been done in \citep[pp. 334-335 ]{2011rcms.book.....K} and it leads to the expression (\ref{pf2b}). It could be expected because we assumed that the the ideal fluid consists of bosons. The metrical and canonical tensors of energy-momentum for the liquid differ, if and only if, the liquid's particles are fermions (see \citep[pp. 331-332]{2011rcms.book.....K} for more detail). We do not consider the fermionic liquids in the present paper.

\subsection{The Lagrangian of the Scalar Field}\la{lsfi}

The Lagrangian of the scalar field $\Psi$ is given by
\be\la{hl4}
\lag^{\rm q}=\sqrt{-g}\le(\frac12 g^{\a\b}\pd_\a\Psi\pd_\b\Psi+W\ri)\;,
\ee
where $W\equiv W\le(\Psi\ri)$ is a potential of the scalar field. We assume that there is no direct coupling between the scalar field and the matter of the ideal fluid. They can interact only through the gravitational field.
Many different potentials of the scalar field are used in cosmology \citep{2010deto.book.....A}. At this step, we do not chose a specific form of the potential which will be selected later.

Metrical tensor of energy-momentum of the scalar field is obtained by taking a variational derivative
\be\la{h15a}
T^{\rm q}_{\a\b}=\frac{2}{\sqrt{-g}}\frac{\d\lag^{\rm q}}{\d g^{\a\b}}\;,
\ee
that yields
\be\la{h15b}
T^{\rm q}_{\a\b}=\pd_\a\Psi\pd_\b\Psi-g_{\a\b}\biggl[g^{\m\n}\pd_\m\Psi\pd_\n\Psi+W(\Psi)\biggr]\;.
\ee
The canonical tensor of energy-momentum of the scalar field is obtained by applying the Neother theorem and leads to the same expression (\ref{h15b}).

One can {\it formally} reduce the tensor (\ref{h15b}) to the form similar to that of the ideal fluid by making use of the following procedure.
First, we define the analogue of the specific enthalpy of the scalar field "fluid"
\be\la{h15c}
\mu_{\rm q}=\sqrt{-g^{\sigma\n}\pd_\sigma\Psi\pd_\n\Psi}\;,
\ee
and the effective four-velocity, $v^\a$, of the "fluid"
\be\la{h15ca}
\mu_{\rm q}v_\a= -\pd_\a\Psi\;.
\ee
The four-velocity $v^\a$ is normalized to $v_\a v^\a=-1$. Therefore, the scalar field enthalpy $\mu_{\rm q}$ can be expressed in terms of the partial derivative from the scalar field
\be\la{h15cb}
\mu_{\rm q}=v^\a\pd_\a\Psi\;.
\ee
Then, we introduce the analogue of the rest mass density $\r_{\rm q}$ of the scalar field "fluid" by defining,
\be\la{h15d}
\r_{\rm q}\equiv\mu_{\rm q}=v^\a\pd_\a\Psi=\sqrt{-g^{\sigma\n}\pd_\sigma\Psi\pd_\n\Psi}\;.
\ee
As a consequence of the above definitions, the energy density, $\e_{\rm q}$ and pressure $p_{\rm q}$ of the scalar field "fluid" can be introduced as follows
\ba\la{h16a}
\e_{\rm q}&\equiv&-\frac12g^{\sigma\n}\pd_\sigma\Psi\pd_\n\Psi+W(\Psi)=\frac12\r_{\rm q}\m_{\rm q}+W(\Psi)\;,\\\la{h16b}
p_{\rm q}&\equiv&-\frac12g^{\sigma\n}\pd_\sigma\Psi\pd_\n\Psi-W(\Psi)=\frac12\r_{\rm q}\m_{\rm q}-W(\Psi)\;.
\ea
One notices that a relationship
\be\la{h16d}
\m_{\rm q}=\frac{\e_{\rm q}+p_{\rm q}}{\r_{\rm q}}\;,
\ee
between the specific enthalpy $\m_{\rm q}$, the density $\r_{\rm q}$, the pressure $p_{\rm q}$ and the energy density $\e_{\rm q}$, of the scalar field "fluid" formally holds on the same form (\ref{pf2}) as in the case of the barotropic ideal fluid.

After applying the above-given definitions in equation (\ref{h15b}), it is formally reduced to the tensor of energy-momentum of an ideal fluid
\be\la{h16c}
T^{\rm q}_{\a\b}=\le(\e_{\rm q}+p_{\rm q}\ri)v_\a v_\b+p_{\rm q}g_{\a\b}\;.
\ee
It is worth emphasizing that the analogy between the tensor of energy-momentum (\ref{h16c}) of the scalar field "fluid" with that of the barotropic ideal fluid (\ref{pf2b}) is rather formal since the scalar field, in the most general case, does not satisfy {\it all} required thermodynamic equations because of the presence of the potential $W=W(\Psi)$ in the energy density $\e_{\rm q}$, and pressure $p_{\rm q}$ of the scalar field.
\subsection{The Lagrangian of the Localized Astronomical System}\la{laloas}

The Lagrangian $\lag^{\rm p}$ of matter of the localized astronomical system which perturbs the geometry of the background manifold of the FLRW universe, can be chosen arbitrary. We shall call the perturbation of the background manifold that is induced by $\lag^{\rm p}$, the {\it bare} perturbation. We assume that the matter of the bare perturbation is described by a set of scalar potentials $\theta$ which are analogues of the Clebsch potential of the matter supporting the background geometry. The Lagrangian density of the bare perturbation is given by ${\cal L}^{\rm p}=\sqrt{-g}L^{\rm p}\le(\theta, g_{\a\b}\ri)$. Tensor of energy-momentum of the matter of the bare perturbation, $\mathfrak{T}_{\a\b}$, is obtained by taking a variational derivative
\be\la{q19z}
\mathfrak{T}_{\a\b}=\frac{2}{\sqrt{-g}}\frac{\d\lag^{\rm p}}{\d g^{\a\b}}\;.
\ee
Tensor $\mathfrak{T}_{\a\b}$ is a source of the bare gravitational perturbation of the background manifold that will be determined by solving Einstein's field equations derived in next sections.

\section{Background Manifold}\la{bma}

\subsection{The Hubble Flow}
We shall consider the background universe as described by the Friedmann-Lem\^itre-Robertson-Walker (FLRW) metric. The functional form of the metric depends on the coordinates introduced on the manifold. Because the FLRW metric describes homogeneous and isotropic spacetime there is a preferred class of coordinates which clearly reveal these properties of the background manifold. These coordinates materialize a special set of freely falling observers, called comoving observers.
These observers are following with the flow of the expanding universe and have constant values of spatial coordinates. The proper distance between the comoving observers increases in proportion to the scale factor $R(T)$.
In the preferred cosmological coordinates, the time coordinate of the FLRW metric is just the proper time as measured by the comoving observers.
A particle moving relative to the local comoving observers has a peculiar velocity with respect to the Hubble flow.
An observer with a non-zero peculiar velocity does not see the universe as isotropic.

For example, the peculiar velocity of the solar system implies the dipole anisotropy of cosmic microwave background (CMBR) radiation corresponding to  $|{\bm v}_{\st\odot}| = 369.0 \pm 0.9$ km$\cdot$s$^{-1}$, towards a point with the galactic coordinates $(l,b) = \le(264^\circ, 48^\circ\ri)$ \citep{wmap_2009,wmap_2011}. Such a solar system's velocity implies a velocity $|{\bm v}_{\sst LG}|= 627 \pm 22$ km$\cdot$s$^{-1}$ toward $(l,b) = \le(276^\circ, 30^\circ\ri)$ for our Galaxy and the Local Group of galaxies relative to the CMBR \citep{1993ApJ...419....1K,1996ApJ...473..576F}. The existence of the preferred frame in cosmology should not be understood as a violation of the Einstein principle of relativity. Indeed, any coordinate chart can be used in order to describe the FLRW universe. A preferred frame exists merely because the FLRW metric admits only six-parametric group (3 spatial translations and 3 spatial rotations) as contrasted with the ten-parametric group of Minkowski (or De Sitter) spacetime which includes the time translation and three Lorentz boosts as well. The metric of FLRW does not remain invariant with respect to the time translation and the Lorentz transformations because its expansion makes different spacelike hypersurfaces non-equivalent. It may lead to some interesting observational predictions of cosmological effects within the solar system \citep{2012PhRvD..86f4004K}.

\subsection{The Friedmann-Lem\^itre-Robertson-Walker metric}\la{frwlm}

In what follows, we shall consider the problem of calculation of the post-Newtonian perturbations in the expanding universe described by the FLRW class of models. The FLRW metric is an exact solution of Einstein's field equations of general relativity that describes a homogeneous, isotropically expanding or contracting universe. The general form of the metric follows from the geometric properties of homogeneity and isotropy of the manifold \citep{1972gcpa.book.....W,weinberg_2008}. Einstein's equations are only needed to derive the scale factor of the universe as a function of time.

The most general form of the FLRW metric is given by
\be\la{bm1}
ds^2=-d{T}^2+R^2\le[\frac{d\rho^2}{1-k\rho^2}+\rho^2\le(d^2\vt+\sin^2\vt d^2\up\ri)\ri] \;,
\ee
where $T$ is the coordinate time, $\{\rho,\vt,\up\}$ are spherical coordinates, $R=R({T})$ is the scale factor depending on time and characterizing the size of the universe compared to the present value of $R=1$. The time ${T}$ has a physical meaning of the proper time of a comoving observer that is being at rest with respect to the cosmological frame of reference. The present epoch corresponds to the value of the time ${T}={T}_0$.
The constant $k$ can take on three different values $k=\{-1,0,+1\}$, where $k=-1$ corresponds to the spatial hyperbolic geometry, $k=0$ does the spatially flat FLRW model, and $k=+1$ does the spatially closed world \citep{mtw}.

The Hubble parameter $\H$ characterizes the rate of the temporal evolution of the universe. It is defined by
\be\la{bm1a}
\H\equiv\frac{\dot R}{R}=\frac1{R}\frac{dR}{d{T}}\;.
\ee
For mathematical reasons, it is convenient to introduce a conformal time, $\eta$, via differential equation
\be\la{coft}
d\eta=\frac{d{T}}{R({T})}\;.
\ee
If the time dependence of the scale factor is known, the equation (\ref{coft}) can be solved, thus, yielding ${T}={T}(\eta)$.
It allows us to re-express the scale factor $R({T})$ in terms of the conformal time, $R\le({T}(\eta)\ri) \equiv a(\eta)$.
The conformal Hubble parameter is, then, defined as
\be\la{bm1b}
{\cal H}\equiv\frac{a'}{a}=\frac{1}{a}\frac{da}{d\eta}\;.
\ee
The two expressions for the Hubble parameters are related by means of equation
\be\la{bm1c} H=\frac{{\cal H}}{a}\;,
\ee
that allows us to link their time derivatives
\ba \la{bm1d}
a^2\dot{H}&=&{\cal H}'-{\cal H}^2\;,\\\la{bm1w}
a^3\ddot{H}&=&{\cal H}''-4{\cal H} {\cal H}'+2{\cal H}^3\;,
\ea
and so on.

It is also convenient to introduce the isotropic Cartesian coordinates ${X}^i=\{{X},{Y},{Z}\}$, by transforming the radial coordinate
\be\la{i1g} \rho=\frac{r}{1+\displaystyle{\frac{k}{4}}{r}^2}\;,
\ee
and defining ${r}^2={X}^2+{Y}^2+{Z}^2=\d_{ij}{X}^i{X}^j$.
In the isotropic coordinates the interval (\ref{bm1}) takes on the following form
\be\la{bm2}
ds^2={G}_{\a\b}d{X}^\a d{X}^\b\;,
\ee
where the coordinates ${X}^\a=\{{X}^0,{X}^1,{X}^2,{X}^3\}=\{\eta,{X},{Y},{Z}\}$, and the
metric has a conformal form
\be\la{bm3}
{G}_{\a\b}=a^2(\eta) {\mathtt g}_{\a\b}\;
\ee
\be\la{bm4}
{\mathtt g}_{00}=-1\;,\qquad {\mathtt g}_{0i}=0\;,\qquad {\mathtt g}_{ij}=\frac{\d_{ij}}{\le(1+\displaystyle{\frac{k}{4}}{r}^2\ri)^2}\;.
\ee
Determinant of the metric $G_{\a\b}$ is $G=a^8{\mathtt g}$, where ${\mathtt g}=-\le(1+kr^2/4\ri)^{-6}$.
The spacetime interval in the isotropic Cartesian coordinates reads
\be\la{bm4aa}
ds^2=a^2(\eta)\le[-d\eta^2+\frac{\d_{ij}d{X}^id{X}^j}{\displaystyle\le(1+\frac{k}{4}{ r}^2\ri)^2}\ri]\;.
\ee
The distinctive property of the isotropic coordinates in the FLRW universe is that the radial coordinate $r$ is defined in such a way that the three-dimensional space looks exactly Euclidean and null cones appear in it as round spheres irrespectively of the value of the space curvature $k$. The isotropic coordinates do not represent proper distances on the sphere, nor does the radial coordinate ${r}$ represents a proper radial distance measured with the help of radar astronomy technique. The proper spatial distance in the isotropic coordinates is $(1+kr^2/4)^{-1}ar$ \citep{1972gcpa.book.....W}.

The FLRW metric presented in the conformal form by equation (\ref{bm4aa}) singles out a preferred cosmological reference frame defined by the congruence of worldlines of the fiducial test particles being at rest with respect to the spatial coordinates ${X}^i$. Four-velocity of a fiducial particle is denoted as $\bar {U}^\alpha=d{X}^\a/d\tau$, where $d\t=-ds$ is the proper time on the worldline of the particle. In the isotropic conformal coordinates, $\bar {U}^\a=(1/a,0,0,0)$. The four-velocity is a unit vector, $\bar {U}^\alpha \bar {U}_\alpha={G}_{\a\b}\bar {U}^\alpha\bar {U}^\beta=-1$. It implies that the covariant components of the four-velocity are $\bar {U}_\a=(-a,0,0,0)$. In the preferred frame the universe looks homogeneous and isotropic. The choice of the isotropic Cartesian coordinates reflects these fundamental properties explicitly in the symmetric form of the metric (\ref{bm3}). However, the set of the fiducial particles is a mathematical idealization. In reality, any isolated astronomical systems (galaxy, binary star, the solar system, etc.) have a peculiar velocity with respect to the preferred cosmological frame formed by the Hubble flow. We have to introduce a locally-inertial coordinate chart which is associated with the isolated system and moves along with it. Transformation from the preferred cosmological frame to the local chart must include the Lorentz boost and a geometric part due to the expansion and curvature of cosmological spacetime. It can take on multiple forms which originate from certain geometric and/or experimental requirements \citep{2006PhRvD..74f4019C,hongya:1920,2005ESASP.576..305K,2010RvMP...82..169C}.

We do not impose specific limitations on the choice of coordinates on the background manifold and keep the overall formalism of the post-Newtonian approximations, covariant. The arbitrary coordinates are denoted as $x^\a=(x^0,x^i)$ and they are related to the preferred isotropic coordinates ${X}^\a=(\eta,{X}^i)$ by the coordinate transformation $x^\a=x^\a\le({X}^\b\ri)$. This transformation has inverse ${X}^\a={X}^\a\le(x^\b\ri)$, at least in some local domain of the background manifold. In this domain, the matrices of the coordinate transformations
\be\la{bub2}
\Lambda^\a{}_\b=\frac{\pd x^\a}{\pd{X}^\b}\;,\qquad\qquad {\rm M}^\a{}_\b=\frac{\pd{X}^\a}{\pd x^\b}\;,
\ee
and they satisfy to the apparent equalities $\Lambda^\a{}_\g{\rm M}^\g{}_\b=\d^\a_\b$ and ${\rm M}^\a{}_\g\Lambda^\g{}_\b=\d^\a_\b$.

Four-velocity of the Hubble observers written in the arbitrary coordinates has the following form
\be\la{bub7}
\bar u^\a=\Lambda^\a{}_\b\bar U^\b=a^{-1}\Lambda^\a{}_0\qquad,\qquad
\bar u_\a={\rm M}^\b{}_\a{U}_\b=-a{\rm M}^0{}_\a\;.
\ee
The background FLRW metric written down in the arbitrary coordinates, $x^\a$, takes on the following form
\be\la{bub5}
\bar g_{\a\b}(x^\a)=a^2\bar{\mathtt f}_{\a\b}(x^\a)\;.
\ee
Here the scalar function $a(x^\a)\equiv a\le[\eta(x^\a)\ri]$, and the conformal metric
\be\la{bub6}
\bar{\mathtt f}_{\a\b}(x^\a)={\rm M}^\m{}_\a{\rm M}^\n{}_\b {\mathtt g}_{\m\n}({X}^i)\;.
\ee

Any metric admits 3+1 decomposition with respect to a congruence of a timelike vector field \citep{mtw}. FLRW universe admits a privileged congruence formed by the four-velocity $\bar u^\a$ of the Hubble observers which is a physically privileged vector field. The 3+1 decomposition of the FLRW metric is applied in arbitrary coordinates and has the following form
\be\la{dec1}
\bar g_{\a\b}=-\bar u_\a\bar u_\b+\bar P_{\a\b}\;,
\ee
where the tensor
\be\la{dec2}
\bar P_{\a\b}=a^2{\rm M}^i{}_\a{\rm M}^j{}_\b {\mathtt g}_{ij}\;,
\ee
describes the metric on the spacelike hypersurface being everywhere orthogonal to the four-velocity $\bar u^\a$ of the Hubble flow. Tensor $\bar P_{\a\b}$ is the operator of projection on this hypersuface. It can be also interpreted as a metric on the hypersurace of orthogonality to the Hubble vector flow. Equation (\ref{dec1}) can be used in order to prove that $\bar P_{\a\b}$ satisfies the following relationship
\be\la{a7a}
{\bar P}^{\b\m} {\bar P}_{\b}{}^\n=\bar P^{\m\n}\;,
\ee
which can be confirmed by inspection. The trace $\bar P^\a{}_\a=\bar g^{\a\b}\bar P_{\a\b}=\bar P^{\a\b}\bar P_{\a\b}=3$.

Now, we consider how to express the partial derivatives of any scalar function $F=F(\eta)$, which depends only on the conformal time $\eta=\eta(x^\a)$, in terms of the four-velocity $\bar u^\a$ of the Hubble flow. Taking into account that $\eta=x^0$ and applying equation (\ref{bub7}), we obtain
\be\la{bub9}
F_{,\a}=\frac{\pd F}{\pd x^\a}=\frac{d F}{d\eta}\frac{\pd \eta}{\pd x^\a}=F'{\rm M}^0{}_\a=-\frac{F'}{a}\bar u_\a=-\dot F\bar u_\a\;.
\ee
In particular, the partial derivative from the scale factor, $a_{,\a}=-\dot a\bar u_\a=-{\cal H}\bar u_\a$, and the partial derivative from the Hubble parameter ${\cal H}_{,\a}=-\dot{\cal H}\bar u_\a$.

\subsection{The Christoffel Symbols and Covariant Derivatives}
In the following sections of the paper we will need to calculate the covariant derivatives from various geometric objects on the background cosmological manifold covered by an arbitrary coordinate chart $x^\a=(x^0,x^i)$. The calculation engages the affine connection $\bar\G^\a_{\b\g}$ of the background manifold which is decomposed into an algebraic sum of two connections (the Christoffel symbols) because of the conformal structure of the FLRW metric \citep{wald}. By definition,
\be\la{bub12}
\bar\G^\a{}_{\b\g}=\frac12\bar g^{\a\n}\le(\bar g_{\n\b,\g}+\bar g_{\n\g,\b}-\bar g_{\b\g,\n}\ri)\;,
\ee
where
\be\la{bu11a}
\bar g_{\a\b,\g}=-2\H\bar g_{\a\b}\bar u_\g+a^2\bar\mm_{\a\b,\g}\;.
\ee
Separating terms in the right side of (\ref{bub12}) yields
\be\la{bm6}
\bar\G^\a{}_{\b\g}=\bar A^\a{}_{\b\g}+\bar B^\a{}_{\b\g}\;,
\ee
where
\be\la{bm7}
\bar A^\a{}_{\b\g}=-\H\le(\d^\a_\b\bar u_\g+\d^\a_\g\bar u_\b-\bar u^\a\bar g_{\b\g}\ri)\;, \ee and \be\la{bm8}
\bar B^\a{}_{\b\g}=\frac12 \bar{\mathtt f}^{\a\m}\le(\bar{\mathtt f}_{\m\b,\g}+\bar{\mathtt f}_{\m\g,\b}-\bar{\mathtt f}_{\b\g,\m}\ri)\;.
\ee
The non-vanishing components of the connections are given in the isotropic Cartesian coordinates $X^\a$ by
\be \la{bm9}
\bar A^\a{}_{0\b}={\cal H}\d^\a_\b\;,\qquad \bar A^0{}_{ij}={\cal H}{\mathtt g}_{ij}\;,\qquad \bar B^i{}_{pq}=-\frac{k}{2}\frac{\d^i_pX_{\rm q}+\d^i_qX_p-\d_{pq}X^i}{1+\displaystyle{\frac{k}{4}}{ r}^2}\;,
\ee
where $X_q\equiv \delta_{qj}X^j$, and all the other components of the connections vanish.

A covariant derivative of a geometric object (scalar, vector, etc.) on the background manifold is denoted in this paper with a vertical bar. For example, the covariant derivative of a vector field $F^\a$ is
\be\la{bm9aa}
F^\a{}_{|\b}=F^\a{}_{,\b}+\bar\Gamma^\a{}_{\b\g}F^\g\;,
\ee
where a comma in front of sub-index $\b$ denotes a partial derivative with respect to coordinate $x^\b$. Equation (\ref{bm9aa}) can be brought to yet another form if we denote the covariant derivative of the affine connection $\bar B^\a{}_{\b\g}$ with a semicolon. Making use of (\ref{bm6}) in equation (\ref{bm9aa})transforms it to the following form
\be\la{bm9ab}
F^\a{}_{|\b}=F^\a{}_{;\b}+\bar A^\a{}_{\b\g}F^\g\;.
\ee
The covariant derivative of a covector $F_\a$ is defined in a similar way,
\be\la{bm9ac}
F_{\a|\b}=F_{\a,\b}-\bar\Gamma^\g{}_{\a\b}F_\g\,
\ee
which is equivalent to
\bsu
\ba\la{bmz1}
F_{\a|\b}&=&F_{\a;\b}-\bar A^\g{}_{\a\b}F_\g\;,\\\la{bmzqw}
F_{\a;\b}&=&F_{\a,\b}-\bar B^\g{}_{\a\b}F_\g\;
\ea
\esu
Equations for tensors of higher rank can be presented in a similar way.
Of course, the covariant derivative of a scalar field $F$ always coincides with its covariant derivative by definition,
\be\la{bm9ad}F_{|\a}=F_{;\a}=F_{,\a}\;.\ee
We also provide an equation for the covariant derivative of the four-velocity of the Hubble flow. Doing calculations in the isotropic coordinates $X^\a$ for the four-velocity $\bar U^\a$, and applying the tensor law of transformation to arbitrary coordinates $x^\a$, results in
\be\la{yui65}
\bar u_{\a|\b}=\H \bar P_{\a\b}\;,\qquad \bar u^\a{}_{|\b}=\H\le(\d^a_\b+\bar u^\a\bar u_\b\ri)\;,\qquad \bar u^{\a|\b}=\H \bar P^{\a\b}\;,
\ee
where the tensor indices are raised and lowered with the metric $\bar g_{\a\b}$.

\subsection{The Riemann Tensor}
The Riemann tensor is defined by
\be\la{a0}
\bar R^\a{}_{\b\m\n}=\bar\Gamma^\a{}_{\b\n,\m}-\bar\Gamma^\a{}_{\b\m,\n}+\bar\Gamma^\a{}_{\m\g}\bar\Gamma^\g{}_{\b\n}-\bar\Gamma^\a{}_{\n\g}\bar\Gamma^\g{}_{\b\m}\;.
\ee
and can be calculated directly from this equation. We prefer a slightly different way by making use of the algebraic decomposition of the Riemann tensor into the irreducible parts
\be\la{a1}
\bar R_{\a\b\m\n}=
\bar C_{\a\b\m\n}+
\frac12\le(\bar S_{\a\m}\bar g_{\b\n}+\bar S_{\b\n}\bar g_{\a\m}-\bar S_{\a\n}\bar g_{\b\m}-\bar S_{\b\m}\bar g_{\a\n} \ri)+
\frac{\bar R}{12} \le(\bar g_{\a\m}\bar g_{\b\n}-\bar g_{\a\n}\bar g_{\b\m} \ri)\;,
\ee
where $\bar C_{\a\b\m\n}$ is the Weyl tensor,
\be\la{a1as}
\bar S_{\mu\nu}=\bar R_{\mu\nu}-\frac14\bar R\bar g_{\mu\nu}\;,
\ee
$\bar R_{\mu\nu}=\bar g^{\a\b}\bar R_{\a\m\b\n}$ is the Ricci tensor, and $R=\bar g^{\a\b}\bar R_{\a\b}$ is the Ricci scalar.
FLRW cosmological metric (\ref{bm1}) has a remarkable property -- it can be always brought up to the conformally-flat form by applying an appropriate coordinate transformation \citep{2007JMP....48l2501I}. However, the Weyl tensor of any conformally-flat spacetime vanishes identically,
\be\la{a3} \bar C_{\a\b\m\n}\equiv 0\;.
\ee
Direct evaluation of other tensors entering (\ref{a1}) by making use of the FLRW metric (\ref{bm3}), (\ref{bm4}) yields
\ba\la{a4}
\bar R_{\m\n}&=&\frac{1}{a^2}\le[{\cal H}'\le(\bar g_{\m\n}-2\bar u_\m\bar u_\n\ri)+2\le({\cal H}^2+k\ri) \le(\bar g_{\m\n}+\bar u_\m\bar u_\n\ri)\ri]\;,
\\\la{a5}
\bar R&=&\frac{6}{a^2}\le[ {\cal H}'+{\cal H}^2+k\ri]
\;.
\ea
Making use of equations (\ref{a3}) -- (\ref{a5}) in the decomposition (\ref{a1}) of the Riemann tensor, yields the following result
\be\la{a6}
\bar R_{\a\b\m\n}=\frac{1 }{a^2}\le[{\cal H}' \le(\bar g_{\a\m}\bar g_{\b\n}-\bar g_{\a\n}\bar g_{\b\m}\ri)-\le( {\cal H}'-{\cal H}^2-k\ri)\le({\bar P}_{\a\m}{\bar P}_{\b\n}-{\bar P}_{\a\n}{\bar P}_{\b\m}\ri)\ri]\;, \ee
where ${\bar P}_{\a\b}=\bar g_{\a\b}+\bar u_\a\bar u_\b$
is the operator of projection that was introduced in (\ref{dec2}).

\subsection{The Friedmann Equations}\la{freq}

The Einstein tensor $\bar{\cal E}_{\a\b}=\bar R_{\a\b}-\bar g_{\a\b}\bar R/2$ of the FLRW cosmological model is derived from equations (\ref{a4}) and (\ref{a5}). It reads
\be\la{a8} \bar{\cal E}_{\a\b}=-\frac1{a^2}\le[2\le( {\cal H}'-{\cal H}^2-k\ri){\bar P}_{\a\b}+3\le({\cal H}^2+k\ri)\bar g_{\a\b}\ri]\;.
\ee
Einstein's field equations on the background spacetime takes on the following form
\be\la{a11}
\bar{\cal E}_{\a\b}=8\pi\bar T_{\a\b}\;,
\ee
where the tensor of energy-momentum of the background spacetime manifold includes the background matter and the scalar field
\be\la{a11a}
\bar T_{\a\b}=\bar T^{\rm m}_{\a\b}+\bar T^{\rm q}_{\a\b}\;.
\ee
Here, tensors of energy-momentum in the right side of Einstein's equations are derived from the Lagrangians (\ref{pzq1}) and (\ref{hl4}), and represent an algebraic sum of tensors (\ref{pf2b}) and (\ref{h15c}).
Each tensor of energy-momentum, $\bar T^{\rm m}_{\a\b}$ and $\bar T^{\rm q}_{\a\b}$, is Lie-invariant with respect to the group of symmetry of the background FLRW metric independently, and each of them have the form of the tensor of energy-momentum of the perfect fluid. Hence, the tensor of energy-momentum $\bar T_{\a\b}$ in the right side of (\ref{a11}) has the form of a perfect fluid as well,
\be\la{a9}
\bar T_{\a\b}=\le(\bar\e+\bar p\ri)\bar u_\a\bar u_\b+\bar p~\bar g_{\a\b}\;.
\ee

It imposes a certain restriction on the effective energy density $\bar\e$ and pressure $\bar p$ which must obey Dalton's law for a partial energy density and pressure of the background matter and the scalar field components \citep{therm_fluid}
\ba\la{a9a}
\bar\e&=&\bar\e_{\rm m}+\bar\e_{\rm q}\;,\\\la{a9b}
\bar p&=&\bar p_{\rm m}+\bar p_{\rm q}\;.
\ea
Here, $\bar\e_{\rm m}$ and $\bar p_{\rm m}$ are the energy density and pressure of the ideal fluid, and $\bar\e_{\rm q}$ and $\bar p_{\rm q}$ are the energy density and pressure of the scalar field which are related to the time derivative $\dot{\bar\Psi}$ of the scalar field and its potential $\bar W=\bar W(\bar\Psi)$ by equations (\ref{h16a}), (\ref{h16b}). On the background spacetime these equations takes on the following form
\ba\la{sq1}
\bar\e_{\rm q}&=&\frac12\bar\r_{\rm q}\bar\m_{\rm q}+\bar W\;,\\\la{sq2}
\bar p_{\rm q}&=&\frac12\bar\r_{\rm q}\bar\m_{\rm q}-\bar W\;,
\ea
where $\bar\m_{\rm q}$ is the background specific enthalpy of the scalar field defined by (\ref{h15c}), and $\bar\rho_{\rm q}=\bar\m_{\rm q}$ is the background density of the scalar field "fluid". It is worthwhile to remind to the reader that due to the homogeneity and isotropy of the FLRW universe, all matter variables on the background manifold are functions of the conformal time $\eta$ only when being expressed in the isotropic Cartesian coordinates.

Einstein's equations (\ref{a11}) can be projected on the direction of the background four-velocity of matter and on the spatial hypersurface being orthogonal to it. It yields two Friedmann equations for the evolution of the scale factor $a$,
\ba\la{a12} {\H}^2&=&~~\frac{8\pi}{3}\bar\e-\frac{k}{a^2}\;,
\\\la{a13}
2{\dot H}+3{\H}^2&=&-8\pi\bar p-\frac{k}{a^2}\;
\ea
where $\bar\e$ and $\bar p$ are the effective energy density and pressure of the mixture of matter and scalar field as defined above.

A consequence of the Friedmann equations (\ref{a12}), (\ref{a13}) is an equation
\be\la{13a}
\dot{\H}-\frac{k}{a^2}=-4\pi\le(\bar\e+\bar p\ri)\;,
\ee
relating the time derivative of the Hubble parameter with the sum of the overall energy density and pressure,
which can be expressed in terms of the density and specific enthalpy of the background components of matter,
\be\la{13b}
\bar\e+\bar p=\bar\r_{\rm m}\bar\m_{\rm m}+\bar\r_{\rm q}\bar\m_{\rm q}\;.
\ee

In order to solve the Friedmann equations (\ref{a12}), (\ref{a13}) we have to employ the equation of state of matter. Customarily, it is assumed that each matter component obeys its own cosmological equation of state,
\be\la{a10}
\bar p_{\rm m}=w_{\rm m}\bar\e_{\rm m}\;,\qquad \bar p_{\rm q}=w_{\rm q}\bar\e_{\rm q}\;,
\ee
where $w_{\rm m}$ and $w_{\rm q}$ are parameters lying in the range from $-1$ to $+1$. In the most simple cosmological models, parameters $w_{\rm m}$ and $w_{\rm q}$ are fixed. More realistic models admit that the parameters of the equation of state may change in the course of the cosmological expansion, that is they may depend on time. The equation of state does not close the system of the Friedmann equations, which have to be complemented with the equations of motion of the scalar field and of the ideal fluid in order to make the system of differential equations for the gravitational and matter field variables complete.

\subsection{The Hydrodynamic Equations of the Ideal Fluid}

The background value of the Clebsch potential of the ideal fluid, $\bar\Phi$, depends only on the conformal time $\eta$ of the FLRW metric. The partial derivative of the potential, taken in arbitrary coordinate chart on the background manifold, can be expressed in accordance with equation (\ref{bub9}) in terms of the background four-velocity $\bar u^\a$ as follows
\be\la{der1}
\bar\Phi_{|\a}=-\dot{\bar\Phi}\bar u_\a\;.
\ee
It allows us to write down the specific enthalpy of the ideal fluid in terms of the Clebsch potential. Taking background value of equation (\ref{pf7a}), we obtain
\be\la{cr4}
\bar\m_{\rm m}\equiv\bar u^\a\bar\Phi_{|\a}=\dot{\bar\p}\;.
\ee

The background equation of continuity for the rest mass density $\bar{\r}_{\rm m}$ of the ideal fluid is
\be\la{cr30}
\le(\bar\r_{\rm m}\bar u^\a\ri)_{|\a}=0\;,
\ee
that is equivalent to
\be\la{cr3}
{\bar\r}_{\rm m|\a}-3\H{\bar\r}_{\rm m}\bar u_\a=0\;,
\ee
where we have used (\ref{yui65}).
The background equation of conservation of energy is
\be\la{cr2}
{\bar\e}_{\rm m|\a}-3\H\left({\bar\e}_{\rm m}+{\bar p}_{\rm m}\right)\bar u_\a=0\;,
\ee
where we have employed definition of the energy (\ref{gug1}), and equation (\ref{cr3}) along with (\ref{pf1}).

\subsection{The Scalar Field Equations}

Background equation for the scalar field $\bar\Psi$ is derived from the action (\ref{qa1}) by taking variational derivatives with respect to $\bar\Psi$. It yields
\be\la{cr0}
\bar g^{\a\b}\bar\Psi_{|\a\b}-\frac{\pd \bar W}{\pd\bar\Psi}=0\;.
\ee
In terms of the time derivatives with respect to the conformal time $\eta$, equation (\ref{cr0}) reads
\be\la{cr1}
\ddot{\bar\Psi}+3{H}\dot{\bar\Psi}+\frac{\pd \bar W}{\pd\bar\Psi}=0\;.
\ee
Here, we have taken into account that the background value of the scalar field, $\bar\Psi$, depends only on time $\eta$, and its derivative (with respect to $\eta$) is proportional to the background four-velocity
\be\la{der2}
\bar\Psi_{|\a}=-\dot{\bar\Psi}\bar u_\a\;.
\ee
If we use definition of the background enthalpy of the scalar field
\be\la{cr4a}
\bar\m_{\rm q}\equiv\bar u^\a\bar\Phi_{|\a}=\dot{\bar\Psi}\;,
\ee
and account for definition (\ref{h16a}) of the specific energy $\e_{\rm q}$ of the scalar field, the equation (\ref{cr1}) will become
\be\la{cr1a}
{\bar\e}_{\rm q|\a}-3\H\left({\bar\e}_{\rm q}+{\bar p}_{\rm q}\right)\bar u_\a=0\;.
\ee
that looks similar to the hydrodynamic equation (\ref{cr3}) of conservation of energy of the ideal fluid. Because of this similarity, the second Friedmann equation (\ref{a13}) can be derived from the first Friedmann equation (\ref{a12}) by taking a time derivative and applying the energy conservation equations (\ref{cr2}) and (\ref{cr1a}).

The background density $\bar\rho_{\rm q}$ of the scalar filed "fluid" is $\bar\rho_{\rm q}=\bar\mu_{\rm q}$ in accordance with (\ref{h15d}).
The equation of continuity for the density $\bar{\r}_{\rm q}$ of the ideal fluid is obtained by differentiating definition of $\bar{\r}_{\rm q}$, and making use of (\ref{cr1}). It yields
\be\la{crq}
\le(\bar\r_{\rm q}\bar u^\a\ri)_{|\a}=-\frac{\pd\bar W}{\pd\bar\Psi}\;,
\ee
or, equivalently,
\be\la{cr3a}
{\bar\r}_{\rm q|\a}-3\H{\bar\r}_{\rm q}\bar u_\a=\frac{\pd\bar W}{\pd\bar\Psi}\bar u_\a\;,
\ee
which shows that the "density" of the scalar field "fluid" is not conserved. We emphasize that there is no any violation of physical laws, since (\ref{cr3a}) is simply another way of writing equation (\ref{cr0}), and the scalar field is not thermodynamically equivalent to the ideal fluid. Equation (\ref{cr3a}) is convenient in the calculations that follows in next sections.

\subsection{Equations of Motion of Matter of the Localized Astronomical System}\la{emmla}

Matter of the localized astronomical system is described by the tensor of energy-momentum $\mathfrak{T}^{\a\b}$ defined in (\ref{q19z}) in terms of the Lagrangian derivative. It can be given explicitly as a function of field variables after we chose a specific form of matter, for example, gas, liquid, solid, or something else. We do not restrict ourselves with a particular form of this tensor, and shall develop a more generic approach that is applicable to any kind of matter comprising the localized astronomical system..

Background equation of motion of matter of the astronomical system is given by the conservation law
\be\la{qq1}
\mathfrak{T}^{\a\b}{}_{|\b}=0\;.
\ee
It can be also written down in terms of a covariant derivative of the conformal metric
\be\la{qq1a}
\le(\sqrt{-\bar g}\mathfrak{T}^{\a\b}\ri)_{;\b}+\sqrt{-\bar g}\bar A^\a{}_{\b\g}\mathfrak{T}^{\b\g}=0\;,
\ee
where the connection $\bar A^\a{}_{\b\g}$ is defined in (\ref{bm7}).

It is natural to write down this equation in 3+1 form by projecting it on the direction of 4-velocity of the Hubble flow, $\bar u^\a$, and on the hypersurface being orthogonal to it. This is achieved by introducing the following projections
\bsu\ba\la{qq4}
\sigma&\equiv&\phantom{-}\bar u^\m\bar u^\n \mathfrak{T}_{\m\n}\;,\\
\la{qq5}
\t&\equiv&\phantom{-}\bar P^{\m\n}\mathfrak{T}_{\m\n}\;,\\
\la{qq6}
\t_\a&\equiv&-\bar P_\a{}^\m\bar u^\n \mathfrak{T}_{\m\n}\;,\\
\la{qq7}
\t_{\a\b}&\equiv&\phantom{-}\bar P_\a{}^\m\bar P_\b{}^\n \mathfrak{T}_{\m\n}\;,
\ea\esu
which corresponds to the kinemetric decomposition of $\mathfrak{T}_{\m\n}$ introduced by \citet{1973SPhD...18..231Z}.
Quantity $\sigma$ is the energy density of matter of the localized system, $t_\a$ is a density of linear momentum of the matter, and $t_{\a\b}$ is the stress tensor of the matter.

Equations of motion (\ref{qq1}) of the localized matter can be rewritten in terms of the kinemetric quantities as follows,
\bsu\ba\la{qq2}
\le(\sigma\bar u^\a+\t^\a\ri)_{|\a}&=&-\H\t\;,\\
\la{qq3}
\le(\t^{\a\b}+\bar u^\b\t^\a\ri)_{|\b}&=&-\H\le(\t^\a-\bar u^\a\t\ri)\;.
\ea\esu
Equation (\ref{qq2}) is equivalent to the law of conservation of energy of matter of the localized system. Equation (\ref{qq3}) is analogues to the Euler equation of motion of fluid or the equation of the force balance in case of solids.

\section{Lagrangian Perturbations of FLRW Manifold}\la{lpsm}

\subsection{The Concept of Perturbations}

In the present paper, FLRW background manifold is defined by the metric $\bar g_{\a\b}$ which dynamics is governed by background matter fields -
the Clebsch potential $\bar\Phi$ of the ideal fluid and the scalar field $\bar\Psi$. We assume that the background metric and the background values of the fields are perturbed by a localized astronomical system which is considered as a {\it bare} perturbation associated with a field variable $\Theta$.
Perturbations of the metric and the matter fields caused by the {\it bare} perturbation are considered to be small so that the perturbed metric and the matter fields can be split in their backgrounds values and the corresponding perturbations,
\be\la{pf8}
g_{\a\b}=\bar g_{\a\b}+\varkappa_{\a\b}\;,\qquad\quad \p=\bar\p+\vp\;,\qquad\quad \Psi=\bar\Psi+\psi\;.
\ee
These equations are exact. We emphasize that all functions entering equation (\ref{pf8}) are taken at one and the same point of the background manifold.
The {\it bare} perturbation does not remain the same in the presence of the perturbations of the metric and the matter fields. Therefore, the field variable $\Theta$ corresponding to the {\it bare} perturbation, is also perturbed
\be\la{ion6}
\qquad\quad \Theta=\bar\Theta+\theta\;.
\ee
We consider the perturbations of the metric - $\varkappa_{\a\b}$, the Clebsch potential - $\vp$, and the scalar field - $\psi$ as being weak with respect to their corresponding background values $\bar g_{\a\b}$, $\bar\p$, and $\bar\Psi$, which dynamics is governed by the background equations that have been explained in section \ref{bma}. Because the field variable $\Theta$ is the source of the {\it bare} perturbation, we postulate that its background value is equal to zero: $\bar\Theta=0$. The perturbations $\varkappa_{\a\b}$, $\vp$, and $\psi$ have the same order of magnitude as $\theta$.

Perturbation of the contravariant component of the metric is determined from the condition $g_{\alpha\gamma}g^{\gamma\beta}=\bar g_{\alpha\gamma}\bar g^{\gamma\beta}=\delta_\alpha^\beta$, and is given by
\begin{equation}\label{pf8q}
g^{\a\b}=\bar g^{\a\b}-\varkappa^{\a\b}+\varkappa^\alpha{}_\gamma \varkappa^{\gamma\beta}+\ldots\;,
\end{equation}
where the ellipses denote terms of the higher order.

It turns out \citep{1984CMaPh..94..379G,1988IJMPA...3.2651P} that a more convenient field variable of the gravitational field in the the theory of Lagrangian perturbations of curved manifolds, is a contravariant (Gothic) metric
\be\la{pf8a}
\gg^{\a\b}=\sqrt{-g}g^{\a\b}\;.
\ee
The convenience of the Gothic metric stems from the fact that it enters the de Donder (harmonic) gauge conditions which significantly simplifies the Einstein equations \citep{LL,wald}. The Gothic metric variable is also indispensable for concise and elegant formulation of dynamic field theories on curved manifolds \citep{Deser_1970GReGr}. Making use of the Gothic metric allows us to significantly reduce the amount of algebra in taking the first and second variational derivatives from the Hilbert Lagrangian and the Lagrangian of the background matter in FLRW universe as explains in the rest of this section.  

The covariant Gothic metric $\gg_{\b\g}$ is defined by means of equation
\be\la{pf8c}
\gg^{\a\b}\gg_{\b\g}=\delta^\a_\g\;,
\ee
that yields $\gg_{\a\b}=g_{\a\b}/\sqrt{-g}$.
We accept that $\gg^{\a\b}$ is expanded around its background value, $\bar\gg^{\a\b}=\sqrt{-\bar g}{\bar g}^{\a\b}$, as follows
\be\la{pf8b}
\gg^{\a\b}=\bar\gg^{\a\b}+\mathfrak{h}^{\a\b}\;,
\ee
which is an exact equation.

Further calculations prompt that it is more suitable to single out $\sqrt{-\bar g}$ from $\mathfrak{h}^{\a\b}$, and operate with a variable
\be\la{mcv1}
l^{\a\b}\equiv\frac{\mathfrak{h}^{\a\b}}{\sqrt{-\bar g}}\;.
\ee
This variable splits the dynamic degrees of freedom of the gravitational perturbations from the background manifold which evolves in according with the unperturbed Friedmann equations.
Tensor indices of $l^{\a\b}$ are raised and lowered with the help of the background metric, for example, $l_{\a\b}\equiv \bar g_{\a\m}\bar g_{\b\n} l^{\m\n}$.
The field variable $l^{\a\b}$ relates to the perturbation $\varkappa_{\a\b}$ of the metric tensor. To establish this relationship, we start from (\ref{pf8a}), substitute equation (\ref{pf8b}) to its left side, and expand its right side in the Taylor series with respect to $\varkappa_{\a\b}$. It results in
\be\la{mcv2}
\mathfrak{h}^{\a\b}=\frac{\pd\bar\gg^{\a\b}}{\pd\bar g_{\m\n}}\,\varkappa_{\m\n}+\frac12\frac{\pd^2\bar\gg^{\a\b}}{\pd\bar g_{\m\n}\pd\bar g_{\rho\sigma}}\,\varkappa_{\m\n}\varkappa_{\rho\sigma}+\ldots\;.
\ee
where the partial derivatives are calculated by successive application of the following rules
\bsu\ba\la{mcv3}
\frac{\pd\bar\gg^{\a\b}}{\pd\bar g_{\m\n}}&=&-\frac12\sqrt{-\bar g}\le(\bar g^{\a\m}\bar g^{\b\n}+\bar g^{\a\n}\bar g^{\b\m}-\bar g^{\a\b}\bar g^{\m\n}\ri)\;,\\\la{kob7}
\frac{\pd\bar g^{\a\b}}{\pd\bar g_{\m\n}}&=&-\frac12\le(\bar g^{\a\m}\bar g^{\b\n}+\bar g^{\a\n}\bar g^{\b\m}\ri)\;,\\\la{mrx4}
\frac{\pd\sqrt{-\bar g}}{\pd\bar g_{\m\n}}&=&+\frac12\sqrt{-\bar g}\bar g^{\m\n}\;,
\ea\esu
which can be easily confirmed by inspection.
Replacing the partial derivatives in (\ref{mcv2}) and making use of the definition (\ref{mcv1}), yields the relationship between $l^{\a\b}$ and $\varkappa^{\a\b}$ as follows
\be\la{ok9}
l^{\a\b}=-\varkappa^{\a\b}+\frac12\bar g^{\a\b}\varkappa+\varkappa^{\mu(\alpha}\varkappa^{\b)}{}_\mu-\frac12 \varkappa^{\a\b}\varkappa-\frac14\bar g^{\a\b}\left(\varkappa^{\mu\nu}\varkappa_{\mu\nu}-\frac12 \varkappa^2\right)+\ldots\;,
\ee
where $\varkappa\equiv \varkappa^\sigma{}_\sigma=\bar g^{\rho\sigma}\varkappa_{\rho\sigma}$, and ellipses denote terms of the cubic and higher order in $\varkappa_{\a\b}$.

Perturbations of four-velocities, $w^\alpha$ and $v^\alpha$, entering definitions of the energy-momentum tensors (\ref{pf2b}), (\ref{h16c}), are fully determined by the perturbations of the metric and the potentials of the matter fields. Indeed, according to definitions (\ref{pf6}) and (\ref{h15d}) the four-velocities are defined by the following equations
\be\la{pf8aa}
w_\alpha=-\frac{\Phi_{,\a}}{\m_{\rm m}}\;,\qquad\qquad v_\alpha=-\frac{\Psi_{,\a}}{\m_{\rm q}}\;.
\ee
where $\m_{\rm m}=\sqrt{-g^{\a\b}\Phi_{,\alpha}\Phi_{,\beta}}$ and $\m_{\rm q}=\sqrt{-g^{\a\b}\Psi_{,\alpha}\Psi_{,\beta}}$ in accordance with (\ref{pf7}) and (\ref{h15c}) respectively.
We define perturbation of the covariant components of the four-velocities as follows
\be\la{pf8s}
w_\a=\bar u_\a+\delta w_\a\;,\qquad\qquad v_\a=\bar u_\a+\delta v_\a\;,
\ee
where the unperturbed values of the four-velocities coincide and are equal to the four-velocity of the Hubble flow due to the requirement of the homogeneity and isotropy of the background FLRW metric. Substituting these expansions to the left side of definitions (\ref{pf8aa}), and expanding its right side by making use of the expansions (\ref{pf8}) and (\ref{pf8q}) of the scalar fields and the metric, yields
\be\la{kio}
\delta w_\a=-\frac{1}{\bar\m_{\rm m}}{\bar P}^\b{}_\a\phi_{|\b}-\frac12\mathfrak{q}\bar u_\a\;,\qquad\qquad
\delta v_\a=-\frac{1}{\bar\m_{\rm q}}{\bar P}^\b{}_\a\psi_{|\b}-\frac12\mathfrak{q}\bar u_\a\;,
\ee
where we have introduced a new notation
\be\la{mt6k}
\mathfrak{q}\equiv-\bar u^\a \bar u^\b \varkappa_{\a\b}\;,
\ee
for the gravitational perturbation of the metric tensor projected on the background four-velocity of the Hubble flow. Making use of $l_{\a\b}$, the previous equation can be recast to
\be\la{mt6b}
\mathfrak{q}\equiv \bar u^\a \bar u^\b l_{\a\b}+\frac{l}{2}\;,
\ee
where $l\equiv l^\alpha{}_\alpha=\bar g^{\alpha\beta}l_{\alpha\beta}$. Remembering that $\bar g^{\a\b}=\bar P^{\a\b}-\bar u^\a\bar u^\b$, we can put equation (\ref{mt6b}) yet to another form
\be\la{mt6t}
\mathfrak{q}\equiv\frac12\le(\bar u^\a \bar u^\b+\bar P^{\a\b}\ri)l_{\a\b}\;,
\ee
which is useful in the calculations that follows.

\subsection{The Perturbative Expansion of the Lagrangian}\la{peot2}

We have introduced the Lagrangian of the theory in section \ref{act}.
The Hilbert Lagrangian of the gravitational field is $\lag^{\rm g}=-\sqrt{-g}R/16\pi$, where $R$ is the Ricci scalar. The Lagrangian density of matter is $\lag^{\rm m}=\sqrt{-g}L^{\rm m}(\p, g_{\a\b})$, and the Lagrangian density of the scalar field $\lag^{\rm q}=\sqrt{-g}L^{\rm q}(\Psi, g_{\a\b})$. The matter, the scalar field as well as the spacetime manifold are perturbed by a matter of N-body system described by a set of field variables $\Theta$ with the Lagrangian density $\lag^{\rm p}=\sqrt{-g}L^{\rm p}(\Theta, g_{\a\b})$.

The action of the unperturbed FLRW universe is a functional
\be\la{pf9} \bar{\cal S}=\int_{\cal M}d^4x\bar\lag\;, \ee
depending on the unperturbed Lagrangian
\be\la{pf12}
\bar\lag=\bar\lag^{\rm g}+\bar\lag^{\rm m}+\bar\lag^{\rm q}\;,
\ee
taken on the background values of the field variables $\bar g_{\a\b}$, $\bar\p$, and $\bar\Psi$.

The presence of a localized astronomical system perturbs the spacetime manifold and the background values of the field variables. The perturbed Lagrangian becomes an algebraic sum of four terms
\be\la{pf10} \lag=\lag^{\rm g}+\lag^{\rm m}+\lag^{\rm q}+\lag^{\rm p}\;,
\ee
where the Lagrangian $\lag^{\rm p}$ describes the {\it bare} perturbation, and $\lag^{\rm g}$, $\lag^{\rm m}$, $\lag^{\rm q}$ are perturbed values of the Lagrangian of the FLRW universe.

The perturbed Lagrangian can be decomposed in a Taylor series with respect to the perturbed values of the field variables. It is achieved by substituting expansions (\ref{pf8}) to the Lagrangian (\ref{pf10}) and expanding it around the background values of the variables. It yields
\be\la{pf11}
\lag=\bar\lag+\lag_1+\lag_2+\lag_3+...\;,
\ee
where $\lag_1$, $\lag_2$, $\lag_3$, \ldots, are the Lagrangian perturbations which are linear, quadratic, cubic, and so on, with respect to the perturbations of the field variables, $h_{\a\b}$, $\phi$, $\psi$, and $\theta$. More specifically \citep{1988IJMPA...3.2651P},
\bsu\ba \la{ia1} \lag_1&=&\mathfrak{h}^{\mu\nu}
\frac{\d \bar\lag} {\d \bar\gg^{\mu\nu}} + \vp \,\frac{\delta \bar\lag}{\delta \bar\p}+\psi \,\frac{\delta \bar\lag}{\delta \bar\Psi}+\lag^{\rm p}\;,\\
\la{ia2}
\lag_2&=&\frac{1}{2!} \mathfrak{h}^{\a\beta} \frac{\d}{\d \bar\gg^{\a\beta}}\le( \mathfrak{h}^{\mu\nu} \frac{\d \bar\lag}
{\d\bar\gg^{\mu\nu}}\ri) + \frac{1}{2!}\vp\, \frac{\d} {\d\bar\p}\le( \vp\, \frac{\delta\bar\lag}{\delta
\bar\p}\ri) + \frac{1}{2!}\psi\, \frac{\d} {\d\bar\Psi}\le( \psi\, \frac{\delta\bar\lag}{\delta
\bar\Psi}\ri) \\\nonumber
&+& \mathfrak{h}^{\a\beta} \frac{\d}{\d \bar\gg^{\a\beta}}\le( \vp\, \frac{\delta\bar \lag} {\delta\bar\p}\ri)+\mathfrak{h}^{\a\beta} \frac{\d}{\d \bar\gg^{\a\beta}}\le( \psi\, \frac{\delta\bar \lag} {\delta\bar\Psi}\ri)+\mathfrak{h}^{\a\b}\frac{\d\lag^{\rm p}}{\d\bar\gg^{\a\b}}\;,
\ea\esu
and so on. Here, the variational derivatives from the Lagrangian density, $\bar\lag$, depending on the field variables and their derivatives,
are defined as follows
\bsu\ba
\la{ia3}
\frac{\d \bar\lag} {\d \bar\gg^{\mu\nu}}&\equiv&\frac{\partial \bar\lag}{\partial\bar\gg^{\mu\nu}}-\frac{\partial}{\partial x^\a}\frac{\partial \bar\lag}{\partial\bar\gg^{\mu\nu}{}_{,\a}}+\frac{\partial^2}{\partial x^\a\partial x^\b}\frac{\partial \bar\lag}{\partial\bar\gg^{\mu\nu}{}_{,\a\b}}\;,\\\la{isa2}
\frac{\d \bar\lag} {\d\bar\p}&\equiv&\frac{\partial \bar\lag}{\partial\bar\p}-\frac{\partial}{\partial x^\a}\frac{\partial \bar\lag}{\partial\bar\p_{,\a}}+\frac{\partial^2}{\partial x^\a\partial x^\b}\frac{\partial \bar\lag}{\partial\bar\p_{,\a\b}}\;,\\\la{isa3}
\frac{\d \bar\lag} {\d\bar\Psi}&\equiv&\frac{\partial \bar\lag}{\partial\bar\Psi}-\frac{\partial}{\partial x^\a}\frac{\partial \bar\lag}{\partial\bar\Psi_{,\a}}+\frac{\partial^2}{\partial x^\a\partial x^\b}\frac{\partial \bar\lag}{\partial\bar\Psi_{,\a\b}}\;,
\ea\esu
The variational derivative with respect to the metric density $\bar\gg^{\mu\nu}$ relates to the derivative with respect to the metric $\bar g^{\mu\nu}$ by an algebraic operator
\be \la{ia5}
\frac{\delta}{\delta \bar\gg^{\mu\nu}}=\frac{\pd \bar{g}^{\a\b}}{\pd\bar \gg^{\mu\nu}}\frac{\delta}{\delta \bar{g}^{\a\b}}=\frac{1}{2\sqrt{-\bar
g}}\le(\delta^\a_\mu\delta^\b_\nu+\delta^\a_\nu\delta^\b_\mu-\bar g_{\mu\nu}\bar g^{\a\b}\ri)\frac{\delta}{\delta \bar{g}^{\a\b}}\;.
\ee

One has to notice that the expansion (\ref{pf11}) is defined up to the terms which are represented as a total covariant derivative from a vector density (the, so-called, divergent terms). For example, the direct Taylor series expansion shows that the Lagrangian $\lag_2$ has a term with cross-coupling of $\phi$ and $\psi$. This term was eliminated from $\lag_2$ because it can be represented as a covariant divergence from a vector that vanishes identically after taking the Lagrangian derivative \citep{1988IJMPA...3.2651P}. The divergency terms can be important in the discussion of the boundary conditions but they do not enter the equations of motion of fields which represent a system of differential equations in partial derivatives for the perturbations of the dynamic (field) variables. Furthermore, it is straightforward to prove that any of the Lagrangian derivatives (\ref{ia3})-(\ref{isa3}), applied to a partial derivative of a geometric object $F=F(\bar\gg^{\a\b};\bar\p;\bar\Psi;\bar\gg^{\a\b}{}_{,\g};\bar\p_{,\g};\bar\Psi_{,\g};\ldots)$, vanishes \cite{1969fpoo.book..326M}
\be
\la{gd+}
\frac{\delta}{\delta \bar\gg^{\a\b}}\le(\frac{\pd F}{\pd x^\a}\ri)= 0\;,\qquad \frac{\delta}{\delta \bar\p}\le(\frac{\pd F}{\pd x^\a}\ri)= 0\;,\qquad \frac{\delta}{\delta \bar\Psi}\le(\frac{\pd F}{\pd x^\a}\ri)= 0\;.
\ee
Equations (\ref{gd+}) does not hold for a covariant derivative \cite{1969fpoo.book..326M}. We shall use equation (\ref{gd+}) for bringing the Lagrangian derivatives to a simpler form.

The field equations are obtained by taking the variational derivatives from the perturbed action with respect to various variables subject to the least action principle. In accordance with this principle, the variational derivatives from the perturbed Lagrangian must vanish,
\be\la{gd2w}
\frac{\d \lag} {\d \gg^{\mu\nu}}=0\;,\qquad
\frac{\d \lag} {\d\p}=0\;,\qquad
\frac{\d \lag} {\d\Psi}=0\;.
\ee
We substitute the Taylor decomposition (\ref{pf11}) of the Lagrangian to equations (\ref{gd2w}) and separate the background value of the derivatives from their perturbed values. We assume that gravitational dynamics the unperturbed universe obeys the background field equations. Then, the perturbed part of the equations represent a series of equations of the first, second, third, etc. order, which can be solved by successive iterations. In this paper we restrict ourselves with the linearized approximation of the first order with respect to the perturbations. It generalizes the first post-Minkowskian approximation to the case of the expanding universe.

\subsection{The Background Field Equations}

The dynamics of the background universe is governed by the variational equations
\begin{subequations}
\ba\la{h1}
\frac{\d\bar\lag^{\rm g}}{\d\bar\gg_{\a\b}}+\frac{\d\bar\lag^{\rm m}}{\d\bar\gg_{\a\b}}+\frac{\d\bar\lag^{\rm q}}{\d\bar\gg_{\a\b}}&=&0\;,\\\la{h1a}
\frac{\d\bar\lag^{\rm m}}{\d\bar\p}&=&0\;,\\\la{h1b}
\frac{\d\bar\lag^{\rm q}}{\d\bar\Psi}&=&0\;.
\ea
\end{subequations}
After performing the derivatives, equation (\ref{h1}) becomes the Einstein equation (\ref{a11}), equation (\ref{h1a}) is reduced to equation of continuity (\ref{cr30}) after taking into account the thermodynamic relationship (\ref{pf3a}), and equation (\ref{h1b}) is equivalent to (\ref{cr0}).   These equations have been thoroughly discussed in section \ref{bma}. Solution of these equations depend on equation of state of background matter. We assume that the solution exists and that the time dependence of the FLRW metric $\bar g_{\a\b}=\bar g_{\a\b}(\eta)$, the Clebsch potential  $\bar\p=\bar\p(\eta)$, and the scalar field $\bar\Psi=\bar\Psi(\eta)$ is explicitly known.

\subsection{The Gravitational Field Perturbations}
The gravitational field perturbations satisfy the following (exact) differential equation
\be \la{mt1}
F_{\m\n} = 8\pi\le(\mathfrak{T}_{\m\n}+t_{\m\n}\ri)\;,
\ee
which generalizes the Einstein field equations of the post-Minkowskian approximations in asymptotically flat spacetime to the case of the expanding universe.
Tensor $F_{\m\n}$ is an algebraic superposition
\be\la{mt1b}
F_{\m\n}\equiv F^{\rm g}_{\m\n}+F^{\rm m}_{\m\n}+F^{\rm q}_{\m\n}\;,
\ee
where the linear operators in the right side are defined through the Lagrangian derivatives as follows,
\bsu\ba\la{mt2} F^{\rm g}_{\m\n} &\equiv& -\frac{16\pi}{\sqrt{-\bar g}}\frac{\d}{\d\bar{g}^{\m\n}} \le(\mathfrak{h}^{\a\b} \frac{\d\bar\lag^{\rm g}}{\d\bar\gg^{\a\b}}\ri)\;,\\
\la{mt3}
F^{\rm m}_{\m\n}&\equiv& -\frac{16\pi}{\sqrt{-\bar g}} \frac{\d}{\d\bar{g}^{\m\n}} \le(\mathfrak{h}^{\a\b} \frac{\d\bar\lag^{\rm m}}{\d\bar \gg^{\a\b}} + \vp\, \frac{\d\bar\lag^{\rm m}}{\d\bar\p}\ri)\;,\\
\la{mt3a}
F^{\rm q}_{\m\n}&\equiv& -\frac{16\pi}{\sqrt{-\bar g}} \frac{\d}{\d\bar{g}^{\m\n}} \le(\mathfrak{h}^{\a\b} \frac{\d\bar\lag^{\rm q}}{\d\bar \gg^{\a\b}} + \psi\, \frac{\d\bar\lag^{\rm q}}{\d\bar\Psi}\ri)\;.
\ea\esu
The right side of equation (\ref{mt1}) contains the tensor of energy-momentum $\mathfrak{T}_{\m\n}$ of the {\it bare} gravitational perturbation which is generated by the matter of the localized astronomical system and can be calculated as the Lagrangian derivative (\ref{q19z}) .
The right side of (\ref{mt1}) also contains the non-linear corrections that are given by
\be \la{mt5} t_{\m\n}=\frac{2}{\sqrt{-\bar g}}\left(\frac{\d\lag_2}{\d\bar{g}^{\m\n}}+\frac{\d\lag_3}{\d\bar{g}^{\m\n}}+\ldots\right)\;.
\ee
In what follows, we shall neglect the contribution of $t_{\m\n}$ as it is of the higher order compared with other terms in (\ref{mt1}).

The differential operator, $F^{\rm g}_{\m\n}$, represents a linearized perturbation of the Ricci tensor, and after calculation of (\ref{mt2}), is given by
\be \la{mt6}
F^{\rm g}_{\m\n}= \frac{1}{2}\le(l_{\m\n}{}^{|\a}{}_{|\a} + \bar {g}_{\m\n} l^{\a\b}{}_{|\a\b} - l_{\a\m|\n}{}^{|\a} - l_{\a\n|\m}{}^{\a}\ri)\;, \ee
where each vertical bar denotes a covariant derivative with respect to the background metric $\bar g_{\m\n}$.

Operators $F^{\rm m}_{\m\n}$ and $F^{\rm q}_{\m\n}$ depend essentially on a particular choice of the Lagrangian of matter and scalar field, and take on different forms depending on the specific analytic dependence of ${\cal L}^{\rm m}$ and $\lag^{\rm q}$ on the field variables. In the particular case of the ideal fluid, the term embraced in the round parentheses in the right side of equation (\ref{mt3}) is
\be\la{io1}
\mathfrak{h}^{\a\b} \frac{\d\bar\lag^{\rm m}}{\d\bar \gg^{\a\b}} + \vp\, \frac{\d\bar\lag^{\rm m}}{\d\bar\p}=
\frac12\mathfrak{h}^{\a\b}\le(\bar T^{\rm m}_{\a\b}-\frac12\bar g_{\a\b}\bar T^{\rm m}\ri)+\phi\,\pd_\a\le(\bar\r_{\rm m}\sqrt{-\bar g}\bar u^\a\ri)\;,
\ee
where $\bar u^\a\equiv -\bar g^{\a\b}\bar\p_{,\b}/\bar\m_{\rm m}$, and $\bar T^{\rm m}_{\a\b}$ is given in (\ref{pf2b}). We emphasize that though the ideal fluid satisfies the equation of continuity (\ref{cr30}), it should not be immediately implemented in (\ref{io1}) because this expression is to be further differentiated with respect to the metric tensor according to (\ref{mt3}).

For the scalar field, the term enclosed to the round parentheses in the right side of (\ref{mt3a}) is
\be\la{io2}
\mathfrak{h}^{\a\b} \frac{\d\bar\lag^{\rm q}}{\d\bar \gg^{\a\b}} + \psi\, \frac{\d\bar\lag^{\rm q}}{\d\bar\Psi}=
\frac12\mathfrak{h}^{\a\b}\le(\bar T^{\rm q}_{\a\b}-\frac12\bar g_{\a\b}\bar T^{\rm q}\ri)+\psi\le[\sqrt{-\bar g}\frac{\pd\bar W}{\pd\bar\Psi}+\pd_\a\le(\bar\r_{\rm q}\sqrt{-\bar g}\bar u^\a\ri)\ri]\;,
\ee
where $\bar u^\a\equiv -\bar g^{\a\b}\bar\Psi_{,\b}/\bar\m_{\rm q}$, $\bar\r_{\rm q}=\bar\m_{\rm q}$, $\bar T^{\rm q}_{\a\b}$ is given in (\ref{h16c}), and the equation of continuity for the scalar field (\ref{crq}) should not be implemented until differentiation with respect to the metric tensor (\ref{mt3a}) is completed.

Taking the variational derivatives with respect to $\bar g^{\m\n}$ from the expressions (\ref{io1}) and (\ref{io2}), and applying thermodynamic equations (\ref{pf5}), allows us to write down the right sides of equations (\ref{mt3}), (\ref{mt3a}) as follows,
\ba\la{mt6a}
F^{\rm m}_{\m\n}&=&-4\pi\left[(\bar p_{\rm m}-\bar\e_{\rm m})l_{\mu\nu}+\left(1-\frac{c^2}{c^2_{\rm s}}\right)(\bar \e_{\rm m}+\bar p_{\rm m}) \mathfrak{q} \bar u_\mu \bar u_\nu\right]
\\\nonumber &&+
8\pi\bar\r_{\rm m}\left\{\bar u_\mu\vp_{,\nu}+\bar u_\nu\vp_{,\mu}+\le[\left(1-\frac{c^2}{c^2_{\rm s}}\right)\bar u_\mu \bar u_\nu-\bar g_{\mu\nu}\ri]\bar u^\a\vp_{,\a}\right\}\;, \\\nonumber\\
\la{mt6c}
F^{\rm q}_{\m\n}&=&-4\pi\left[\le(p_{\rm q}-\e_{\rm q}\ri) l_{\mu\nu}-2\bar g_{\m\n}\frac{\pd\bar W}{\pd\bar\Psi}\,\psi\right]+8\pi\bar\r_{\rm q}\left(\bar u_\m\psi_{,\n}+\bar u_\n\psi_{,\m}-\bar g_{\mu\nu}\bar u^\a\psi_{,\a}\right)\;, \ea
where $\bar\r_{\rm q}\equiv\dot{\bar\Psi}/a$ in accordance with definition (\ref{h15d}). The potential energy of the scalar field, $\bar W=\bar W(\bar\Psi)$, remains arbitrary as yet.

It is important to emphasize that in the most general case the ratio $c^2_{\rm s}/c^2$ of the speed of sound in fluid to the fundamental speed $c$, is not equal to the parameter $w_{\rm m}$ of the equation of state (\ref{a10}), that is $w_{\rm m}\not=(c_{\rm s}/c)^2$. Indeed, the speed of sound is defined as a partial derivative of pressure $p_{\rm m}$ with respect to the energy density $\e_{\rm m}$ taken under the condition of a constant entropy $s_{\rm m}$,
\be\la{a8a}
\frac{c^2_{\rm s}}{c^2}=\le(\frac{\partial p_{\rm m}}{\partial\e_{\rm m}}\ri)_{s_{\rm m}={\rm const.}}\;.
\ee
This equation is equivalent to the following relationship
\be\la{a8b}
\frac{c^2_{\rm s}}{c^2}=\frac{\le(\partial p_{\rm m}/\partial\m_{\rm m}\ri)_{s_{\rm m}={\rm const.}}}{\le(\partial \e_{\rm m}/\partial\m_{\rm m}\ri)_{s_{\rm m}={\rm const.}}}\;,
\ee
which is a consequence of thermodynamic relationships and definition of a partial derivative. The ratio of the partial derivatives in (\ref{a8b}) is not reduced to $w_{\rm m}$ in case when $w_{\rm m}$ depends on some other thermodynamic parameters implicitly depending on the specific enthalpy. For example, in case of an ideal gas the equation of state $p_{\rm m}=w_{\rm m}\epsilon_{\rm m}$, where $w_{\rm m}=kT/mc^2$, $k$ is the Boltzmann constant, $m$ - mass of a particle of the ideal fluid, and $T$ is the fluid temperature. The speed of sound $c^2_{\rm s}=c^2\le(\partial p_{\rm m}/\partial\e_{\rm m}\ri)_{s_{\rm m}={\rm const.}}=\Gamma w_{\rm m}>w_{\rm m}=p_{\rm m}/\e_{\rm m}$, where $\Gamma>1$ is the ratio of the heat capacities of the gas taken for the constant pressure and the constant volume respectively.

The scalar field with the potential function $W(\Psi)\not=0$ does not bear all thermodynamic properties of an ideal fluid. Nevertheless, we can formally define the speed of "sound" $\hat c_{\rm s}$ propagating in the scalar field "fluid", by equation being similar to (\ref{a8b}). More specifically,
\be\la{a8c}
\frac{\hat c^2_{\rm s}}{c^2}=\frac{\le(\partial p_{\rm q}/\partial\m_{\rm q}\ri)_{\Psi={\rm const.}}}{\le(\partial \e_{\rm q}/\partial\m_{\rm q}\ri)_{\Psi={\rm const.}}}\;.
\ee
Simple calculation reveals that the speed of "sound" of the scalar field is always equal to the fundamental speed
\be\la{a8d}
\hat c_{\rm s}=c\;,
\ee
irrespectively of the value of the potential function $W(\Psi)$. It explains why the terms being proportional to the factor $1-c^2/\hat c^2_{\rm s}$, do not appear in the expression (\ref{mt6c}) as contrasted with (\ref{mt6a}).

\subsection{The Ideal Fluid Perturbations}
 The perturbed field equations for the ideal fluid are obtained by taking the variational derivatives with respect to the field $\p$ from the Lagrangian (\ref{pf10}) - it corresponds to the middle equation in (\ref{gd2w}). Taking into account the background equation (\ref{h1a}) yields the equation of sound waves propagating in the fluid as small perturbations,
\be \la{mt7}
F^{\rm m} = 8\pi\Sigma^{\rm m}\;, \ee
where the linear differential operator
\be \la{mt8} F^{\rm m}\equiv -\frac{1}{\sqrt{-\bar g}}\frac{\d}{\d\bar\p}\le( \mathfrak{h}^{\m\n}\frac{\d\bar\lag^{\rm m}}{\d\bar\gg^{\m\n}} + \vp \frac{\d\bar\lag^{\rm m}}{\d\bar\p}\ri)\;,
\ee
and the source term
\be
\la{mt9}
\Sigma^{\rm m} \equiv\frac{1}{8\pi\sqrt{-\bar g}}\left(\frac{\d\lag^{\rm m}_2}{\d\bar\p}+\frac{\d\lag^{\rm m}_3}{\d\bar\p}+...\right)\;.
\ee
In the case of a single-component ideal fluid, the Lagrangian (\ref{pzq1}) depends merely on the derivative of the Clebsch potential $\p$ and on the metric tensor. Therefore, the explicit form of the linear operator $F^{\rm m}$ is reduced to a covariant divergence
\be \la{mt10}
F^{\rm m}=Y^\a{}_{|\a}\;,
\ee
where a vector field
\be\la{vf1}
Y^\a\equiv\frac{\pd}{\pd\bar\p_{,\a}}\le[\le(l^{\m\n}-\frac12l\bar g^{\m\n}\ri)\le(\frac{\pd\bar L^{\rm m}}{\pd\bar g^{\m\n}}-\frac12 g_{\m\n} \bar L^{\rm m}\ri) + \vp_{,\b} \frac{\pd\bar L^{\rm m}}{\pd\bar\p_{,\b}}\ri]\;,
\ee
where the partial derivatives are taken from the Lagrangian $L^{\rm m}$, but not from its density $\lag^{\rm m}=\sqrt{-g}L^{\rm m}$.
More specifically, calculations yield
\be\la{mt10a}
Y^\a\equiv\frac{\bar\r_{\rm m}}{\bar\mu_{\rm m}}~\vp^{|\a}-\bar\r_{\rm m} l^{\a\b}\bar u_\b+\left(1-\frac{c^2}{c^2_{\rm s}}\right)\le(\frac{\bar\r_{\rm m}}{\bar\mu_{\rm m}}\bar u^\a \bar u^\b\vp_{|\b} -\frac12\bar\r_{\rm m} \bar u^\a \mathfrak{q}\ri)\;.
\ee
Similar equations were derived by Lukash \citep{1980JETPL..31..596L} who used the variational method to analyze production of sound waves in the early universe.

\subsection{The Scalar Field Perturbations}
Equations for the scalar field perturbations are derived by taking the variational derivative from the Lagrangian (\ref{pf10}) with respect to the field variable $\Psi$ - see the last equation in (\ref{gd2w}). Subtracting the background equation (\ref{h1b}) leads to
\be \la{mt7q}
F^{\rm q} = 8\pi\Sigma^{\rm q}\;, \ee
where the linear differential operator
\be \la{mt8q}
F^{\rm q}\equiv  -\frac{1}{\sqrt{-\bar g}}\frac{\d}{\d\bar\Psi}\le( \mathfrak{h}^{\m\n}\frac{\d\bar\lag^{\rm q}}{\d\bar\gg^{\m\n}} + \psi \frac{\d\bar\lag^{\rm q}}{\d\bar\Psi}\ri)\;,
\ee
and the source term
\be
\la{mt9q}
\Sigma^{\rm q} \equiv\frac{1}{8\pi\sqrt{-\bar g}}\left(\frac{\d\lag^{\rm q}_2}{\d\bar\Psi}+\frac{\d\lag^{\rm q}_3}{\d\bar\Psi}+...\right)\;.
\ee
According to equation (\ref{hl4}), the Lagrangian density of the scalar field $\lag^{\rm q}=\sqrt{-g}L^{\rm q}$ depends on both the field $\Psi$ and its first derivative, $\Psi_{,\a}$. For this reason, the differential operator $F^{\rm q}$ is not reduced to the covariant derivative from a vector field as the partial derivative of the Lagrangian with respect to $\Psi$ does not vanish. We have
\be\la{mt10c}
F^{\rm q}\equiv Z^\a{}_{|\a}-\frac{l}{2}\frac{\pd\bar W}{\pd\bar\Psi}-\psi\frac{\pd^2\bar W}{\pd\bar\Psi^2}
\ee
where $l\equiv \bar g^{\a\b}l_{\a\b}$, and vector field
\be\la{mt11}
Z^\a\equiv\frac{\pd}{\pd\bar\Psi_{,\a}}\le[\le(l^{\m\n}-\frac12l\bar g^{\m\n}\ri)\le(\frac{\pd\bar L^{\rm q}}{\pd\bar g^{\m\n}}-\frac12 \bar g_{\m\n} \bar L^{\rm q}\ri) + \psi_{,\b} \frac{\pd\bar L^{\rm q}}{\pd\bar\Psi_{,\b}}\ri]\;.
\ee
Performing the partial derivatives in equation (\ref{mt11}), yields a  rather simple expression
\be\la{mt12}
Z^\a\equiv\psi^{|\a}-\bar\r_{\rm q}l^{\a\b}\bar u_\b\;,
\ee
where we have used equation $\bar\Psi_{|\a}=-\bar u^\b\bar\Psi_{|\b}\bar u_\a=-\bar\r_{\rm q}\bar u_\a$. The reader is invited to compare equation (\ref{mt12}) with (\ref{mt10a}) to observe the differences between the Lagrangian perturbations of the ideal fluid and the scalar field. One may observe that (\ref{mt10a}) becomes identical with (\ref{mt12}) in the limit $c_{\rm s}\rightarrow c$, and $\bar\rho_{\rm m}\rightarrow \bar\mu_{\rm m}$. This corresponds to the case of an extremely rigid equation of state $w_{\rm m}=1$ in equation (\ref{a10}). According to the discussion following equations (\ref{a8c}), (\ref{a8d}) the speed of 'sound' $\hat c_{\rm s}$ in the scalar field 'fluid' is always equal to $c$. However, it does not assume that the parameter $w_{\rm q}$ of the equation of state of the scalar field, $\bar p_{\rm q}=w_{\rm q}\bar\e_{\rm q}$, in (\ref{a10}) is equal to unity. This is because the scalar field is not completely equivalent to the ideal fluid in the sense of thermodynamic \citep{2010deto.book.....A}.
.

\subsection{The Lagrangian Equations for Field Variables}

\subsubsection{Equations for the metric tensor perturbations}

Linearized equations for gravitational field variables, $l_{\m\n}$, are obtained from (\ref{h1}) after neglecting in its right side the non-linear source $t_{\m\n}$, and rendering a series of transformations which re-arrange and sort out similar terms.
First, let us make use of Einstein's equations (\ref{mt6a}) and (\ref{mt6c}) to find
\ba
\la{qa1a}
F^{\rm m}_{\m\n}+F^{\rm q}_{\m\n}&=&
4\pi\le(\bar\e-\bar p\ri)l_{\m\n}\\\nonumber
&+&8\pi\bar\r_{\rm m}\left[\bar u_\mu\vp_{,\nu}+\bar u_\nu\vp_{,\mu}-\bar g_{\m\n}u^\a\vp_{,\a}+ \left(1-\frac{c^2}{c^2_{\rm s}}\right)\le(\bar u^\a\vp_{,\a}-\frac12\bar\m_{\rm m}\mathfrak{q}\ri)\bar u_\mu \bar u_\nu\right]\\\nonumber
&+&8\pi\bar\r_{\rm q}\left[\bar u_\mu\psi_{,\nu}+\bar u_\nu\psi_{,\mu}-\bar g_{\m\n}u^\a\psi_{,\a}+ \bar g_{\m\n}\frac{\pd\bar W(\bar\Psi)}{\pd\bar\Psi}\frac{\psi}{\bar\m_{\rm q}}\right]
\;.
\ea

Second step is to transform the linear differential operator $F^{\rm g}_{\m\n}$ in (\ref{mt6}) to a more convenient form that will allow us to single out the gauge-dependent terms denoted by
\be\la{qe1}
{A}^\m\equiv l^{\m\n}{}_{|\n}\;.
\ee
Changing the order of the covariant derivatives in (\ref{mt6}) and taking into account that the commutator of the second covariant derivatives is proportional to the Riemann tensor, we can recast (\ref{mt6}) to the following form,
\be \la{qe2}
F^{\rm g}_{\m\n}\equiv \frac{1}{2}\le(l_{\m\n}{}^{|\a}{}_{|\a} + \bar {g}_{\m\n} {A}^\a{}_{|\a} - {A}_{\m|\n} - {A}_{\n|\m}\ri)-\bar R^\a{}_{(\m} l_{\n)\a}-\bar R_{\m\a\b\n}l^{\a\b}\;,
\ee
where the round brackets around indices denote symmetrization. The terms with the Ricci and Riemann tensors can be expressed in terms of the total background energy and pressure of the ideal fluid and scalar field by making use of equations (\ref{a1}), (\ref{a3}) and Einstein's equations (\ref{a11}). It yields
\be\la{qe3}
%%%%%%%%%%%%%%%%%%% R^\a{}_{(\m} l_{\n)\a}+\bar R_{\m\a\b\n}l^{\a\b}\equiv\frac{1}{a^2}\le\{\dot H\le(4l_{\m\n}-\bar g_{\m\n} l\ri)-\le(\dot H-H^2-k\ri)\le[\le({\bar P}_{\a\m}{\bar P}_{\b\n}-{\bar P}_{\m\n}{\bar P}_{\a\b}\ri)l^{\a\b}+2{\bar P}^\a{}_{(\m}l_{\n)\a}\ri]\ri\}\;,%%%%%%%%%%%%%%%%%%%%%
R^\a{}_{(\m} l_{\n)\a}+\bar R_{\m\a\b\n}l^{\a\b}=
4\pi\le[\le(\frac{5\bar\e}{3}-\bar p\ri)l_{\m\n}+\frac{l}2\le(\bar p-\frac{\bar\e}{3}\ri)\bar g_{\m\n}+\le(\bar\e+\bar p\ri)\le(2\bar u^\a\bar u_\m l_{\n\a}+2\bar u^\a\bar u_\n l_{\m\a}-\bar u_\m\bar u_\n l-\bar g_{\m\n}\mathfrak{q}\ri)\ri]\,.
\ee
Finally, substituting equations (\ref{qa1a}), (\ref{qe2}) and (\ref{qe3}) to (\ref{mt1}) results in
\ba\la{qe3a}
l_{\m\n}{}^{|\a}{}_{|\a} + \bar {g}_{\m\n} {A}^\a{}_{|\a} - {A}_{\m|\n} - {A}_{\n|\m}&&\\\nonumber
-16\pi\le[ \frac{\bar\e}{3} l_{\m\n}+\frac{l}4\le(\bar p-\frac{\bar\e}{3}\ri)\bar g_{\m\n}+\le(\bar\e+\bar p\ri)\le(\bar u^\a\bar u_\m l_{\n\a}+\bar u^\a\bar u_\n l_{\m\a}-\frac12\bar u_\m\bar u_\n l-\frac12\bar g_{\m\n}\mathfrak{q}\ri) \ri]&&\\\nonumber
+16\pi\bar\r_{\rm m}\left[\bar u_\mu\vp_{,\nu}+\bar u_\nu\vp_{,\mu}-\bar g_{\m\n}u^\a\vp_{,\a}+ \left(1-\frac{c^2}{c^2_{\rm s}}\right)\le(\bar u^\a\vp_{,\a}-\frac12\bar\m_{\rm m}\mathfrak{q}\ri)\bar u_\mu \bar u_\nu\right]&&\\\nonumber
+16\pi\bar\r_{\rm q}\left[\bar u_\mu\psi_{,\nu}+\bar u_\nu\psi_{,\mu}-\bar g_{\m\n}u^\a\psi_{,\a}+ \bar g_{\m\n}\frac{\pd\bar W(\bar\Psi)}{\pd\bar\Psi}\frac{\psi}{\bar\m_{\rm q}}\right]&=&16\pi \mathfrak{T}_{\m\n}\;,
\ea
where the non-linear term $t_{\m\n}$ was neglected.

The first term in (\ref{qe3a}) is a covariant Laplace-Beltrami operator, $l_{\m\n}{}^{|\a}{}_{|\a}\equiv\bar g^{\a\b}l_{\m\n|\a\b}$, that is a rather complicated geometric object. Its explicit expression can be developed by making use of the Christoffel symbols given in (\ref{bm6}). Tedious but straightforward calculation yields
\ba\la{qe4}
l_{\m\n}{}^{|\a}{}_{|\a}&=&l_{\m\n}{}^{;\a}{}_{;\a}+2{\H}\bar u^\a l_{\m\n;\a}-2\le({\H}\bar u^\a l_{\a\m}\ri)_{|\n}
-2\le({\H}\bar u^\a l_{\a\n}\ri)_{|\m}\\\nonumber&&+~2{\H}\le(\bar u_{\m}{A}_\n+\bar u_\n {A}_\m\ri) +2{\H}'\le(l_{\m\n}-\bar u^\a\bar u_\m l_{\n\a}-\bar u^\a\bar u_\n l_{\m\a}\ri)\\\nonumber&&+~2{\H}^2\le(2l_{\m\n}+3\bar u_\m\bar u^\a l_{\a\n}+3\bar u_\n\bar u^\a l_{\a\m}-\bar g_{\m\n}\bar u^\a\bar u^\b l_{\a\b}-\bar u_\m\bar u_\n l \ri) \;,
\ea
where the semicolon denotes a covariant derivative that is calculated with the Christoffel symbols $B^\a{}_{\m\n}$ like in (\ref{bmzqw}), and the differential operator $l_{\m\n}{}^{;\a}{}_{;\a}\equiv\bar g^{\a\b}l_{\m\n;\a\b}$.

%%%%%%%%%%  \be\la{qe5}\Box_{\scriptscriptstyle f} l_{\m\n}\equiv \bar{\mathtt f}^{\a\b}l_{\m\n;\a\b}\;.\ee     %%%%%%%%%%%%%%%%%%%%%

Further derivation of the differential equations for linearized metric tensor perturbations can be significantly simplified if we re-define the gauge function, ${A}^\a\equiv l^{\m\n}{}_{|\n}$, in the following form
\be
\la{qe6}
%%%%%%%%%%%%%%%           {A}^\a=-2\H l^{\a\b}\bar u_\b+16\pi\le(\beta_{\rm m}\bar\r\bar u^\a\vp+\beta_{\rm q}\frac{\dot{\bar\Psi}}{a}\bar u^\a\psi\ri)\;,          %%%%%%%%%%%%%%%%%%%%%%%%%
{A}^\a=-2\H l^{\a\b}\bar u_\b+16\pi\le(\bar\r_{\rm m}\vp+\bar\r_{\rm q}\psi\ri)\bar u^\a+{B}^\a\;,
\ee
where ${B}^\a$ is an arbitrary gauge function. This choice of the gauge function ${A}^\a$ allows us to eliminate two terms in equation (\ref{qe4}) which depend on the first covariant derivatives with respect to the background metric $\bar g_{\a\b}$. Moreover, it allows to eliminate a number of terms depending on the first derivatives of the fields $\phi$ and $\psi$ in equation (\ref{qe3a}). Since we keep the gauge function ${B}^\a$ arbitrary, the equation (\ref{qe6}) does not fix any gauge. The choice of the gauge is controlled by the gauge function ${B}^\a$.

One substitutes the gauge function (\ref{qe6}) to equations (\ref{qe4}) and (\ref{qe3a}) and make use of the background Friedmann equations (\ref{a12}), (\ref{a13}) to replace the background values of the energy density, $\bar\e$, and pressure, $\bar p$, with the Hubble parameter $\H$ and its time derivative ${\H}'$. It brings about equation (\ref{qe3a}) to the following form
\ba
\la{qe7}
l_{\m\n}{}^{;\a}{}_{;\a}+2{\H}\bar u^\a l_{\m\n;\a}
+2\le(\H'+{\H}^2\ri)\le(l_{\m\n}+\bar u_\m\bar u^\a l_{\a\n}+\bar u_\n\bar u^\a l_{\a\m}-l\bar u_\m\bar u_\n\ri)&&\\\nonumber
-\frac{2k}{a^2}\le[l_{\m\n}+2\bar u_\m\bar u^\a l_{\a\n}+2\bar u_\n\bar u^\a l_{\a\m}-l\bar u_\m\bar u_\n-\le(\mathfrak{q}+\frac{l}{2}\ri)\bar g_{\m\n}\ri]&&\\\nonumber
+16\pi\bar u_\m\bar u_\n\le[\bar\r_{\rm m}\le(1-\frac{c^2}{c^2_{\rm s}}\ri)\le(\bar u^\a\phi_{,\a}-\frac12\bar\m_{\rm m}\mathfrak{q}\ri)-2\frac{\pd\bar W}{\pd\bar\Psi}\psi-4{\H}\le(\bar\r_{\rm m}\phi+\bar\r_{\rm q}\psi\ri)\ri]&&\\\nonumber
+\bar g_{\m\n}{B}^\a{}_{|\a}-{B}_{\m|\n}-{B}_{\n|\m}
+2{\H}\le(\bar u_\m{B}_\n+\bar u_\n{B}_\m-\bar g_{\m\n}\bar u^\a{B}_\a\ri)
&=&16\pi \mathfrak{T}_{\m\n}\;.
\ea
This equation is fully covariant and is valid in any gauge and/or coordinate chart. It clarifies the advantage in the choice of the gauge function (\ref{qe6}).

Indeed, if one works in the isotropic coordinates associated with the Hubble flow, where $\bar U^\a=(1/a,0,0,0)$, it allows us to fully decouple the differential equations for $l_{00}$, $l_{0i}$ and $l_{ij}$ components of the metric tensor perturbations. Let us assume, for simplicity, $B^\a=0$ that is an analogue of the harmonic gauge in asymptotically-flat spacetime. Then, different tensor components of equation (\ref{qe7}) become
\bsu\ba\la{qe7a}
\Box q+2{\cal H} q_{;0}+4k q-4\pi\le(1-\frac{c^2}{c^2_{\rm s}}\ri)\bar\r_{\rm m}\bar\m_{\rm m}q
&=&8\pi\le(\mathfrak{T}_{00}+\mathfrak{T}_{kk}\ri)-\\\nonumber
&&8\pi a\le[\bar\r_{\rm m}\le(1-\frac{c^2}{c^2_{\rm s}}\ri)\phi_{,0}-2a\frac{\pd\bar W}{\pd\bar\Psi}\psi-4{\cal H}\le(\bar\r_{\rm m}\phi+\bar\r_{\rm q}\psi\ri)\ri]\;,\\\la{qe7b}
\Box l_{0i}+2{\cal H} l_{0i;0}+2k l_{0i}&=&16\pi \mathfrak{T}_{0i}\;,\\\la{qe7c}
\Box l_{<ij>}+2{\cal H} l_{<ij>;0}+2\le({\cal H}'-k\ri) l_{<ij>} &=&16\pi \mathfrak{T}_{<ij>}\;,\\\la{qe7d}
\Box l_{kk}+2{\cal H} l_{kk;0}+2\le({\cal H}'+2k\ri) l_{kk} &=&16\pi \mathfrak{T}_{kk}\;.
\ea\esu
Here, we denoted $\Box l_{\m\n}=\bar\mm^{\a\b}l_{\m\n;\a\b}$, $q=\le(l_{00}+l_{kk}\ri)/2$, $l_{kk}=l_{11}+l_{22}+l_{33}$, $l_{<ij>}=l_{ij}-(1/3)\d_{ij}l_{kk}$, and the same index notations are applied to the tensor of energy-momentum $\mathfrak{T}_{ij}$ of the localized astronomical system. These equations are clearly decoupled from one another, thus, demonstrating the advantage of the gauge condition (\ref{qe6}) used along with $B^\a=0$.

Equations (\ref{qe7b})--(\ref{qe7d}) can be solved independently if the initial and boundary conditions are known, and the tensor of energy-momentum of the localized astronomical system is well-defined. Equation (\ref{qe7a}) for scalar $q$ besides knowledge of $\mathfrak{T}_{\a\b}$, demands to know the scalar field perturbations, $\phi$ and $\psi$, that contribute to the source of $q$ standing in the right side of (\ref{qe7a}). Equations for these perturbations are obtained in the following text.

\subsubsection{Equations for the ideal fluid perturbations}

The ideal fluid perturbations, $\phi$, evolve in accordance with the Lagrangian equation  (\ref{mt7}). In the linear approximation we can neglect the source term $\Sigma^{\rm m}$ in its right side. The covariant derivative in the definition of the linear operator $F^{\rm m}$ given by (\ref{mt10}), can be explicitly performed that yields equation for the Clebsch potential
\be\la{qe8}
\vp^{|\a}{}_{|\a}-2\bar\m_{\rm m}\H \mathfrak{q}+16\pi\bar\m_{\rm m}\le(\bar\r_{\rm m}\vp+\bar\r_{\rm q}\psi\ri)+\le(1-\frac{c^2}{c^2_{\rm s}}\ri)\le(\bar u^\a\bar u^\b\vp_{|\a\b}-\frac12\bar\m_{\rm m}\bar u^\a \mathfrak{q}_{,\a}\ri)=\bar\m_{\rm m}\bar u^\a{B}_\a\;,
\ee
where the gauge function (\ref{qe6} has been used. The gauge ${B}^\a$ remains yet unspecified so that equation (\ref{qe8}) is covariant and is valid in any coordinate chart.

\subsubsection{Equations for the scalar field perturbations}
Linearized equation for the scalar field perturbations, $\psi$, is obtained from the Lagrangian equation (\ref{mt7q}) after neglecting the (non-linear) source term $\Sigma^{\rm q}$. After performing the covariant differentiation in equation (\ref{mt10c}), we get equation for the scalar field perturbation
\be\la{qe9}
\psi^{|\a}{}_{|\a}-\le(2\bar\m_{\rm q}\H +\frac{\pd\bar W}{\pd\bar\Psi}\ri)\mathfrak{q}+16\pi\bar\m_{\rm q}\le(\bar\r_{\rm m}\vp+\bar\r_{\rm q}\psi\ri)-\frac{\pd^2\bar W}{\pd\bar\Psi^2}\,\psi=\bar\m_{\rm q}\bar u^\a{B}_\a\;,
\ee
where equation (\ref{qe6}) has been used along with the equality $\bar\r_{\rm q}=\bar\m_{\rm q}$. The gauge function ${B}^\a$ is kept unspecified so that equation (\ref{qe8}) is covariant and is valid in any coordinates.

\section{Gauge-invariant Field Equations in 3+1 Formalism}\la{gife3}

\subsection{Algebraic Decomposition of the Metric Perturbations.}\la{alge}
We have derived the system of coupled differential equations (\ref{qe7}), (\ref{qe8}), (\ref{qe9}) for the field variables $l_{\a\b}$, $\phi$ and $\psi$, describing perturbations of the gravitational field, the ideal fluid, and the scalar field respectively. These system of equations can be split into a set of gauge-invariant differential equations for the scalar, vector, and tensor parts which is convenient for theoretical study of the evolution of the perturbations in arbitrary coordinates. This 3+1 split is achieved by making use of the operator of projection $\bar P_{\a\b}$ onto a hypersurface being orthogonal to the congruence of worldlines of the Hubble flow.

The theory under development admits four, algebraically-independent scalar perturbations. Two of them are the Clebsch potential of the ideal fluid $\phi$ and the scalar field $\psi$. The two other scalars characterize the scalar perturbations of the gravitational field. They can be chosen, for example, as a projection of the metric tensor perturbation on the direction of the background four-velocity field, $\bar u^\a\bar u^\b l_{\a\b}$, and the trace of the metric tensor perturbation, $l=\bar g^{\a\b}l_{\a\b}$. However, it is more convenient to work with two other scalars, defined as their linear combinations,
\bsu\ba\la{sp1}
\mathfrak{q}&\equiv&\frac12\le(\bar u^\a \bar u^\b +\bar P^{\a\b}\ri)l_{\a\b}\;,\\\la{sp1a}
\mathfrak{p}&\equiv&\le(\bar u^\a \bar u^\b +\bar g^{\a\b}\ri)l_{\a\b}\;,
\ea\esu
Notice that the scalar $\mathfrak{q}$ has been introduced earlier in (\ref{mt6c}). The scalar $\mathfrak{p}$ is, in fact, projection of $l_{\a\b}$ onto the space-like hypersurface being orthogonal everywhere to the worldlines of fiducial observers moving with the background four-velocity $\bar u^\a$ of the Hubble flow. Indeed, after accounting for definition (\ref{dec1}), equation (\ref{sp1a}) can be written as
\be\la{sp1c}
\mathfrak{p}={\bar P}^{\a\b}l_{\a\b}\;.
\ee

Vectorial gravitational perturbations are defined by a spacial-temporal projection
\be\la{veq1}
\mathfrak{p}_\a\equiv -{\bar P}_{\a}{}^\b\bar u^\g l_{\b\g}\;,
\ee
where minus sign was taken for the sake of mathematical convenience. Due to its definition, vector $\mathfrak{p}^\a=\bar g^{\a\b}\mathfrak{p}_\b$ is orthogonal to the four-velocity $\bar u^\a$, that is $\bar u^\a\mathfrak{p}_\a=0$. Hence, it describes a space-like vector-like gravitational perturbations with three algebraically-independent components.

Tensorial gravitational perturbations are associated with the projection
\be\la{tp1}
\mathfrak{p}^{\sst\intercal}_{\a\b}\equiv \mathfrak{p}_{\a\b}-\frac13 {\bar P}_{\a\b}\mathfrak{p}\;,
\ee
where
\be\la{tp1a}
\mathfrak{p}_{\a\b}\equiv {\bar P}_{\a}{}^\m {\bar P}_{\b}{}^\n l_{\m\n}\;.
\ee
Here, the tensor $\mathfrak{p}_{\a\b}$ is a double projection of $l_{\a\b}$ onto space-like hypersurface being orthogonal to the worldlines of fiducial observers moving with the four-velocity $\bar u^\a$ of the Hubble flow. The trace of this tensor coincides with the scalar $\mathfrak{p}$. Indeed,
\be\la{tp1b}
\bar g^{\a\b}\mathfrak{p}_{\a\b}=\bar g^{\a\b}{\bar P}_{\a}{}^\m {\bar P}_{\b}{}^\n l_{\m\n}={\bar P}^{\b\m} {\bar P}_{\b}{}^\n l_{\m\n}=\bar P^{\m\n}l_{\m\n}=\mathfrak{p}\;,
\ee
where the property of the projection tensor ${\bar P}^{\b\m} {\bar P}_{\b}{}^\n=\bar P^{\m\n}$ has been used. Equation (\ref{tp1b}) makes it clear that  tensor $\mathfrak{p}^{\sst\intercal}_{\a\b}$ is traceless, that is $\bar g^{\a\b}\mathfrak{p}^{\sst\intercal}_{\a\b}=0$. Because of this property, and four orthogonality conditions, $\bar u^\a\mathfrak{p}^{\sst\intercal}_{\a\b}=0$, the symmetric tensor $\mathfrak{p}^{\sst\intercal}_{\a\b}$ has only five, algebraically-independent components.

Gravitational perturbation $l_{\a\b}$ can be decomposed into the algebraically-irreducible scalar, vector and tensor parts as follows
\be\la{yu1}
l_{\a\b}=\mathfrak{p}^{\sst\intercal}_{\a\b}+\bar u_\a\mathfrak{p}_\b+\bar u_\b\mathfrak{p}_\a+\le(\bar u_\a\bar u_\b+\frac13 {\bar P}_{\a\b}\ri)\mathfrak{p}+2\bar u_\a\bar u_\b\le(\mathfrak{q}-\mathfrak{p}\ri)\;.
\ee
One should not confuse the pure algebraic decomposition of the metric tensor perturbation with its decomposition in a functional (Hilbert) space. This decomposition was pioneered by \citet{1964SvPhU...6..495L} and later on, structured by Arnowitt, Deser and Misner \citet{1959PhRv..116.1322A} (see also \citep{mtw}). It is commonly used in the research on the relativistic theory of formation of the large-scale structure in the universe. The functional ADM decomposition of the metric tensor perturbations is done with respect to the direction of propagation of weak gravitational waves and singles out the longitudinal (L), transversal (T) and transverse-traceless (TT) parts of the perturbations. In other words, the functional decomposition make sure that the vector $\mathfrak{p}_\a$ and the tensor parts of the gravitational perturbation, $\mathfrak{p}^{\sst\intercal}_{\a\b}$, are further decomposed in the functionally-irreducible components that are reduced to two more scalars, and two transverse vectors each of which has only two, functionally-independent components. The remaining part of the tensor perturbations, $\mathfrak{p}^{\sst\intercal}_{\a\b}$, is transverse-trackless and has only two functionally-independent components denoted as $\mathfrak{p}^{\sst\rm TT}_{\a\b}$. The ADM decomposition of the metric tensor is a powerful technique  in the theory of gauge-invariant cosmological perturbations \citep{1988eur..book..563B}. However, it is not convenient in the development of the post-Newtonian dynamics of celestial bodies in cosmology, and shall not be used in the present paper.

Our next step is derivation of the field equations for the algebraically-irreducible components of matter and gravitational field.
Before doing this derivation, let us discuss the gauge transformations of the corresponding field variables.

\subsection{The Gauge Transformation of the Field Variables}
Gauge invariance is a cornerstone of the modern theoretical physics with a long and interesting history \citep{2001RvMP...73..663J}. Gauge invariance should be distinguished from the coordinate invariance or the general covariance because, by definition, a gauge transformation changes only field variables of the theory under consideration but not coordinates. Discussion of gauge transformation and invariance requires introducing a gauge field and a new geometric object -- an affine connection -- on a fibre bundle manifold describing the intrinsic degrees of freedom of corresponding field variables of the gauge field theory.

The present paper discusses physical perturbations of the field variables $l_{\a\b}$, $\p$, $\Psi$ on the cosmological spacetime manifold in the framework of general relativity.
The affine connection on the spacetime manifold of general relativity is represented by the Christoffel symbols while the gauge transformation is generated by a flow of an arbitrary vector (gauge) field $\xi^\a$ that maps the manifold into itself. Gauge transformation of the fields on a curved manifold is associated with a Lie transport of the fields along the vector flow $\xi^\a$ \citep{LL,1972gcpa.book.....W}. Infinitesimal gauge transformation is a Lie derivative of the field taken at the value of the parameter on the curves of the vector flow equal to 1 \citep[chapter 3.6]{2011rcms.book.....K}.

Let us consider a mapping of spacetime manifold into itself induced by a vector flow, $\xi^\a=\xi^\a(x^\b)$. It means that each point of the manifold with coordinates $x^\a$ is mapped to another point with coordinates
\be
\la{qe11}
\hat x^\a=x^\a-\xi^\a(x)\;.
\ee
This mapping of the manifold into itself can be interpreted as a local diffeomorphism which transforms the field variables in accordance to their tensor properties. The transformed value of the field variable is pulled back to the point of the manifold having the original coordinates $x^\a$, and is compared with the value of the field at this point. The difference between the transformed and the original value of the field, generated by the diffeomorphism (\ref{qe11}) is the gauge transformation of the field that is given by the Lie derivative taken along the vector field $\xi^\a$ at the point of the manifold with coordinates $x^\a$.

Let us denote the transformed values of the field variables with a hat. In the linearized perturbation theory of the cosmological manifold, the gauge transformations of the field variables are given by equations
\bsu\ba\la{qme1}
\hat\varkappa_{\a\b}&=&\varkappa_{\a\b}+\xi_{\a|\b}+\xi_{\b|\a}\;,
\\\la{qe12}
\hat l_{\a\b}&=&l_{\a\b}-\xi_{\a|\b}-\xi_{\b|\a}+\bar g_{\a\b}\xi^\g{}_{|\g}\;,
\\\la{qe13}
\hat\vp&=&\vp+\bar\p_{|\a}\xi^\a\;,
\\\la{qe14}
\hat\psi&=&\psi+\bar\Psi_{|\a}\xi^\a\;,
\ea\esu
where the hat above each symbol denotes a new value of the field variable after applying the gauge transformation, and all functions are calculated at the same value of coordinates  $x^\a$. The gauge transformations of the field variables are expressed in terms of the covariant derivatives on the manifold and, thus, are coordinate-independent.
Equation (\ref{qe12}) is derived from the Lie transformation (\ref{qme1}) of the metric tensor perturbation, and the relationship (\ref{ok9}) between $\varkappa_{\a\b}$ and $l_{\a\b}$.

Gauge invariance of the Lagrangian perturbation theory means that the gauge transformations of the field variables do not change the content of the theory. In other words, the equations for the field variables must be invariant with respect to the gauge transformations (\ref{qme1})-(\ref{qe14}). However, direct inspection of equations (\ref{qe7}), (\ref{qe8}), (\ref{qe9}) shows that they do depend on the choice of the gauge in the form of the gauge function $B^\a$ introduced in equation (\ref{qe6}). To find out the gauge-invariant content of the theory one should search for the gauge-invariant field variables and to derive the gauge-invariant equations for them. This program has been completed by Bardeen \citep{1988eur..book..563B} who used the functional 3+1 decomposition of the metric tensor perturbations and the vector field $\xi^\a$ to build the gauge-invariant variables out of the various projections of the metric tensor components on space an time. Various modifications of Bardeen's approach can be found, for example, in \citep{ellis1,ellis2,1997GReGr..29..733E,1992CQGra...9..921B,mukh,1980JETPL..31..596L} and in the book by Mukhanov \citep{2005pfc..book.....M}.
We use algebraic 3+1 decomposition of the metric tensor perturbations (\ref{yu1}) that allows us to build gauge-invariant scalars. Vector and tensor perturbations remain gauge-dependent in this approach. In order to suppress the gauge degrees of freedom in these variables we impose a particular gauge condition $B^\a=0$ in equation (\ref{qe6}). This limits the freedom of the gauge field $\xi^\a$ by a particular set of differential equations which are discussed in section (\ref{rgfr}).

\subsection{The Gauge-invariant Scalars}\la{gis}

The existence of the preferred four-velocity, $\bar u^\a$, of the Hubble flow in the expanding universe provides a natural way of separating the perturbations of the field variables in scalar, vector, and tensor components. This section discusses how to build the gauge-invariant scalars. Vector and tensor perturbations are discussed afterwards.

The gauge-invariant scalar perturbations can be build from the perturbation of the Clebsch potential, $\phi$, the perturbation of the scalar field $\psi$, and two scalars associated with the trace of the metric tensor, $\mathfrak{q}$, and its projection on the worldlines of the Hubble flow, $\mathfrak{q}$. To build the first gauge-invariant scalar, we introduce the scalar perturbations
\be\la{sp2}
\chi_{\rm m}\equiv\frac{\phi}{\bar\m_{\rm m}}\;,\qquad\qquad \chi_{\rm q}\equiv\frac{\psi}{\bar\m_{\rm q}}\;,
\ee
that normalize perturbations of the Clebsch potential $\phi$ and that of the scalar field $\psi$ to the corresponding background values of the specific enthalpy, $\bar\m_{\rm m}$ and $\bar\m_{\rm q}$. The gauge transformations for the three scalars $\mathfrak{q}$, $\chi_{\rm m}$, and $\chi_{\rm q}$ are obtained from (\ref{qe12})--(\ref{qe14}), and read
\bsu\ba\la{sp3}
\hat{\mathfrak{q}}&=&\mathfrak{q}-2\bar u^\a\bar u^\b\xi_{\a|\b}\;,
\\\la{sp4}
\hat\chi_{\rm m}&=&\chi_{\rm m}-\bar u_\a\xi^\a\;,
\\\la{sp5}
\hat\chi_{\rm q}&=&\chi_{\rm q}-\bar u_\a\xi^\a\;,
\ea\esu
where we have used the definition of the background four-velocity $\bar u^\a=-\bar\p_{|\a}/\bar\m_{\rm m}=-\bar\Psi_{|\a}/\bar\m_{\rm q}$ in terms of the partial derivatives of the background values of the scalar fields $\Phi$ and $\Psi$. Equations (\ref{sp4}), (\ref{sp5}) immediately reveal that the linear combination
\be\la{sp6}
\chi\equiv\chi_{\rm m}-\chi_{\rm q}\;,
\ee
is gauge-invariant, $\hat\chi=\chi$, that is the diffeomorphism (\ref{qe11}) does not change the value of the scalar variable $\chi$.

Two other gauge-invariant scalars are defined by the following equations,
\bsu\ba\la{sp7}
V_{\rm m}&\equiv&\bar u^\a \chi_{{\rm m}|\a}-\frac{\mathfrak{q}}{2}\;,\\
\la{sp8}
V_{\rm q}&\equiv&\bar u^\a \chi_{{\rm q}|\a}-\frac{\mathfrak{q}}{2}\;,
\ea\esu
or, more explicitly,
\bsu\ba\la{qe10a}
V_{\rm m}&=&\frac{1}{\bar\m_{\rm m}}\bar u^\a\vp_{|\a}-\frac{\mathfrak{q}}{2}+3\frac{c^2_{\rm s}}{c^2}\H \chi_{\rm m}\;,\\
\la{sp9}
V_{\rm q}&=&\frac{1}{\bar\m_{\rm q}}\bar u^\a\psi_{|\a}-\frac{\mathfrak{q}}{2}+3\H \chi_{\rm q}+\frac{\chi_{\rm q}}{\m_{\rm q}}\frac{\pd\bar W}{\pd\bar\Psi}\;,
\ea\esu
where the last terms in the right side of these equations were obtained by making use of thermodynamic relationships (\ref{pf5}), the equality $\bar\r_{\rm q}=\bar\m_{\rm q}$, and the equations of continuity (\ref{cr3}) and (\ref{cr3a}) for the density of the ideal fluid, $\bar\r_{\rm m}$, and that of the scalar field, $\bar\r_{\rm q}$, respectively.

One can easily check that both scalars, $V_{\rm m}$ and $V_{\rm q}$ remain unchanged after making the infinitesimal coordinate transformation (\ref{qe11}). Indeed, the gauge transformation of the derivatives
\bsu\ba\la{sp10}
\hat\chi_{\rm m|\a}&=&\chi_{\rm m|\a}-\H {\bar P}_{\a\b}\xi^\b-\bar u_\b\xi^\b{}_{|\a}\;,\\
\la{sp11}
\hat\chi_{\rm q|\a}&=&\chi_{\rm q|\a}-\H {\bar P}_{\a\b}\xi^\b-\bar u_\b\xi^\b{}_{|\a}\;,
\ea\esu
where ${\bar P}_{\a\b}=\bar g_{\a\b}+\bar u_\a\bar u_\b$ is the operator of projection on the hypersurface being orthogonal to the Hubble flow of four-velocity $\bar u^\a$. Making the coordinate transformation (\ref{qe11}), and substituting the transformations of functions $\mathfrak{q}$, $\chi_{\rm m}$ and $\chi_{\rm q}$ to the definitions of $V_{\rm m}$ and $V_{\rm q}$ we find
\be\la{qe15}
\hat V_{\rm m}=V_{\rm m}\;,\qquad\qquad \hat V_{\rm q}=V_{\rm q}\;,
\ee
that proves the gauge-invariant property of the scalars $V_{\rm m}$ and $V_{\rm q}$.

Physical meaning of the gauge-invariant quantity $V_{\rm m}$ can be understood as follows. We consider the perturbation of the specific enthalpy $\mu_{\rm m}$ defined in equation (\ref{pf7}). Substituting the decomposition (\ref{pf8}) of the field variables to equation (\ref{pf7}) and expanding, we obtain
\be\la{qe16}
\m_{\rm m}=\bar\m_{\rm m}+\d\m_{\rm m}\;,
\ee
where the perturbation $\d\m_{\rm m}$ of the specific enthalpy is defined (in the linearized order) by
\be\la{qe17}
\d\m_{\rm m}=\bar u^\a\vp_{|\a}-\frac{1}{2}\bar\m_{\rm m} \mathfrak{q}\;.
\ee
It helps us to recognize that
\be\la{qe18}
V_{\rm m}=\frac{\d\m_{\rm m}}{\bar\m_{\rm m}}+3\frac{c^2_{\rm s}}{c^2}\H \chi_{\rm m}\;.
\ee
Fractional perturbation of the specific enthalpy can be re-written with the help of thermodynamic equations (\ref{pf5}) in terms of the perturbation $\d\e_{\rm m}$ of the energy density of the ideal fluid,
\be\la{qe19}
\frac{\d\m_{\rm m}}{\bar\m_{\rm m}}=\frac{c^2_{\rm s}}{c^2}\frac{\d\e_{\rm m}}{\bar\e_{\rm m}+\bar p_{\rm m}}\;,
\ee
or, by making use of equation (\ref{pf3a}), in terms of the perturbation $\d\r_{\rm m}$ of the density of the ideal fluid
\be\la{qe20}
\frac{\d\m_{\rm m}}{\bar\m_{\rm m}}=\frac{c^2_{\rm s}}{c^2}\frac{\d\r_{\rm m}}{\bar\r_{\rm m}}\;.
\ee
This allows us to write down equation (\ref{qe18}) as follows
\be\la{qe21}
V_{\rm m}=\frac{c^2_{\rm s}}{c^2}\le(\frac{\d\r_{\rm m}}{\bar\r_{\rm m}}+3\H \chi_{\rm m}\ri)\;,
\ee
which elucidates the relationship between the gauge-invariant variable $V_{\rm m}$ and the perturbation $\d\r_{\rm m}$ of the rest mass density of the ideal fluid. More specifically, $V_{\rm m}$ is an algebraic sum of two scalar functions, $\d\r_{\rm m}$ and $\chi_{\rm m}$ neither of each is gauge-invariant. The gauge transformation of the ideal-fluid density perturbation is
\be\la{qe22}
\d\hat\r_{\rm m}=\d\r_{\rm m}-\bar\r_{{\rm m}|\a}\xi^\a=\d\r_{\rm m}+3\H \bar\r_{\rm m}\bar u_\a\xi^\a\;,
\ee
and the gauge transformation of the variable $\chi_{\rm m}$ is given by (\ref{sp4}). Their algebraic sum in equation (\ref{qe21}) does not change under the diffeomorphism (\ref{qe11}) showing that $V_{\rm m}$ is the gauge-invariant density fluctuation that does not depend on a particular choice of coordinates on spacetime manifold.

Similar considerations, applied to function $V_{\rm q}$ reveals that it can be represented as an algebraic sum of the perturbation, $\d\r_{\rm q}$, of the density of the scalar field, and the function $\chi_{\rm q}$,
\be\la{qe23c}
V_{\rm q}=\frac{\d\r_{\rm q}}{\bar\r_{\rm q}}+3\H \chi_{\rm q}\;.
\ee
It is easy to check that each term in the right side of this equation is not gauge-invariant but their linear combination does. The reader should notice that standard textbooks on cosmological theory (see, for example, \citep{1972gcpa.book.....W,weinberg_2008,lnder,1980lssu.book.....P}) derive equations for the density perturbations $\d\r/\bar\r$ but those equations are not gauge-invariant and, hence, their solutions should be interpreted with care (see discussion in section \ref{deos}).

\subsection{Field Equations for the Scalar Perturbations.}
\subsubsection{Equation for a scalar $\mathfrak{q}$.}

Function $\mathfrak{q}$ was defined in (\ref{sp1}). In order to derive a differential equation for $\mathfrak{q}$, we apply the covariant Laplace-Beltrami operator to $\mathfrak{q}$, and make use of the covariant equations (\ref{qe3a}) and (\ref{qe6}). Straightforward but fairly long calculation yields
\ba\la{qe23}
\mathfrak{q}^{|\a}{}_{|\a}-2\le(\dot H+\H^2-\frac{2k}{a^2}\ri)\mathfrak{q}
+8\pi\bar\r_{\rm m}\bar\m_{\rm m}\le[\le(1-\frac{c^2}{c^2_{\rm s}}\ri)V_{\rm m}-\le(1+3\frac{c^2_{\rm s}}{c^2}\ri)\H\chi_{\rm m}\ri]&&\\\nonumber
-16\pi\bar\r_{\rm q}\le(\frac{\pd\bar W}{\pd\bar\Psi}+2\H\bar\m_{\rm q}\ri)\chi_{\rm q}
-2\bar u^\a\bar u^\b{B}_{\a|\b}-4\H\bar u^\a{B}_\a
&=&8\pi\le(\sigma+\tau\ri)\;,
\ea
where the source density $\s+\t$ for the field $\mathfrak{q}$ is
\be\la{qe23a}
\sigma+\t =\le(\bar u^\a\bar u^\b+\bar P^{\a\b}\ri)\mathfrak{T}_{\a\b}\;,
\ee
in accordance with the definitions introduced in (\ref{qq4}), (\ref{qq5}). The reader should notice that equation (\ref{qe23}) depends on the gauge function $B^\a$ which remains arbitrary so far.

\subsubsection{Equation for a scalar $\mathfrak{p}$.}
Function $\mathfrak{p}$ was defined in (\ref{sp1a}). In order to derive equation for $\mathfrak{p}$, we apply the covariant Laplace-Beltrami operator to the definition of $\mathfrak{p}$, and make use of the covariant equations (\ref{qe3a}) and (\ref{qe6}). It results in a wave equation
\be\la{sp1b}
\mathfrak{p}^{|\a}{}_{|\a}+\frac{4k}{a^2}\mathfrak{p}+{B}^\a{}_{|\a}-2\bar u^\a\bar u^\b{B}_{\a|\b}-6\H\bar u^\a{B}_\a=16\pi\t\;.
\ee
where the source density $\t$ has been defined in (\ref{qq5}). Equation (\ref{sp1b}) depends on the arbitrary gauge function $B^\a$.

\subsubsection{Equation for a scalar $\chi$.}

Equation for the gauge invariant scalar, $\chi=\chi_{\rm m}-\chi_{\rm q}$, is derived from the definitions (\ref{sp2}) and the field equations (\ref{qe8}), (\ref{qe9}). Replacing $\phi$ and $\psi$ in those equations with $\chi_{\rm m}$ and $\chi_{\rm q}$, and making use of equations (\ref{13a}), (\ref{13b}) for reshuffling some terms, yields
\bsu\ba\la{feq1}
\chi_{\rm m}^{|\a}{}_{|\a}+2{\H}\bar u^\alpha\chi_{{\rm m}|\alpha}-\le({\dot H}-\frac{4k}{a^2}\ri)\chi_{\rm m}+4{\H}V_{\rm m}
+\le(1-\frac{c^2}{c_{\rm s}^2}\ri)\bar u^\a V_{{\rm m}|\a}-16\pi\bar\r_{\rm q}\bar\m_{\rm q}\chi&=&\bar u^\a{B}_\a\;,
\\
\la{feq2}
\chi_{\rm q}^{|\a}{}_{|\a}+2{\H}\bar u^\alpha\chi_{{\rm q}|\alpha}-\le({\dot H}-\frac{4k}{a^2}\ri)\chi_{\rm q}+4{\H}V_{\rm q}
+\frac{2}{\bar\m_{\rm q}}\frac{\pd\bar W}{\pd\bar\Psi}V_{\rm q}+16\pi\bar\r_{\rm m}\bar\m_{\rm m}\chi&=&\bar u^\a{B}_\a\;.
\ea\esu
Subtracting (\ref{feq2}) from (\ref{feq1}) cancels the gauge-dependent term, $\bar u^\a{B}_\a$, and brings about the field equation for $\chi$,
\be\la{feq3}
\chi^{|\a}{}_{|\a}
%%%%%%%%%%%%%%%%%%%%%%%%+\le(1-\frac{c^2}{c_{\rm s}^2}\ri)\bar u^\a\bar u^\b \chi_{|\a\b}
+6\H\bar u^\a\chi_{|\a}+3{\dot H}\chi=\frac{2}{\bar\m_{\rm q}}\frac{\pd\bar W}{\pd\bar\Psi}V_{\rm q}-\le(1-\frac{c^2}{c_{\rm s}^2}\ri)\bar u^\a V_{{\rm m}|\a}\;.
\ee
This equation is apparently gauge-invariant since any dependence on the arbitrary gauge function $B^\a$ disappeared. It is also covariant that is valid in any coordinates.

Equation (\ref{feq3}) can be recast to the form of an inhomogeneous wave equation:
\be\la{feq3a}
\le(\r_{\rm m}\chi\ri)^{|\a}{}_{|\a}
%%%%%%%%%%%%%%%%%%%%%%%%+\le(1-\frac{c^2}{c_{\rm s}^2}\ri)\bar u^\a\bar u^\b \chi_{|\a\b}
=2\frac{\bar\r_{\rm m}}{\bar\r_{\rm q}}\frac{\pd\bar W}{\pd\bar\Psi}V_{\rm q}-\le(1-\frac{c^2}{c_{\rm s}^2}\ri)\bar\r_{\rm m}\bar u^\a V_{{\rm m}|\a}\;.
\ee
Yet another form of equation (\ref{feq3}) is obtained in terms of the variable $\r_{\rm q}\chi$. By simple inspection we can check that equation (\ref{feq3}) is transformed to
\be\la{feq3b}
\le(\r_{\rm q}\chi\ri)^{|\a}{}_{|\a}-\frac{\pd^2\bar W}{\pd\bar\Psi^2}\le(\r_{\rm q}\chi\ri)=
2\frac{\pd\bar W}{\pd\bar\Psi}V_{\rm m}-\le(1-\frac{c^2}{c_{\rm s}^2}\ri)\bar\r_{\rm q}\bar u^\a V_{{\rm m}|\a}\;.
\ee
This is an inhomogeneous Klein-Gordon equation for the field $\le(\r_{\rm q}\chi\ri)$ governed by $V_{\rm m}$. The 'mass' of the scalar field $\r_{\rm q}\chi$ depends on the second derivative of the potential function $\bar W$ which defines the 'coefficient of elasticity' of the background scalar field $\bar\Psi$.

Inhomogeneous equations (\ref{feq3}), (\ref{feq3a}), (\ref{feq3b})  have the source terms that is determined by variables $V_{\rm m}$ and $V_{\rm q}$. We derive differential equations for these field variables in the next sections.

\subsubsection{Equation for a scalar $V_{\rm m}$.}

Equation for the field variable $V_{\rm m}$ is derived from the equations for functions $\chi_{\rm m}$ and $q$ that enter its definition (\ref{sp7}). By applying the Laplace-Beltrami operator to function $V_{\rm m}$ we get
\be\la{feq4}
V_{\rm m}^{|\a}{}_{|\a}=\bar u^\b\left(\chi^{|\a}_{\rm m}{}_{|\a}\ri)_{|\b}+2\H\chi^{|\a}_{\rm m}{}_{|\a}-\frac12\mathfrak{q}^{|\a}{}_{|\a}+\bar u^\b\bar R^\a{}_\b\chi_{{\rm m}|\a}+2\H\bar u^\a\le(V_{\rm m}+\frac12\mathfrak{q}\ri)_{|\a}+3\H^2\le(V_{\rm m}+\frac12\mathfrak{q}\ri)\;.
\ee
The Laplace-Beltrami operator for function $\chi_{\rm m}$ is given in equation (\ref{feq1}) which is not gauge-invariant. Taking the covariant derivative from this equation and contracting it with $\bar u^\a$ brings about the first term in the right side of equation (\ref{feq4}),
\ba\la{feq5}
\bar u^\b\left(\chi^{|\a}_{\rm m}{}_{|\a}\ri)_{|\b}&=&-\le(1-\frac{c^2}{c^2_{\rm s}}\ri)\bar u^\a\bar u^\b V_{{\rm m}|\a\b}-6\H\bar u^\a V_{{\rm m}|\a}-\le(5\dot H+\frac{4k}{a^2}\ri)V_{\rm m}\\\nonumber
&&-\H\bar u^\a\mathfrak{q}_{|\a}-\le(\frac12\dot H+\frac{2k}{a^2}\ri)\mathfrak{q}
-3\H\le[\le(1+\frac{c^2_{\rm s}}{c^2}\ri)\dot H-\le(3+\frac{c^2_{\rm s}}{c^2}\ri)\frac{k}{a^2}\ri]\chi_{\rm m}\\\nonumber
&&+8\pi\bar\r_{\rm q}\frac{\pd\bar W}{\pd\bar\Psi}\le(4\chi_{\rm q}-3\chi_{\rm m}\ri)+16\pi\bar\r_{\rm q}\bar\m_{\rm q}\le[\bar u^\a\chi_{|\a}-6\chi+\frac34\H\le(1-\frac{c^2_{\rm s}}{c^2}\ri)\chi_{\rm m}\ri]+\bar u^\a\bar u^\b{B}_{\a|\b}\;.
\ea
The Laplace-Beltrami operator for function $\mathfrak{q}$ has been derived in (\ref{qe23}).
Now, we make use of equations (\ref{qe23}), (\ref{feq1}), (\ref{feq5}) in calculating the right side of (\ref{feq4}). After a significant amount of algebra, we find out that all terms explicitly depending on $q$ and the gauge functions ${B}^\a$ cancel out, so that equation for $V_{\rm m}$ becomes
\ba\la{qe25}
V_{\rm m}^{|\a}{}_{|\a}+\le(1-\frac{c^2}{c_{\rm s}^2}\ri)\bar u^\a\bar u^\b V_{{\rm m}|\a\b}+
2\le(3-\frac{c^2}{c_{\rm s}^2}\ri){\H}\bar u^\a V_{{\rm m}|\a}
&+&\\\nonumber
\le[2\le({\dot H}+3\H^2+\frac{2k}{a^2}\ri)-4\pi\bar\r_{\rm m}\bar\m_{\rm m}\le(1-\frac{c^2}{c_{\rm s}^2}\ri)\ri] V_{\rm m}
&-&\\\nonumber
16\pi\bar\r_{\rm q}\bar\m_{\rm q}\le[\bar u^\a\chi_{|\a}-3\le({\H}+\frac{1}{2\bar\m_{\rm q}}\frac{\pd\bar W}{\pd\bar\Psi}\ri)\chi\ri]
%&=&-8\pi\le(\bar u^\a\bar u^\b\tau_{\a\b}+\frac{\tau}{2}\ri)\;,
&=&-4\pi\le(\sigma+\t\ri)\;.
\ea

Second-order covariant derivatives in this equation read
\ba\la{nn12}
\le[\bar g^{\a\b}+\le(1-\frac{c^2}{c_{\rm s}^2}\ri)\bar u^\a\bar u^\b \ri]V_{{\rm m}|\a\b}\equiv\le(-\frac{c^2}{c_{\rm s}^2}\bar u^\a\bar u^\b +{\bar P}^{\a\b}\ri)V_{{\rm m}|\a\b}\;,
\ea
and they form a hyperbolic-type operator describing propagation of sound waves in the expanding universe from the source of the sound waves towards the field point with the constant velocity $c^2_{\rm s}$. Additional terms in the left side of equation (\ref{qe25}) depend on the Hubble parameter $\H$, and take into account the expansion of the universe. Equation (\ref{qe25}) contains only gauge-invariant scalars, $V_{\rm m}$ and $\chi$. Moreover, it does not depend on the choice of coordinates on the background manifold. It also becomes clear that the field variables $V_{\rm m}$ and $\chi$ are coupled through the differential equations (\ref{feq3b}) and (\ref{qe25}) which should be solved simultaneously in order to determine these variables. Solution of the coupled system of differential equations is a very complicated task which cannot be rendered analytically in the most general case. Only in some simple cases, the equations can be decoupled. We discuss such cases in section \ref{dspn}.

\subsubsection{Equation for a scalar $V_{\rm q}$.}
The field variable $V_{\rm q}$ is not independent since it relates to $V_{\rm m}$ and $\chi$ by a simple relationship
\begin{equation}\label{feq5a}
V_{\rm q}=V_{\rm m}-\bar u^\alpha\chi_{|\alpha}\;,
\end{equation}
which is obtained after subtraction of equation (\ref{sp7}) from (\ref{sp8}). Equation for $V_{\rm q}$ is derived directly from (\ref{feq5a}) and equations (\ref{qe25}) and (\ref{feq3})  for $V_{\rm m}$ and $\chi$ respectively. We obtain,
\ba
\la{feq6}
V_{\rm q}^{|\a}{}_{|\a}+
4\le({\H}+\frac{1}{2\bar\m_{\rm q}}\frac{\pd\bar W}{\pd\bar\Psi}\ri)\bar u^\a V_{{\rm q}|\a}&&\\\nonumber
+\le[2\le({\dot H}+3{\H}^2+\frac{2k}{a^2}\ri)-4\pi\bar\r_{\rm m}\bar\m_{\rm m}\le(1-\frac{c^2}{c_{\rm s}^2}\ri)+\frac{2}{\bar\m_{\rm q}}\le(5{\H}+\frac{1}{\bar\m_{\rm q}}\frac{\pd\bar W}{\pd\bar\Psi}\ri)\frac{\pd\bar W}{\pd\bar\Psi}+2\frac{\pd^2\bar W}{\pd\bar\Psi^2}\ri] V_{\rm q}&&\\\nonumber
+4\pi\bar\r_{\rm m}\bar\m_{\rm m}\le(3+\frac{c^2}{c_{\rm s}^2}\ri)\le(\bar u^\a\chi_{|\a}
-3\frac{c_{\rm s}^2}{c^2}{\H}\chi\ri)
%&=&-8\pi\le(\bar u^\a\bar u^\b\tau_{\a\b}+\frac{\tau}{2}\ri)
&=&-4\pi\le(\sigma+\t\ri)\;.
\ea
This equation can be also derived by the procedure being similar to that used in the previous subsection in deriving equation for $V_{\rm m}$. We followed this procedure and confirm that it leads to (\ref{feq6}) as expected. Equation (\ref{feq6}) is clearly gauge-invariant. It couples with the variable $\chi$ and should be solved along with equation (\ref{feq3}).

\subsection{Field Equations for Vector Perturbations}\la{feveper}
Vector perturbations of the ideal fluid and scalar field are gradients, $\phi_{|\a}$ and $\psi_{|\a}$. However, they are insufficient to build a gauge-invariant vector perturbation out of the vector perturbation of the metric tensor $\mathfrak{p}_\a$. Field equations for vector $\mathfrak{p}_\a$ can be derived by applying the covariant Laplace-Beltrami operator to both sides of definition (\ref{veq1}) and making use of equation (\ref{qe7}). After performing the covariant differentiation and a significant amount of algebra, we derive the field equation
\be\la{veq2}
\mathfrak{p}_\a{}^{|\b}{}_{|\b}-2\H\bar u_\a\mathfrak{p}_\b{}^{|\b}-\le(2\dot H+3\H^2-\frac{2k}{a^2}\ri)\mathfrak{p}_\a+{\bar P}_\a{}^\b\bar u^\g\le({B}_{\b|\g}+{B}_{\g|\b}+2\H\bar u_\g{B}_\b\ri)=16\pi\t_\a\;,
\ee
where the matter current $\s_\a$ is defined in (\ref{qq6}). This equation is apparently gauge-dependent as shown by the appearance of the gauge function $B^\a$. This equation reduces to a much simpler form
\be\la{vewe2}
\mathfrak{p}_\a{}^{|\b}{}_{|\b}-2\H\bar u_\a\mathfrak{p}_\b{}^{|\b}-\le(2\dot H+3\H^2-\frac{2k}{a^2}\ri)\mathfrak{p}_\a=16\pi\t_\a\;,
\ee
in a special gauge $B^\a$=0 which imposes a restriction on the divergence of the metric tensor perturbation in equation (\ref{qe6}). 

Equation (\ref{veq2}) points out that the vector perturbations are generated by the current of matter $\t_a$ existing in the localized astronomical system which physical origin may be a relict of the primordial perturbations. We do not discuss this interesting scenario in the present paper as it would require a non-conservation of entropy and non-isentropic background fluid -- the case which we have intentionally excluded in order to focus on derivation of cosmological generalization of the post-Newtonian equations of relativistic celestial mechanics \citep{2011rcms.book.....K}.   

\subsection{Field Equations for Tensor Perturbations}
Field equations for traceless tensor $\mathfrak{p}^{\sst\intercal}_{\a\b}$ can be derived by applying the covariant Laplace-Beltrami operator to the definition (\ref{tp1}) and making use of equation (\ref{qe7}) along with a tedious algebraic transformations. This yields the following equation
\be\la{veq4}
\mathfrak{p}^{\sst\intercal}_{\a\b}{}^{|\g}{}_{|\g}-2\H\le(\bar u_\a\mathfrak{p}^{\sst\intercal}_{\b\g}{}^{|\g}+\bar u_\b\mathfrak{p}^{\sst\intercal}_{\a\g}{}^{|\g}\ri)
-2\le(\H^2+\frac{k}{a^2}\ri)\mathfrak{p}^{\sst\intercal}_{\a\b}-{\bar P}_\a{}^\m {\bar P}_\b{}^\n \le({B}_{\m|\n}+{B}_{\n|\m}\ri)+\frac23{\bar P}_{\a\b}{\bar P}^{\m\n}{B}_{\m|\n}=16\pi\t^{\sst\intercal}_{\a\b}\;.
\ee
Here the transverse and traceless tensor source of the tensor perturbations is
\be\la{veq5}
\t^{\sst\intercal}_{\a\b} \equiv \t_{\a\b}-\frac13\bar P_{\a\b}\;\t\;,
\ee
where $\t_{\a\b}$ has been introduced in (\ref{qq7}), and $\t=\bar P^{\a\b}\t_{\a\b}$ in accordance with equation (\ref{qq5}). Tensor
 $\t^{\sst\intercal}_{\a\b} $ is traceless, that is $\bar g^{\a\b}\t^{\sst\intercal}_{\a\b}=\bar P^{\a\b}\t^{\sst\intercal}_{\a\b}=0$.

 Equation (\ref{veq4}) is gauge-dependent. The gauge freedom is significantly reduced by imposing the gauge condition $B^\a=0$ which brings equation (\ref{veq4}) to the following form,
 \be\la{vety4}
 \mathfrak{p}^{\sst\intercal}_{\a\b}{}^{|\g}{}_{|\g}-2\H\le(\bar u_\a\mathfrak{p}^{\sst\intercal}_{\b\g}{}^{|\g}+\bar u_\b\mathfrak{p}^{\sst\intercal}_{\a\g}{}^{|\g}\ri)
 -2\le(\H^2+\frac{k}{a^2}\ri)\mathfrak{p}^{\sst\intercal}_{\a\b}=16\pi\t^{\sst\intercal}_{\a\b}\;.
 \ee

\subsection{The Residual Gauge Freedom}\la{rgfr}

The gauge freedom of the theory under discussion is associated with the gauge function $B^\a$ appearing in equation (\ref{qe6}). The most favourable choice of the gauge condition is
\be\la{pn9}
{B}^\a=0\;,
\ee
which drastically simplifies the above equations for vector and tensor gravitational perturbations. The gauge (\ref{pn9}) is a generalization of the harmonic (de Donder) gauge condition used in the gravitational wave astronomy and in the post-Newtonian dynamics of extended bodies. This choice of the gauge establishes differential relationships between the algebraically-independent metric tensor components introduced in section \ref{alge}. Indeed, substituting the algebraic decomposition (\ref{yu1}) of the metric tensor perturbations to equation (\ref{qe6}) and imposing the condition (\ref{pn9}) yields
\be\la{ppn1}
\mathfrak{p}^{\sst\intercal}_{\a\b}{}^{|\b}+\bar u_\a\mathfrak{p}_\b{}^{|\b}+\bar u_\b\mathfrak{p}_\a{}^{|\b}-\le(\bar u_\a\bar u_\b-\frac13\bar P_{\a\b}\ri)\mathfrak{p}^{|\b}+2\bar u_\a\bar u_\b\mathfrak{q}^{|\b}+2\H\mathfrak{p}_\a+2\H\mathfrak{q}\bar u_\a=16\pi\le(\bar\r_{\rm m}\bar\m_{\rm m}\chi_{\rm m}+\bar\r_{\rm q}\bar\mu_{\rm q}\chi_{\rm q}\ri)\bar u_\a\;.
\ee
Projecting this relationship on the direction of the background 4-velocity, $\bar u^\a$, and on the hypersurface being orthogonal to it, we derive two algebraically-independent equations between the perturbations of metric tensor components and of the matter variables. They are
\bsu\ba\la{ppn2}
\mathfrak{p}_\b{}^{|\b}+\bar u_\b\le(2\mathfrak{q}-\mathfrak{p}\ri)^{|\b}+2\H\mathfrak{q}&=&16\pi\le(\bar\r_{\rm m}\bar\m_{\rm m}\chi_{\rm m}+\bar\r_{\rm q}\bar\mu_{\rm q}\chi_{\rm q}\ri)\;,\\\la{ppn3}
\mathfrak{p}^{\sst\intercal}_{\a\b}{}^{|\b}+\bar u_\b\mathfrak{p}_\a{}^{|\b}+\frac13\bar P_{\a\b}\mathfrak{p}^{|\b}+2\H\mathfrak{p}_\a&=&0\;.
\ea\esu

The gauge (\ref{pn9}) does not fix the gauge function $\xi^\a$ uniquely. The residual gauge freedom is described by the gauge transformations that preserve equations (\ref{ppn2}), (\ref{ppn3}). Substituting the gauge transformation (\ref{qe12}) of the gravitational field perturbation $l_{\a\b}$ to equation (\ref{qe6}) and holding on the gauge condition (\ref{pn9}), yields the differential equation for the  vector function $\xi^\a$
\be\la{jk1}
\xi^{\a|\b}{}_{|\b}+\bar g^{\a\g}\le(\xi^\b{}_{|\g\b}-\xi^\b{}_{|\b\g}\ri)+2\H\le(\xi^{\a|\b}\bar u_\b+\xi^{\b|\a}\bar u_\b-\xi^{\b}{}_{|\b}\bar u^\a\ri)-16\pi\le(\bar\r_{\rm m}\bar\m_{\rm m}+\bar\r_{\rm q}\bar\m_{\rm q}\ri)\xi^\b\bar u_\b\bar u^\a=0\;,
\ee
which can be further recast to
\be\la{jk2}
\xi^{\a|\b}{}_{|\b}+2\H\le(\xi^{\a|\b}\bar u_\b+\xi^{\b|\a}\bar u_\b-\xi^{\b}{}_{|\b}\bar u^\a\ri)+2\le(\dot H-\frac{k}{a^2}\ri)\xi^\b \bar u_\b\bar u^\a+\le(\dot H+3\H^2+\frac{2k}{a^2}\ri)\xi^\a=0\;.
\ee
The gauge function $\xi^\a$ can be decomposed in time-like, $\xi\equiv -\xi^\b\bar u_\b$, and space-like, $\zeta^\a\equiv \bar P^\a{}_\b\xi^\b$, components,
\be\la{jk2a}
\xi^\a=\zeta^\a+\bar u^\a\xi\;.
\ee
Calculating covariant derivatives from $\xi$ and $\zeta^\a$ and making use of equation (\ref{jk2}), yield equations
\bsu\ba\la{jk3}
\xi^{|\b}{}_{|\b}+2\H\bar u^\b\xi_{|\b}-\le(\dot H-\frac{4k}{a^2}\ri)\xi\phantom{^\a}&=&0\;,\\
\la{jk4}
\zeta^{\a|\b}{}_{|\b}+2\H\le(\bar u^\b\zeta^\a{}_{|\b}-\bar u^\a\zeta^\b{}_{|\b}\ri)+\le(\dot H+\H^2+\frac{2k}{a^2}\ri)\zeta^\a&=&0\;.
\ea\esu
These equations have non-trivial solutions which describe the residual gauge freedom in choosing the coordinates on the background manifold subject to the gauge condition (\ref{pn9}). It is remarkable that equations (\ref{jk3}), (\ref{jk4}) are decoupled and can be solved separately. It means that the residual gauge transformations along the worldlines of the Hubble flow are functionally independent of those performed on the hypersurface being orthogonal to the Hubble flow.

\section{Post-Newtonian Field Equations in a Spatially-Flat Universe}\la{pnfe}

\subsection{Cosmological Parameters and Scalar Field Potential}
Equations of the field perturbations given in the previous section are generic and valid for any model of the FLRW universe. They neither specify the equation of state of matter, nor that of the scalar field, nor the parameter of the space curvature $k$. By choosing a specific model of matter and picking up a value of $k=-1,0,+1$, we can solve, at least, in principle the field equations governing the time evolution of the background cosmological manifold. Realistic models of the cosmological matter are rather sophisticated and, as a rule, include several components. It leads to the system of the coupled field equations which can be solved only numerically. However, the large scale structure of the universe is formed at rather late stages of the cosmological evolution being fairly close to the present epoch. Therefore, the study of the impact of cosmological expansion on the post-Newtonian dynamics of isolated astronomical systems is based on the recent and present states of the universe.

Radiometric observations of the relic CMB radiation and photometry of type Ia supernova explosions reveal that at the present epoch the space curvature of the universe, $k=0$, and the evolution of the universe is primarily governed by the dark energy and dark matter, which make up to 74\% and 24\% of the total energy density of the universe respectively, while 4\% of the energy density of the universe belongs to visible matter (baryons), and a tiny fraction of the energy density occupies by the CMBR radiation \citep{1996ApJ...473..576F,1993ApJ...419....1K,wmap_2009,wmap_2011}.
It means that we can neglect the effects of the baryonic matter and CMB radiation field in consideration of the post-Newtonian dynamics of astronomical systems in the expanding universe.

We shall assume that the dark matter is made of an ideal fluid and the dark energy is represented by a scalar field with a potential function $\bar W$ which structure should be further specified. In doing this, we shall follow discussion in \citep{2010deto.book.....A} assuming that the spatial curvature $k=0$, and the potential, $\bar W$, of the scalar field relates to its derivative by a simple equation
\be\la{pn24}
\frac{\pd\bar W}{\pd\bar\Psi}=-\sqrt{8\pi}\lambda\bar W\;,
\ee
where the time-dependent parameter, $\lambda=\lambda(\bar\Psi)$, characterizes the slope of the field potential $\bar W$. The time evolution of the background universe can be described in terms of the parameter $\lambda$ and two other parameters, $\mathrm{x}_1=\mathrm{x}_1(\bar\Psi)$ and $\mathrm{x}_2=\mathrm{x}_2(\bar\Psi)$, which are functions of the density, $\bar\r_{\rm q}=\bar\m_{\rm q}=\bar\Psi$, of the background scalar field, and the potential, $\bar W$, scaled to the Hubble parameter, $\H$. These parameters are defined more specifically as follows,
\ba\la{pn26}
\bar\r_{\rm q}^2&=&\frac{3\H^2}{4\pi}\,\mathrm{x}_1\;,
\\\la{pnw1}
\bar W&=&\frac{3\H^2}{8\pi}\,\mathrm{x}_2\;.
\ea
The energy density of the scalar field, $\bar\e_{\rm q}$, is expressed in terms of the parameters $\mathrm{x}_1$ and $\mathrm{x}_2$ and the parameter $\Omega_{\rm q}\equiv 8\pi\bar\e_{\rm q}/3\H^2$, by a simple relationship
\be\la{oku}
\Omega_{\rm q}=\mathrm{x}_1+\mathrm{x}_2\;.
\ee

The time evolution of the parameters $\mathrm{x}_1$ and $\mathrm{x}_2$ is given by the system of two ordinary differential equations which are obtained by differentiating the definitions (\ref{pn26}), (\ref{pnw1}) and making use of the equations (\ref{cr3a}) along with the Friedmann equation (\ref{13a}) with $k=0$. It yields
\bsu\ba\la{pn27}
\frac{d\mathrm{x}_1}{d\omega}&=&-6\mathrm{x}_1+\lambda\sqrt{6\mathrm{x}_1}\mathrm{x}_2+3\mathrm{x}_1\le[\le(1-w_{\rm m}\ri)\mathrm{x}_1+\le(1+w_{\rm m}\ri)\le(1-\mathrm{x}_2\ri)\ri]\;,\\
\la{pn28}
\frac{d\mathrm{x}_2}{d\omega}&=&-\lambda\sqrt{6\mathrm{x}_1}\mathrm{x}_2+3\mathrm{x}_2\le[\le(1-w_{\rm m}\ri)\mathrm{x}_1+\le(1+w_{\rm m}\ri)\le(1-\mathrm{x}_2\ri)\ri]\;,
\ea\esu
where $\omega\equiv\ln a$ is the logarithmic scale factor characterizing the number of e-folding of the universe, $w_{\rm m}$ is the parameter entering the hydrodynamic equation of state (\ref{a10}), and the parameters $\mathrm{x}_1$ and $\mathrm{x}_2$ are restricted by the condition imposed by the Friedmann equation (\ref{a12}), that is
\be\la{pon}
\mathrm{x}_1+\mathrm{x}_2=1-\Omega_{\rm m}\;,
\ee
where $\Omega_{\rm m}\equiv 8\pi\bar\e_{\rm m}/3\H^2$.

The parameter $\lambda$ obeys the following equation
\be\la{pn29a}
\frac{d\lambda}{d\omega}=-\sqrt{6\mathrm{x}_1}\lambda^2\le(\G_{\rm q}-1\ri)\;,
\ee
where
\be\la{pn30}
\Gamma_{\rm q}=\frac{\displaystyle\pd^2\bar W/\pd\bar\Psi^2}{\displaystyle\le(\pd\bar W/\pd\bar\Psi\ri)^2}\,\bar W\;,
\ee
If $\Gamma_{\rm q}=1$, the parameter $\lambda$ is constant, and equation (\ref{pn24}) can be integrated yielding an exponential potential
\be\la{pn25}
\bar W(\bar\Psi)=\bar W_0\exp(-\sqrt{8\pi}\lambda\bar\Psi)\;.
\ee
In this case, and under assumption that, $w_{\rm m}={\rm const.}$, the system of two differential equations (\ref{pn27}), (\ref{pn28}) is closed. If $\Gamma_{\rm q}\not=1$, three equations (\ref{pn27})--(\ref{pn29a}) must be solved in order to describe the evolution of the background cosmological manifold.

In the general case, derivatives of the potential $\bar W$ are expressed in terms of the parameters under discussion. Namely,
\be\la{wer1}
\frac{\pd\bar W}{\pd\bar\Psi}=-\frac{3\lambda}{\sqrt{8\pi}}\H^2 \mathrm{x}_2\qquad,\qquad
\frac{\pd^2\bar W}{\pd\bar\Psi^2}=3\Gamma_{\rm q}\lambda^2\H^2 \mathrm{x}_2\;.
\ee
It is also useful to express the products $\bar\r_{\rm q}\bar\m_{\rm q}$ and $\bar\r_{\rm m}\bar\m_{\rm m}$ in terms of the parameters $\mathrm{x}_1$ and $\mathrm{x}_2$.
For $\bar\m_{\rm q}=\bar\r_{\rm q}$, one can use definition (\ref{pn26}) to obtain
\be\la{pn26s}
\bar\r_{\rm q}\bar\m_{\rm q}=\frac{3\H^2}{4\pi}\,\mathrm{x}_1\;.
\ee
The product $\bar\r_{\rm m}\bar\m_{\rm m}=\bar\e_{\rm m}+\bar p_{\rm m}$, so that making use of the matter equation of state, $\bar p_{\rm m}=w_{\rm m}\bar\e_{\rm m}$, and equation (\ref{pon}), we derive
\be\la{wwr}
\bar\r_{\rm m}\bar\m_{\rm m}=\frac{3\H^2}{8\pi}(1+w_{\rm m})\Omega_{\rm m}\;,
\ee
where $\Omega_{\rm m}=1-\mathrm{x}_1-\mathrm{x}_2$.
These equations allow us to recast equation (\ref{13a}) for the time derivative of the Hubble parameter to the following form
\be\la{wer3}
\dot H=-\frac32\le(1+w_{\rm eff}\ri)\H^2\;,
\ee
where
\be\la{bvc}
w_{\rm eff}\equiv w_{\rm m}+(1-w_{\rm m})\mathrm{x}_1-(1+w_{\rm m})\mathrm{x}_2\;,
\ee
is the (time-dependent) parameter of the effective equation of state of the mixture of the ideal fluid and the scalar field.

\subsection{Conformal Cosmological Perturbations}\la{ccpe}

The FLRW metric (\ref{bub5}) is a product of the scale factor $a$ and a conformal metric $\bar\mm_{\a\b}$. The conformal spacetime is comoving with the Hubble flow and is not globally expanding. In case of the flat spatial curvature, $k=0$, the conformal spacetime becomes equivalent to the Minkowski spacetime which is used as a starting point in the standard theory of the post-Newtonian or post-Minkowskian approximations \citep{1987thyg.book..128D}. Therefore, it is instructive to formulate the field equations for cosmological perturbations in the conformal spacetime.

Let us define the cosmological perturbations, $h_{\a\b}$, of gravitational field in the conformal spacetime with the background metric $\bar{\mathtt f}_{\a\b}$ as follows,
\be\la{pn1}
\varkappa_{\a\b}=a^2(\eta)h_{\a\b}\;,
\ee
where perturbations $\varkappa_{\a\b}$ has been defined in (\ref{pf8}), and $a(\eta)$ is the scale factor of the FLRW universe. Perturbation $l_{\a\b}$ relates to $\varkappa_{\a\b}$ by equation (\ref{ok9}), and can be also represented in the conformal form
\be\la{pop3a}
l_{\a\b}=a^2(\eta)\gamma_{\a\b}\;,
\ee
where
\be\la{pn1a}
\gamma_{\a\b}=-h_{\a\b}+\frac12\bar{\mathtt f}_{\a\b}h\;,
\ee
with $h\equiv \bar{\mathtt f}^{\a\b}h_{\a\b}$. In what follows, tensor indices of geometric objects in the conformal spacetime are raised and lowered with the help of the conformal metric $\bar\mm_{\a\b}$.

We assume that the scale factor $a$ of the universe remains unperturbed. This assumption is justified since we can always include the perturbation of the scale factor to that of the conformal metric. Thus, the perturbed physical spacetime interval, $ds$, of the FLRW universe relates to the perturbed conformal spacetime interval, $d\tilde s$, by the conformal transformation
\be\la{pn2}
ds^2=a^2(\eta)d{\tilde s}^2\;.
\ee
The perturbed conformal spacetime interval reads
\be\la{pn2a}
d{\tilde s}^2=f_{\a\b}dx^\a dx^\b\;,
\ee
where
\be\la{pn2x}
f_{\a\b}=\bar{\mathtt f}_{\a\b}+h_{\a\b}\;,
\ee
is the perturbed conformal metric. Here, $\bar{\mathtt f}_{\a\b}$ is the unperturbed conformal metric defined in (\ref{bub6}), $h_{\a\b}$ is the perturbation of the conformal metric, and $x^\a=(x^0,x^i)$ are arbitrary coordinates which are the same as in the physical spacetime manifold.

It is worth emphasizing that in case of the space curvature $k=0$, the background conformal metric, ${\mathtt g}_{\a\b}(\eta,X^i)$, expressed in the isotropic Cartesian coordinates $(\eta,X^i)$, is the diagonal Minkowski metric, ${\mathtt g}_{\a\b}(\eta,X^i)=\eta_{\a\b}={\rm diag}(-1,1,1,1)$. Therefore, the background metric $\bar{\mathtt f}_{\a\b}$ remains the Minkowski metric with the components expressed in arbitrary coordinates by means of tensor transformation
\be\la{pn2z}
\bar{\mathtt f}_{\a\b}={\rm M}^\m{}_\a{\rm M}^\n{}_\b \eta_{\m\n}\;,
\ee
where the matrix of transformation has been defined in (\ref{bub2}).
If the matrix of transformation, ${\rm M}^\m{}_\a$, is the Lorentz boost, the conformal metric, $\bar{\mathtt f}_{\a\b}$, remains flat, $\bar{\mathtt f}_{\a\b}=\eta_{\a\b}$. It is worth noticing that, in general, the unperturbed conformal metric can be chosen flat even in case of $k=-1,+1$ \citep{2007JMP....48l2501I}. It means that our formalism is applicable to FLRW universe with any space curvature. However, the conformal factor in this case is not merely the scale factor $a(\eta)$ of the FLRW universe but a more complicated function of coordinates. Though it is not difficult to handle all three cases of $k=-1,0,+1$ but it burdens equations for the field perturbations and we restrict ourselves only to the case of the spatially flat universe with $k=0$.

Similarly to (\ref{yu1}) the conformal metric perturbation, $\gamma_{\a\b}$, can be split in the algebraically-irreducible components
\be\la{pn3}
\gamma_{\a\b}={p}^{\sst\intercal}_{\a\b}+\bar{\mathtt v}_\a{p}_\b+\bar{\mathtt v}_\b{p}_\a+\le(\bar{\mathtt v}_\a\bar{\mathtt v}_\b+\frac13 {{\bar\pi}}_{\a\b}\ri){p}+2\bar{\mathtt v}_\a\bar{\mathtt v}_\b\le({q}-{p}\ri)\;,
\ee
where the four-velocity $\bar{\mathtt v}^\a=a\bar u^\a$, $\bar{\mathtt v}_\a=\bar{\mathtt f}_{\a\b}\bar{\mathtt v}^\b=a^{-1}\bar g_{\a\b}\bar u^\b=a^{-1}\bar u_\a$, and
\be\la{opp3}
{{\bar\pi}}_{\a\b}=\bar{\mathtt f}_{\a\b}+\bar{\mathtt v}_\a\bar{\mathtt v}_\b\;,
\ee
is the operator of projection on the conformal space which represents a hypersurface being everywhere orthogonal to the congruence of worldlines of four-velocity $\bar{\mathtt v}^\a$. Four-velocity $\bar{\mathtt v}^\a$ is an analogue of the Hubble flow in the conformal spacetime. We also notice that $\bar P_{\a\b}=a^2{{\bar\pi}}_{\a\b}$.

Different pieces of the conformal metric perturbation, $\gamma_{\a\b}$, are related to those of the physical metric perturbation, $l_{\a\b}$, by the powers of the scale factor,
\be\la{pn4}
\mathfrak{p}^{\sst\intercal}_{\a\b}=a^2{p}^{\sst\intercal}_{\a\b}\;,\qquad \mathfrak{p}_{\a}=a{p}_{\a}\;,\qquad \mathfrak{p}={p}\;,\qquad \mathfrak{q}={q}\;.
\ee
More specifically,
\bsu\ba\la{pn5}
{q}&=&\frac12\le(\bar{\mathtt v}^\m\bar{\mathtt v}^\n+{\bar\pi}^{\m\n}\ri)\gamma_{\m\n}\;,\\
\la{pn6}
{p}&=&\phantom{-}{\bar\pi}^{\m\n}\gamma_{\m\n}\;,\\
\la{pn7}
{p}_\a&=&-{{\bar\pi}}_\a{}^\b \bar{\mathtt v}^\g\gamma_{\b\g}\;,\\
\la{pn8}
{p}^{\sst\intercal}_{\a\b}&=&\phantom{-}p_{\a\b}-\frac13{{\bar\pi}}_{\a\b} p\;,
\ea\esu
where
\be\la{pn8a}
p_{\a\b}={{\bar\pi}}_\a{}^\m {{\bar\pi}}_\b{}^\n\gamma_{\m\n}\;.
\ee
The trace of the gravitational perturbation, $\gamma=\bar{\mathtt f}^{\a\b}\gamma_{\a\b}=2(p-q)$.
The components $h_{\a\b}=-\gamma_{\a\b}+\bar{\mathtt f}_{\a\b}\gamma/2$ are used in calculating dynamical behavior of particles and light in the conformal spacetime as well as in matching theory with observables. The components of $h_{\a\b}$ are
\be\la{pn5a}
h_{\a\b}=-{p}^{\sst\intercal}_{\a\b}-\bar{\mathtt v}_\a{p}_\b-\bar{\mathtt v}_\b{p}_\a+\frac23 {{\bar\pi}}_{\a\b}{p}-\le(\bar{\mathtt v}_\a\bar{\mathtt v}_\b+{{\bar\pi}}_{\a\b}\ri){q}\;,
\ee
and $h=\bar{\mathtt f}^{\a\b}h_{\a\b}=2(p-q)=\gamma$.

It turns out that the conformal Hubble parameter, ${\cal H}=a'/a$
is more convenient in the conformal spacetime than $\H=\dot R/R=R^{-1}dR/dT$, where $T$ is the cosmological time (see section \ref{frwlm}). Relationships between ${\cal H}$ and $\H$, and their derivatives are shown in equations (\ref{bm1c})--(\ref{bm1w}). These relationships along with equations (\ref{bm1d}) and (\ref{wer3}) are employed in order to express the time derivative, ${\cal H}'$, of the conformal Hubble parameter in terms of ${\cal H}^2$ and the parameter $w_{\rm eff}$ of the effective equation of state
\be\la{frg}
{\cal H}'=-\frac12(1+3w_{\rm eff}){\cal H}^2\;.
\ee
We shall use this expression in the calculations that follows.

\subsection{The Post-Newtonian Field Equations in Conformal Spacetime}
The set of the post-Newtonian field equations in cosmology consists of equations for perturbations of the background matter and gravitational field. Perturbations of matter are described by four scalars, $V_{\rm m}$, $V_{\rm q}$, $\chi_{\rm m}$ and $\chi_{\rm q}$ but only three of them are functionally-independent because of equality (\ref{feq5a}), that is
\be\la{frg1}
V_{\rm m}-V_{\rm q}=\bar u^\a\le(\chi_{\rm m}-\chi_{\rm q}\ri)_{|\a}\;.
\ee
Depending on a particular situation, any of the three scalars can be taken as independent variables.

The gravitational field perturbations are $q$, $p$, ${p}_\a$,  ${p}^{\sst\intercal}_{\a\b}$ but among them the scalar $q$ is not independent and can be expressed either in terms of $\chi_{\rm m}$ and $V_{\rm m}$ in accordance with (\ref{sp7}),
\be\la{frg2}
q=-2\bigl(V_{\rm m}-\bar u^\a\chi_{{\rm m},\a}\bigr)\;, \qquad
\ee
where we have also used the equality $\mathfrak{q}=q$ as follows from (\ref{pn4}). The scalar $q$ can be also expressed in terms of $\chi_{\rm q}$ and $V_{\rm q}$ in accordance with  (\ref{sp8}).
Hence, as soon as the pairs, $V_{\rm m}$ and $\chi_{\rm m}$ or $V_{\rm q}$ and $\chi_{\rm q}$ are known, the scalar gravitational perturbation $q$ can be easily calculated from (\ref{frg2}). Functions $p$, ${p}_\a$,  ${p}^{\sst\intercal}_{\a\b}$ are independent and decouple both from each other and from the other perturbations. Thus, the most difficult part of the theory is to find out solutions of the scalar perturbations which are coupled one to another.

The field equations in the conformal spacetime for variables $\chi_{\rm m}$, $\chi_{\rm q}$, $V_{\rm m}$ and for $p$, ${p}_\a$,  ${p}^{\sst\intercal}_{\a\b}$ are derived from the equations of the previous section by transforming all functions and operators from physical to  conformal spacetime. The important part of the transformation technique is based on formulas converting the covariant Laplace-Beltrami wave operators, defined on the background spacetime manifold, to their conformal spacetime counterparts.

\subsubsection{The Laplace-Beltrami operator}

Let $F$ be an arbitrary scalar, $F_\a$ - an arbitrary covector, and $F_{\a\b}$ - an arbitrary covariant tensor of the second rank. We have three Laplace-Beltrami operators on the curved background manifold: scalar - $F^{|\m}{}_{|\m}$, vector - $F_\a{}^{|\m}{}_{|\m}$, and tensor - $F_{\a\b}{}^{|\m}{}_{|\m}$ types where the covariant derivatives are taken with respect to the affine connection $\bar\Gamma^\a{}_{\b\g}$ being compatible with the metric $\bar g_{\a\b}$ (see equation \ref{bub12}). Covariant derivatives are the most convenient for the invariant description of differential equations of mathematical physics on curved manifolds. For practical purposes of finding solutions of the differential equations, the covariant operators must be expressed in terms of partial derivatives with respect to coordinates chosen for solving the equations.

Transformation of the covariant Laplace-Beltrami operators to the partial derivatives is achieved after writing down the covariant derivatives for scalar, vector and tensor in explicit form by making use of the Christoffel symbols given in (\ref{bm6})--(\ref{bm8}). Tedious but straightforward calculations of the covariant derivatives yield the scalar, vector and tensor Laplace-Beltrami operators in the following form
\bsu\ba\la{wer6}
F^{|\m}{}_{|\m}&=&\frac{1}{a^2}\Bigl[\Box F-2{\cal H}\bar{\mathtt v}^\m F_{;\m}\Bigr]\;,\\
\la{wer7}
F_\a{}^{|\m}{}_{|\m}&=&\frac{1}{a^2}\Bigl[\Box F_\a-2{\cal H}\bar{\mathtt v}^\m F_{\m;\a}+2{\cal H}\bar{\mathtt v}_\a \bar{\mathtt f}^{\m\n}F_{\m;\n}+\le({\cal H}'+2{\cal H}^2\ri)F_\a-2{\cal H}^2\bar{\mathtt v}_\a\bar{\mathtt v}^\m F_\m\Bigr]\;,\\
\la{wer8}
F_{\a\b}{}^{|\m}{}_{|\m}&=& \frac{1}{a^2}\Bigl[\Box F_{\a\b}+2{\cal H}\bar{\mathtt v}^\m F_{\a\b;\m}-2{\cal H}\bar{\mathtt v}^\m F_{\m\a;\b}-2{\cal H}\bar{\mathtt v}^\m F_{\m\b;\a}+
2{\cal H} \bar{\mathtt f}^{\m\n}\le(\bar{\mathtt v}_\a F_{\b\m;\n}+\bar{\mathtt v}_\b F_{\a\m;\n}\ri)\\\nonumber
&&\phantom{\frac{1}{a^2}}+2\le({\cal H}'+{\cal H}^2\ri)F_{\a\b}-4{\cal H}^2\Bigl(\bar{\mathtt v}^\m\bar{\mathtt v}_\a F_{\b\m}+\bar{\mathtt v}^\m\bar{\mathtt v}_\b F_{\a\m}-\frac12\bar{\mathtt v}_\a\bar{\mathtt v}_\b \bar{\mathtt f}^{\m\n}F_{\m\n}-\frac12\bar{\mathtt f}_{\a\b}\bar{\mathtt v}^\m\bar{\mathtt v}^\n F_{\m\n}\Bigr)\Bigr]\;,
\ea\esu
where we have introduced notations
\be\la{wer9}
\Box F\equiv \bar{\mathtt f}^{\m\n}F_{;\m\n}\;,\qquad\qquad\Box F_\a\equiv \bar{\mathtt f}^{\m\n}F_{\a;\m\n}\;,\qquad\qquad\Box F_{\a\b}\equiv \bar{\mathtt f}^{\m\n}F_{\a\b;\m\n}\;,\qquad\qquad
\ee
of the wave operators for scalar, vector and tensor fields in the conformal spacetime and in arbitrary coordinates. Notice that although the conformal spacetime coincides, in case of $k=0$, with the Minkowski spacetime, the metric $\bar{\mathtt f}_{\a\b}$ is not the diagonal Minkowski metric $\eta_{\a\b}$ unless the coordinates are Cartesian. Of course, the covariant derivative from a scalar must be understood as a partial derivative, that is $F_{;\a}=F_{,\a}$.

We will need several other equations to complete the transformation of the Laplace-Beltrami operators to the conformal spacetime since the wave operator $\Box$ acts on functions like (\ref{pn4}) which are made of a product of some power $n$ of the scale factor, $a=a(\eta)$, with a geometric object, $\digamma=\digamma(x^\a)$, which can be a scalar, a vector or a tensor of the second rank (we have suppressed the tensor indices of $\digamma$ since they do not interfere with the derivation of the equations which follow). These equations are
\bsu\ba\la{wer12}
\le(a^n\digamma\ri)_{;\m}&=&a^n\le(\digamma_{;\m}-n{\cal H}\bar{\mathtt v}_\m \digamma\ri)\;,\\\la{wer13}
\le(a^n\digamma\ri)_{;\m\n}&=&a^n\le[\digamma_{;\m\n}-n{\cal H}\le(\bar{\mathtt v}_\m \digamma_{;\n}+\bar{\mathtt v}_\n \digamma_{;\m}\ri)+n\le( {\cal H}'+n{\cal H}^2\ri)\bar{\mathtt v}_\m\bar{\mathtt v}_\n\ri]\;,
\ea\esu
and they allow us to write down the wave operator from the product of $a^n$ and $\digamma$ in the following form
\be\la{wer10}
\Box\le(a^n\digamma\ri)=a^n\Bigl[\Box \digamma-2n{\cal H}\bar{\mathtt v}^\m \digamma_{;\m}-n\le({\cal H}'+n{\cal H}^2\ri)\digamma\Bigr]\;,
\ee
It is easy to confirm that contraction of (\ref{wer13}) with the conformal four-velocity, $\bar{\mathtt v}^\a$, brings about another differential operator
\be\la{wer11}
\bar{\mathtt v}^\m\bar{\mathtt v}^\n\le(a^n\digamma\ri)_{;\m\n}=a^n\Bigl[\bar{\mathtt v}^\m\bar{\mathtt v}^\n \digamma_{;\m\n}+2n{\cal H}\bar{\mathtt v}^\m \digamma_{;\m}+n\le({\cal H}'+n{\cal H}^2\ri)\digamma\Bigr]\;.
\ee
We remind that if the object $\digamma$ is a scalar, the covariant derivative is reduced to a partial derivative, $\digamma_{;\a}=\digamma_{,\a}$. In case, when $\digamma$ is either a vector or a tensor, the covariant derivative must be calculated with taking into account the affine connection $\bar B^\a{}_{\b\g}$ defined in (\ref{bm8}).

It is also interesting to notice that in the expanding universe the conformal Laplace operator, $\Delta\digamma\equiv\bar\pi^{\m\n}\digamma_{;\m\n}$ is the scale invariant in the sense that
\be\la{wer10a}
\Delta\le(a^n\digamma\ri)=a^n\Delta\digamma\;,
\ee
where $\digamma$ is a tensor of an arbitrary rank. Equation (\ref{wer10a}) can be proven by adding up (\ref{wer10}) and (\ref{wer11}), and accounting for definition (\ref{opp3}) of the projection operator on the hypersurface being orthogonal to $\bar{\mathtt v}^\a$.

Now, we are ready to formulate the field equations for cosmological perturbations in the conformal spacetime.

\subsubsection{Equations for the matter perturbations}\la{emp}

We accept the gauge condition imposed by equations (\ref{qe6}), (\ref{pn9}) and convert the covariant derivatives taken with respect to the background metric, $\bar g_{\a\b}$, to the partial derivatives of the conformally-flat metric, $\bar{\mathtt f}_{\a\b}$, in equation (\ref{qe25}) for scalar $V_{\rm m}$. We use equation (\ref{wer6}) for the Laplace-Beltrami operator, and expressions for various cosmological parameters given in section \ref{ccpe}. After arranging terms with respect to the powers of the Hubble parameter ${\cal H}$, we obtain the sound-wave equation for function $V_{\rm m}$ describing perturbations of the ideal fluid,
\ba\la{pn10}
\Box V_{\rm m}+\le(1-\frac{c^2}{c_{\rm s}^2}\ri)\bar{\mathtt v}^\a\bar{\mathtt v}^\b V_{{\rm m};\a\b}+
\le(3-\frac{c^2}{c_{\rm s}^2}\ri){\cal H}\bar{\mathtt v}^\a V_{{\rm m},\a}&&\\\nonumber
+3\le[1-w_{\rm eff}-\frac12(1+w_{\rm m})\le(1-\frac{c^2}{c^2_{\rm s}}\ri)\Omega_{\rm m}\ri]{\cal H}^2V_{\rm m}&&\\\nonumber
+12{\cal H}^2\le[\bar{\mathtt v}^\a\chi_{,\a}-3\le(1-\sqrt{\frac{3}{8\mathrm{x}_1}}\lambda \mathrm{x}_2\ri){\cal H}\chi\ri]\frac{\mathrm{x}_1}{a}
&=&-4\pi a^2\le(\sigma+\t\ri)\;.\ea
Similar procedure applied to equation (\ref{feq6}) leads to a wave equation for function $V_{\rm q}$ describing perturbations of the scalar field,
\ba\la{vq1}
\Box V_{\rm q}+2\le(1-\sqrt{\frac{3}{2\mathrm{x}_1}}\lambda \mathrm{x}_2\ri){\cal H}\bar{\mathtt v}^\m V_{{\rm q},\m}&&\\\nonumber
+3\le[1-w_{\rm eff}-\frac12(1+w_{\rm m})\le(1-\frac{c^2}{c^2_{\rm s}}\ri)\Omega_{\rm m}\ri]{\cal H}^2V_{\rm q}&&\\\nonumber
+\lambda \mathrm{x}_2\le[3\lambda\le(2\Gamma_{\rm q}+\frac{\mathrm{x}_2}{\mathrm{x}_1}\ri)-5\sqrt{\frac{6}{\mathrm{x}_1}}\ri]{\cal H}^2V_{\rm q}&&\\\nonumber
+\frac32 {\cal H}^2\le(1+w_{\rm m}\ri)\le(3+\frac{c^2}{c^2_{\rm s}}\ri)\le[\bar{\mathtt v}^\m\chi_{,\m}-3\frac{c^2_{\rm s}}{c^2}{\cal H}\chi\ri]\frac{\Omega_{\rm m}}{a}
&=&-4\pi a^2\le(\sigma+\t\ri)\;.
\ea
Equations (\ref{pn10}) and (\ref{vq1}) contains function $\chi$ which obeys equation (\ref{feq3}). Making use of the same transformations as above, we recast (\ref{feq3}) to a wave equation for $\chi$,
\be
\la{pn11}
\Box\chi+4{\cal H}\le(1-\sqrt{\frac{3}{8\mathrm{x}_1}}\lambda \mathrm{x}_2\ri)\bar{\mathtt v}^\a\chi_{,\a}-\frac92\le(1+w_{\rm eff}\ri){\cal H}^2\chi
=-a\le[\sqrt{\frac{6}{\mathrm{x}_1}}\lambda \mathrm{x}_2 {\cal H}V_{\rm m}+\le(1-\frac{c^2}{c_{\rm s}^2}\ri)\bar{\mathtt v}^\a V_{{\rm m},\a}\ri]\;.
\ee
Equations (\ref{pn10})-(\ref{pn11}) are closed with respect to the variables $V_{\rm m}$, $V_{\rm q}$ and $\chi$. The gauge-invariant scalar, $V_{\rm m}$ describes propagation of sound waves in the ideal fluid filling up the expanding universe. It can be found from solving two equations (\ref{pn10}) and (\ref{pn11}) simultaneously after imposing a certain (cosmological) boundary conditions. As soon as the gauge-invariant scalar $\chi$ is known, the potential, $V_{\rm q}$, can be determined as a particular solution of the inhomogeneous equation (\ref{vq1}) or, more simple, from equation (\ref{feq5a}).

We also need equations for the normalized Clebsch and scalar potentials, $\chi_{\rm m}$ and $\chi_{\rm q}$. These potentials are required to determine the gravitational perturbation, $q$, with the help of (\ref{frg2}) and/or to get the check on self-consistency of the solutions of equations in the matter sector of the perturbation theory. Conformal-spacetime equations for $\chi_{\rm m}$ and $\chi_{\rm q}$ are derived from their definition (\ref{sp2}) and the field equations (\ref{qe8}) and (\ref{qe9}). They are
\ba\la{frg7}
\Box\chi_{\rm m}+\frac32\le(1+w_{\rm eff}\ri){\cal H}^2\chi_{\rm m}=12{\cal H}^2\mathrm{x}_1\chi-a\le[4{\cal H}V_{\rm m}+\le(1-\frac{c^2}{c^2_{\rm s}}\ri)\bar{\mathtt v}^\a V_{{\rm m},\a}\ri]\;,\\
\la{frg8}
\Box\chi_{\rm q}+\frac32\le(1+w_{\rm eff}\ri){\cal H}^2\chi_{\rm q}=-6{\cal H}^2(1+w_{\rm m})\Omega_{\rm m}\chi-a\le(4-\sqrt{\frac{6}{{\mathtt x}_1}}\lambda{\mathtt x}_2 \ri){\cal H}V_{{\rm q}}\;.
\ea
By subtracting one of these equations from another, we get back to equation (\ref{pn11}).

\subsubsection{Equations for the metric perturbations}

Post-Newtonian equations for gravitational perturbations in physical spacetime are (\ref{qe23}), (\ref{sp1b}), (\ref{veq2}) and (\ref{veq4}). We remind to the reader that the gauge conditions (\ref{qe6}), (\ref{pn9}) has been imposed. In this gauge, equations for the conformal metric tensor perturbations become
\bsu\ba
\la{pou1}
\Box q-2{\cal H}\bar{\mathtt v}^\a {q}_{,\a}+\le(1+3w_{\rm eff}\ri){\cal H}^2{q}&=&8\pi a^2\le(\sigma+\t\ri)-24{\cal H}^2\le[\sqrt{\frac{3{\mathtt x}_1}{8}}\lambda{\mathtt x}_2-{\cal H}{\mathtt x}_1\ri]\frac{\chi_{\rm q}}{a}\\\nonumber
&&-3\le(1+w_{\rm eff}\ri){\cal H}^2\Omega_{\rm m}\le[\le(1-\frac{c^2}{c^2_{\rm s}}\ri)V_{\rm m}-{\cal H}\le(1+3\frac{c^2_{\rm s}}{c^2}\ri)\frac{\chi_{\rm m}}{a}\ri]\;,\\
\la{pn17}
\Box{p}-2{\cal H}\bar{\mathtt v}^\a{p}_{,\a}&=&16\pi a^2\t\;,\\
\la{pn18}
\Box{p}_\a-2{\cal H}\bar{\mathtt v}^\b{p}_{\a;\b}+\le(1+3w_{\rm eff}\ri){\cal H}^2{p}_\a&=&16\pi a\t_\a\;,\\
\la{pn19}
\Box{p}^{\sst\intercal}_{\a\b}-2{\cal H}\bar{\mathtt v}^\g{p}_{\a\b;\g}&=&16\pi\t^{\sst\intercal}_{\a\b}\;.
\ea\esu
The reader can observe that equations (\ref{pou1})-(\ref{pn19}) for linearized metric perturbations are decoupled from each other. Moreover, equations (\ref{pn17}-(\ref{pn19}) are decoupled from the matter perturbations $V_{\rm m}$, $\chi_{\rm m}$, etc. Only equation (\ref{pou1}) for $q$ is coupled with the matter perturbations governed by equations (\ref{pn10}), (\ref{frg7}), (\ref{frg8}) so that these equations should be solved together. As we have mentioned above, function $q$ is a linear combination of $V_{\rm m}$ and $\chi_{\rm m}$ according to (\ref{frg2}). Hence, in order to determine $q$ it is, in fact, sufficient to solve (\ref{pn10}) and (\ref{frg7}). Nevertheless, it is convenient to present the differential equation (\ref{pou1}) for $q$ explicitly for the sake of mathematical completeness and rigour. It can be used for independent validation of the solution of the system of equations (\ref{pn10}), (\ref{frg7}) and (\ref{frg2}). Unfortunately, these equations are strongly mixed up and cannot be solved analytically in the most general situation of a multi-component background universe governed by the dark energy and dark matter. Solution of (\ref{pn10})-(\ref{frg8}) and (\ref{pou1})-(\ref{pn19}) would require an application of the methods of numerical integration.

It would be instrumental to get better insight to the post-Newtonian theory of cosmological perturbations by making some simplifying assumptions about the background model of the expanding universe in order to decouple the system of the post-Newtonian equations and to find their analytic solution explicitly. We discuss these assumptions and the corresponding system of the decoupled post-Newtonian equations in the section \ref{dspn}.

\subsection{The Residual Gauge Freedom in the Conformal Spacetime}

The gauge conditions (\ref{qe6}), (\ref{pn9}) in the physical space are given by (\ref{ppn2}), (\ref{ppn3}). After transforming to the conformal spacetime the gauge conditions read
\bsu\ba\la{gh1}
{p}^\b{}_{;\b}+\bar{\mathtt{v}}^\b\le(2q-p\ri)_{,\b}+2{\cal H}{q}&=&16\pi a\le(\bar\r_{\rm m}\bar\m_{\rm m}\chi_{\rm m}+\bar\r_{\rm q}\bar\mu_{\rm q}\chi_{\rm q}\ri)\;,\\\la{gh2}
{p}^{{\sst\intercal}\a\b}{}_{;\b}+\bar{\mathtt{v}}^\b{p}^\a{}_{;\b}+\frac13\bar \pi^{\a\b}{p}_{,\b}+2{\cal H}{p}^\a&=&0\;.
\ea\esu

The residual gauge freedom in the conformal spacetime is described by two functions, $\zeta\equiv\xi/a$ and $\zeta^\a$, where $\xi$ and $\zeta^\a$ have been defined in section \ref{rgfr}. Differential equations for $\zeta$ and $\zeta^\a$ are obtained by making transformation of equations (\ref{jk3}), (\ref{jk4}) to the conformal spacetime. The calculation is straightforward and results in
\bsu\ba\la{jk3a}
\Box\zeta-2{\cal H}\bar{\mathtt v}^\b\zeta_{,\b}+\le(1+3w_{\rm eff}\ri){\cal H}^2\zeta\phantom{^\a}&=&0\;,\\
\la{jk4a}
\Box\zeta^\a-2{\cal H}\bar{\mathtt v}^\b\zeta^\a{}_{;\b}&=&0\;.
\ea\esu
Solutions of equations (\ref{pou1})--(\ref{pn19}) are determined up to the gauge transformations
\bsu\ba\la{cvb4}
\hat q&=&q+2\bar{\mathtt v}^\a\zeta_{,\a}+2{\cal H}\zeta\;,\\\la{cvb5}
\hat p&=&p+\zeta^\a{}_{;\a}+3\bar{\mathtt v}^\a\zeta_{,\a}+6{\cal H}\zeta\;,\\\la{cvb6}
\hat p_\a&=&p_\a+\bar\pi_{\a\b}\le(\bar{\mathtt v}^\g\zeta^\b{}_{;\g}-\zeta^{,\b}+2{\cal H}\zeta^\b\ri)\;,\\\la{cvb7}
\hat p_{\a\b} &=&p_{\a\b}-\le(\bar\pi_{\m\a}\bar\pi_{\b}{}^\n+\bar\pi_{\m\b}\bar\pi_{\a}{}^\n\ri)\zeta^\m{}_{;\n}+\bar\pi_{\a\b}\le(\zeta^\a{}_{;\a}+\bar{\mathtt v}^\a\zeta_{,\a}+2{\cal H}\zeta\ri)\;,
\ea\esu
where the gauge functions $\zeta$, $\zeta^\a$ are solutions of the differential equations (\ref{jk3a}), (\ref{jk4a}).

\section{The Decoupled Systems of the Post-Newtonian Field Equations}\la{dspn}
\subsection{The Universe Governed by the Ideal Fluid and Cosmological Constant}
\subsubsection{Case 1: Arbitrary equation of state of the ideal fluid}
Let us consider a special case of the dark energy represented by the cosmological constant $\Lambda$. In this case, the equation of state of the scalar field is $w_{\rm q}=1$, and we have $\bar\r_{\rm q}\bar\m_{\rm q}=\bar\e_{\rm q}+\bar p_{\rm q}=0$. The parameter $\mathrm{x}_1=0$, and $\mathrm{x}_2=\Lambda/(3\H^2)$. It yields the parameter $\Omega_{\rm q}=\mathrm{x}_2$, and $\Omega_{\rm m}=1-\mathrm{x}_2$. Since the cosmological constant corresponds to a constant potential $\bar W$ of the scalar field, we get for its derivative $\pd\bar W/\pd\bar\Psi=0$, and equation (\ref{pn24}) points out that the parameter $\lambda=0$.

In the universe governed by the ideal fluid and the cosmological constant the parameter of the effective equation of state
\be\la{pes1}
w_{\rm eff}=w_{\rm m}-(1+w_{\rm m})\frac{\Lambda}{3\H^2}\;.
\ee
Hence, the time derivative of the Hubble parameter defined in (\ref{wer3}), is reduced to a more simple expression,
\be\la{ui8}
\dot H=\frac12\le(1+w_{\rm m}\ri)\le(\Lambda-3\H^2\ri)\;.
\ee
On the other hand, equation (\ref{13a}) tells us that in this model of the universe
\be\la{hjoq2}
\dot H=-4\pi\bar\r_{\rm m}\bar\m_{\rm m}\;,
\ee
The field equation (\ref{pn10}) for scalar $V_{\rm m}$ is reduced to that describing the time evolution of the perturbation of the ideal fluid density, $\delta\r_{\rm m}$. Indeed, the scalar $V_{\rm m}$ defined by equation (\ref{sp7}), can be recast to the form given by equation (\ref{qe21}), that is
\be\la{pn1a2}
V_{\rm m}=\frac{c^2_{\rm s}}{c^2}\,\delta_{\rm m}\;,
\ee
where the gauge-invariant scalar perturbation
\be\la{rtq1}
\delta_{\rm m}\equiv\frac{\delta\rho_{\rm m}}{\bar\r_{\rm m}}+3\H\chi_{\rm m}\;,
\ee
is a linear combination of the perturbation of the mass density of the fluid and the normalized Clebsch potential.
Replacing expression (\ref{pn1a2}) in equation (\ref{pn10}), yields the exact equation for $\delta_{\rm m}$ that is
\ba\la{pnra}
\le(1-\frac{c_{\rm s}^2}{c^2}\ri)\bar{\mathtt v}^\a\bar{\mathtt v}^\b \delta_{{\rm m};\a\b}-\frac{c_{\rm s}^2}{c^2}\Box\delta_{\rm m}+\le(1-3\frac{c_{\rm s}^2}{c^2}\ri){\cal H}\bar{\mathtt v}^\a\delta_{{\rm m},\a}&&\\\nonumber-
\frac32\le[\le(1-3w_{\rm m}\ri)\frac{c_{\rm s}^2}{c^2}+\le(1+w_{\rm m}\ri)\ri]{\cal H}^2\delta_{\rm m}+
\frac12\le(1+w_{\rm m}\ri)\le(1-3\frac{c_{\rm s}^2}{c^2}\ri)a^2\Lambda\delta_{\rm m} \,
&=&4\pi a^2\le(\sigma+\t\ri)\;.
\ea
This equation describes propagation of the ideal fluid density perturbation $\delta_{\rm m}$ in the form of sound waves with velocity $c_{\rm s}$.

Equation (\ref{pn11}) for potential $\chi$ makes no sense since the normalized perturbation $\chi_{\rm q}=\psi/\bar\m_{\rm q}$ of the scalar field diverges due to the condition $\m_{\rm q}=\r_{\rm q}=0$. Equation for the perturbation of the scalar field $\psi$ itself is obtained from (\ref{qe9}) and is reduced to a homogeneous wave equation
\be\la{rtq2}
\Box\psi-2{\cal H}\bar{\mathtt v}^\m\psi_{,\m}=0\;.
\ee

Equation for the normalized Clebsch potential, $\chi_{\rm m}$, is derived from equation (\ref{frg7}). In the case of the universe under consideration this equation reads
\be\la{rtq3}
\Box\chi_{\rm m}+\frac12\le(1+w_{\rm m}\ri)\le(3{\cal H}^2-a^2\Lambda\ri)\chi_{\rm m}=\le(1-\frac{c_{\rm s}^2}{c^2}\ri)a\bar{\mathtt v}^\m \d_{{\rm m},\m}-4a{\cal H}\frac{c_{\rm s}^2}{c^2}\d_{\rm m}\;.
\ee
This is an inhomogeneous equation that can be solved as soon as one knows $\d_{\rm m}$ from equation (\ref{pnra}).

Gravitational potential $q$ can be determined directly from equation (\ref{frg2}) after solving equations (\ref{pnra}) and (\ref{rtq3}) or by solving equation (\ref{pou1}) which takes on the following form,
\bsu\be
\la{pnq2}
\Box{q}-2{\cal H}\bar{\mathtt v}^\m{q}_{,\m}+\le[\le(1+3w_{\rm m}\ri){\cal H}^2-\le(1+w_{\rm m}\ri)a^2\Lambda\ri]{q}
=8\pi a^2\le\{\sigma+\t+\bar\r_{\rm m}\bar\m_{\rm m}\le[\le(1-\frac{c^2_{\rm s}}{c^2}\ri)\d_{\rm m}+{\cal H}\le(1+3\frac{c^2_{\rm s}}{c^2}\ri)\frac{\chi_{\rm m}}{a}\ri]\ri\}\;.
\ee
Equations for the remaining gravitational perturbations are found from (\ref{pn17})-(\ref{pn19}) which read
\ba
\la{pn21}
\Box{p}-2{\cal H}\bar{\mathtt v}^\m{p}_{,\m}&=&16\pi a^2\t\;,\\
\la{pn22}
\Box{p}_\a-2{\cal H}\bar{\mathtt v}^\m{p}_{\a;\m}+\le[\le(1+3w_{\rm m}\ri){\cal H}^2-\le(1+w_{\rm m}\ri)a^2\Lambda\ri]{p}_\a&=&16\pi a\t_\a\;,\\
\la{pn23}
\Box{p}^{\sst\intercal}_{\a\b}-2{\cal H}\bar{\mathtt v}^\m{p}^{\sst\intercal}_{\a\b;\m}&=&16\pi\t^{\sst\intercal}_{\a\b}\;.
\ea\esu
Equations given in this section are valid for arbitrary cosmological equation of state of the ideal fluid, $\bar p_{\rm m}=w_{\rm m}\bar\epsilon_{\rm m}$, that is physically reasonable. The parameter $w_{\rm m}$ of the equation of state should not be replaced with the ratio of $c^2_{\rm s}/c^2$ which characterizes the derivative of pressure $\bar p_{\rm m}$ with respect to the energy density $\bar\epsilon$. This is because the parameter $w_{\rm m}$ can depend in the most general case on the other thermodynamic quantities (like enthropy, etc.) which may implicitly depend on $\bar\epsilon$. Equations (\ref{pnra})--(\ref{pn23}) are decoupled in the sense that all of them can be solved one after another starting from solving equation (\ref{pnra}) for $\delta_{\rm m}$, which is a primary equation.

\subsubsection{Case 2: Dust equation of state}\la{deos}

Equations of the previous section can be further simplified for some particular equations of state of the ideal fluid. For example,
in the case when the ideal fluid is made of dust, the background pressure of matter drops to zero making parameter of the equation of state $w_{\rm m}=0$. Sound waves do not propagate in dust. Hence, the speed of sound $c_{\rm s}=0$. For this reason all terms being proportional to $c^2_{\rm s}$ and $w_{\rm m}$ vanish in equation (\ref{pnra}). Moreover, dust has the specific enthalpy, $\m_{\rm m}=1$ making the energy density of dust equal to its rest mass density $\bar\e_{\rm m}=\bar\r_{\rm m}$, and the normalized perturbation $\chi_{\rm m}$ of the Clebsch potential of dust is equal to the perturbation $\phi$ of the Clebsch potential itself, $\chi_{\rm m}=\phi$. The Friedmann equation (\ref{a12}) tells us that
\be\la{rtq4}
{\cal H}^2=\frac{a^2}{3}\le(8\pi\r_{\rm m}+\Lambda\ri)\;.
\ee
Accounting for this result in equation (\ref{pnra}), and neglecting all terms being proportional to the speed of sound, $c_{\rm s}$, we obtain
\be\la{ecr1}
\bar{\mathtt v}^\a\bar{\mathtt v}^\b \delta_{{\rm m};\a\b}+{\cal H}\bar{\mathtt v}^\a\delta_{{\rm m},\a}-4\pi a^2\bar\r_{\rm m}\d_{\rm m}=4\pi a^2\le(\s+\t\ri)\;,
\ee
where the terms depending on the cosmological constant, $\Lambda$, have cancelled out.
This equation is more familiar when is written down in the preferred FLRW frame, where $\bar{\mathtt v}^\a=(1,0,0,0)$.  Equation (\ref{ecr1}) assumes the ``canonical'' form
\be\la{pn14}
\ddot\d_{\rm m}+{\cal H}\dot\d_{\rm m}-4\pi a^2\bar\r_{\rm m}\d_{\rm m}=4\pi a^2\le(\s+\t\ri)\;,
%\ddot\d_{\rm m}+{\cal H}\dot\d_{\rm m}-\frac{c_{\rm s}^2}{c^2}\Delta\d_{\rm m}-4\pi a^2\bar\r_{\rm m}\d_{\rm m}=4\pi a^2\sigma\;,
\ee
which can be found in many textbooks on cosmology \citep{1972gcpa.book.....W,weinberg_2008,1980lssu.book.....P,2005pfc..book.....M,lnder}.

Equation (\ref{pn14}) has been derived by previous researchers without resorting to the concept of the Clebsch potential of the ideal fluid. For this reason, the density contrast, $\delta_{\rm m}$, was interpreted as the ratio of the perturbation of the dust density to its background value, $\delta=\d\r_{\rm m}/\bar\r_{\rm m}$, without taking into account the perturbation, $\phi$, of the Clebsch potential. However, the quantity $\delta$ is not gauge-invariant which was considered as a drawback. The scrutiny analysis of the underlying principles of hydrodynamics in the expanding universe given in the present paper, reveals that equation (\ref{pn14}) is, in fact, valid for the gauge-invariant density perturbation $\d_{\rm m}$ defined above in (\ref{rtq1}).   Another distinctive feature of equation (\ref{pn14}) is the presence of the source of a {\it bare} perturbation in its right side. The {\it bare} perturbation is caused by the effective density $\sigma+\tau$ of the matter which comprises the isolated astronomical system and initiates the growth of instability in the cosmological matter that, in its own turn, induces formation of the large scale structure of the universe \citep{1980lssu.book.....P,weinberg_2008}. Standard approach to cosmological perturbation theory always set $\sigma+\t=0$ and operates with the spectrum of the primordial perturbation of the density $\delta\rho_{\rm m}/\rho_{\rm m}$ (but not with the spectrum for $\delta_{\rm m}$).

Equation (\ref{rtq3}) in case of dust reads,
\be\la{pn15}
\Box\chi_{\rm m}+\frac12\le(3{\cal H}^2-a^2\Lambda\ri)\chi_{\rm m}=a\bar{\mathtt v}^\a\delta_{{\rm m},\a}\;,
\ee
where $\chi_{\rm m}$ is reduced to the perturbation of the Clebsch potential, $\chi_{\rm m}=\phi$, for the reason that has been mentioned above.

If equations (\ref{pn14}) and (\ref{pn15}) are solved, the gravitational perturbations can be found from equations (\ref{pnq2})--(\ref{pn23}), which take on the following form
\bsu\ba\la{pn20}
\Box{q}-2{\cal H}\bar{\mathtt v}^\a{q}_{,\a}+\le({\cal H}^2-a^2\Lambda\ri){q}
&=&8\pi a^2\le[\s+\t+\bar\r_{\rm m}\le(\d_{\rm m}+{\cal H}\frac{\chi_{\rm m}}{a}\ri)\ri]\;,\\
\la{pn211}
\Box{p}-2{\cal H}\bar{\mathtt v}^\a{p}_{,\a}&=&16\pi a^2\t\;,\\
\la{pn221}
\Box{p}_\a-2{\cal H}\bar{\mathtt v}^\b{p}_{\a;\b}+\le({\cal H}^2-a^2\Lambda\ri){p}_\a&=&16\pi a\t_\a\;,\\
\la{pn231}
\Box{p}^{\sst\intercal}_{\a\b}-2{\cal H}\bar{\mathtt v}^\g{p}^{\sst\intercal}_{\a\b;\g}&=&16\pi\t^{\sst\intercal}_{\a\b}\;.
\ea\esu
It is interesting to notice that besides the {\it bare} density perturbation, $\sigma+\tau$, the source for the scalar gravitational perturbation, $q$, contains the {\it induced} density perturbation $\bar\rho_{\rm m}\le(\delta_{\rm m}+{\cal H}\chi_{\rm m}/a\ri)=\delta\rho_{\rm m}+H\bar\rho_{\rm m}\phi$ in the right side of equation (\ref{pn20}). This {\it induced} density perturbation changes the initial mass of the isolated astronomical system in the course of the evolution (expansion) of the universe. This explains the origin of the time-dependence of the central point-like mass in the cosmological solution found by McVittie \citep{1933MNRAS..93..325M} (see also discussion in \citep{2010RvMP...82..169C}).

\subsection{The Universe Governed by a Scalar Field}

In this section we explore the case of the universe governed primarily by a scalar field with all other matter variables being unimportant. In this case, the time evolution of the background universe is defined exceptionally by equations (\ref{pn27}), (\ref{pn28}). The most general solution of (\ref{pn27}), (\ref{pn28}) is complicated and can not be achieved analytically. Numerical analysis shows that the solution evolves in the phase space of the two variables $\{\mathrm{x}_1,\mathrm{x}_2\}$ from an unstable to a stable fixed point by passing through a saddle point \citep{2010deto.book.....A}. The cosmic acceleration is realized by the stable point with the values of $\mathrm{x}_1=\lambda^2/6$ and $\mathrm{x}_2=1-\lambda^2/6$, which is equivalent to the equations of state (\ref{a10}) with the values of the parameters, $w_{\rm m}=0$, and, $w_{\rm q}=-1+\lambda^2/3$. It also requires the energy density of the background matter $\bar\e_{\rm m}=0$, that is $\Omega_{\rm m}=0$.
In such a universe the derivatives of the potential of the scalar field are
\be\la{pn26a}
\frac{1}{\bar\m_{\rm q}}\frac{\pd\bar W}{\pd\bar\Psi}=-\frac32\H\le(1-w_{\rm q}\ri)\qquad,\qquad
\frac{\pd^2\bar W}{\pd\bar\Psi^2}=\frac92\H^2\le(1-w_{\rm q}^2\ri)\;.
\ee
Moreover, because $\bar\r_{\rm m}\bar\m_{\rm m}=\bar\e_{\rm m}+\bar p_{\rm m}=0$, the time derivative of the Hubble parameter is
\be\la{pn28az}
\dot H=-4\pi\bar\r_{\rm q}\bar\m_{\rm q}=-\frac32\H^2\le(1+w_{\rm q}\ri)\;.
\ee

In the point of the attractor of the scalar field, perturbations of the ideal fluid are fully suppressed that is the Clebsch potential of the fluid, $\chi_{\rm m}=0$. It makes the function $V_{\rm m}=q/2$, that is reduced to the perturbation of the scalar component of the gravitational field only.  Perturbations of the scalar field are described by the scalar field variable, $\chi_{\rm q}$. In particular, after substituting the derivatives (\ref{pn26a}) of the scalar field potential along with the derivative (\ref{pn28az}) of the Hubble parameter, in equation (\ref{feq6}), one obtains the post-Newtonian equation for function $V_{\rm q}$,
\be\la{pn29}
\Box V_{\rm q}-
\le(1-3w_{\rm q}\ri){\cal H}\bar{\mathtt v}^\m V_{{\rm q},\m}+\frac32{\cal H}^2\le(1-w_{\rm q}\ri)\le(1+3w_{\rm q}\ri) V_{\rm q}\,
=-4\pi a^2\le(\s+\t\ri)\;.
\ee
Field equation for the perturbation of the scalar field, $\chi_{\rm q}$, is reduced to
\be\la{pn30w}
\Box \tilde\chi_{\rm q}-2{\cal H}\bar{\mathtt v}^\m\tilde\chi_{{\rm q},\m}+{\cal H}^2\le(1+3w_{\rm q}\ri)\tilde\chi_{\rm q}
=-\le(1+3w_{\rm q}\ri){\cal H} V_{\rm q}\;,
\ee
where the variable, $\tilde\chi_{\rm q}\equiv\chi_{\rm q}/a$, has been used for the notational convenience.

Post-Newtonian equations for gravitational perturbations are (\ref{pou1})--(\ref{pn19}). After substituting the values of the parameters $\mathrm{x}_1,\mathrm{x}_2, w_{\rm eff}$, etc., corresponding to the model of the universe governed by the scalar field alone, the post-Newtonian equations for the metric perturbations become
\bsu\ba\la{pn31}
\Box{q}-2{\cal H}\bar{\mathtt v}^\m{q}_{,\m}+\le(1+3w_{\rm q}\ri){\cal H}^2q
&=&8\pi a^2\le(\s+\t\ri)+3\le(1+w_{\rm q}\ri)\le(1+3w_{\rm q}\ri){\cal H}^3\hat\chi_{\rm q}\;,\\
\la{pn32}
\Box{p}-2{\cal H}\bar{\mathtt v}^\m{p}_{,\m}&=&16\pi a^2\t\;,\\
\la{pn33}
\Box{p}_\a-2{\cal H}\bar{\mathtt v}^\m{p}_{\a;\m}+\le(1+3w_{\rm q}\ri){\cal H}^2{p}_\a&=&16\pi a\t_\a\;,\\
\la{pn34}
\Box{p}^{\sst\intercal}_{\a\b}-2{\cal H}\bar{\mathtt v}^\m{p}^{\sst\intercal}_{\a\b;\m}&=&16\pi\t^{\sst\intercal}_{\a\b}\;.
\ea\esu
One can see that the field equations for the scalar field and metric perturbations are decoupled, and can be solved separately starting from the primary equation (\ref{pn29}).

\subsection{Post-Newtonian Potentials in the Linearized Hubble Approximation}
\subsubsection{The metric tensor perturbations}
The post-Newtonian equations for cosmological perturbations of gravitational and matter field variables crucially depend on the equation of state of the matter fields in the background universe. It determines the time evolution of the scale factor $a=a(\eta)$ and the Hubble parameter ${\cal H}={\cal H}(\eta)$ which are described by the wide range of elementary and special functions of mathematical physics (see, for example, the books by \citet{2003esef.book.....S,2010deto.book.....A,2001essf.book.....M} and references therein). It is not the goal of the present article to provide the reader with an exhaustive list of the mathematical solutions of the perturbed equations which requires a meticulous development of cosmological Green's function (see, for example, \citep{lif,1963AdPhy..12..185L,2005CQGra..22S.739H,2002PhLB..532....1R}).

In this section we shall focus on the observation that the post-Newtonian equations for the field perturbations have identical mathematical structure if all terms that are quadratic with respect to the Hubble parameter, ${\cal H}$, are neglected. In such a linearized Hubble approximation the differential equations for cosmological perturbations are not only decoupled from one another, but their generic solution can be found irrespectively of the equation of state governing the background universe. Indeed, if we neglect all quadratic with respect to ${\cal H}$ terms, the field equations for the conformal metric perturbations are reduced to the following set,
\bsu\ba\la{pnc1}
\Box{q}-2{\cal H}\bar{\mathtt v}^\a{q}_{,\a}&=&8\pi a^2\le(\s+\t\ri)\;,\\
\la{pnc2}
\Box{p}-2{\cal H}\bar{\mathtt v}^\a{p}_{,\a}&=&16\pi a^2\t\;,\\
\la{pnc3}
\Box{p}_\a-2{\cal H}\bar{\mathtt v}^\b{p}_{\a;\b}&=&16\pi a\t_\a\;,\\
\la{pnc4}
\Box{p}^{\sst\intercal}_{\a\b}-2{\cal H}\bar{\mathtt v}^\g{p}^{\sst\intercal}_{\a\b;\g}&=&16\pi\t^{\sst\intercal}_{\a\b}\;,
\ea\esu
where the wave operator $\Box$ has been defined in (\ref{wer9}), and the source of the perturbation is the tensor of energy-momentum of a localized astronomical system with the matter having a bounded support in space -- see section (\ref{emmla}).
The differential structure of the left side of equations (\ref{pnc1})--(\ref{pnc4}) is the same for all functions. The equations differ from each other only in terms of the order of ${\cal H}^2$ which have been omitted.

In order to bring equations (\ref{pnc1})-(\ref{pnc4}) to a solvable form, we resort to relationship (\ref{wer10}) which reveals that in the linearized Hubble approximation, the equations can be reduced to the form of a wave equation
\bsu\ba\la{pnc5}
\Box{\le(aq\ri)}&=&8\pi a^3\le(\s+\t\ri)\;,\\
\la{pnc6}
\Box{\le(ap\ri)}&=&16\pi a^3\t\;,\\
\la{pnc7}
\Box\le(ap_\a\ri)&=&16\pi a^2\t_\a\;,\\
\la{pnc8}
\Box\le(ap^{\sst\intercal}_{\a\b}\ri)&=&16\pi a\t^{\sst\intercal}_{\a\b}\;.
\ea\esu
So far, we did not impose any limitations on the curvature of space that can take three values: $k=\{-1,0,+1\}$. Solution of wave equations (\ref{pnc5})-(\ref{pnc8}) can be given in terms of special functions in case of the Riemann ($k=+1)$ or the Lobachevsky $(k=-1)$ geometry  \citep{lif,1963AdPhy..12..185L}. The case of the spatial Euclidean geometry ($k=0)$ is more manageable, and will be discussed below.

If the FLRW universe is spatially-flat universe, $k=0$, and we chose the Cartesian coordinates $x^\a$ related to the isotropic coordinates $X^\a$ of the FLRW universe by a Lorentz transformation, the operator $\Box$ becomes a wave operator in the Minkowski spacetime,
\be\la{vic4}
\Box=\eta^{\m\n}\pd_{\m\n}\;.
\ee
In this case, equations (\ref{pnc5})-(\ref{pnc8}) are reduced to the inhomogeneous wave equations which
solution depends essentially on the boundary conditions imposed on the metric tensor perturbations at conformal past-null infinity $\cal J^-$ of the cosmological manifold \citep{mtw}. We shall assume a no-incoming radiation condition also known as Fock-Sommerfeld's condition \citep{ZAMM19650450133,1987thyg.book..128D}
\be\la{pnc9}
\lim_{r\rightarrow +\infty\atop t+r={\rm const.}}n^\g\pd_\g\le[a(\eta)rl_{\a\b}(x^\g)\ri] =0\;,
\ee
where $x^\g=(x^0,x^i)$, $\eta=\eta(x^\g)$, the null vector $n^\a=\{1,x^i/r\}$, and $r=\delta_{ij}x^ix^j$ is the radial distance.
This condition ensures that there is no infalling gravitational radiation arriving to the localized astronomical system from the future null infinity $\cal J^+$. Effectively, it singles out the retarded solution of the wave equation.

A particular solution of the wave equations satisfying condition (\ref{pnc9}), is the retarded integral \citep{LL}
\bsu\ba\la{pnc10}
a q(t,{\bm x})&=&-2\int_{\cal V}\frac{a^3\le[\eta\le(s,{\bm x}'\ri)\ri]\le[\s\le(s,{\bm x}'\ri)+\t\le(s,{\bm x}'\ri)\ri]d^3x'}{|{\bm x}-{\bm x}'|}\;,\\
\la{pnc11}
a p(t,{\bm x})&=&-4\int_{\cal V}\frac{a^3\le[\eta(s,{\bm x}')\ri]\t\le(s,{\bm x}'\ri)d^3x'}{|{\bm x}-{\bm x}'|}\;,\\
\la{pnc12}
a p_\a(t,{\bm x})&=&-4\int_{\cal V}\frac{a^2\le[\eta(s,{\bm x}')\ri]\t_\a\le(s,{\bm x}'\ri)d^3x'}{|{\bm x}-{\bm x}'|}\;,\\
\la{pnc13}
a p^{\sst\intercal}_{\a\b}(t,{\bm x})&=&-4\int_{\cal V}\frac{a\le[\eta(s,{\bm x}')\ri]\t^{\sst\intercal}_{\a\b}\le(s,{\bm x}'\ri)d^3x'}{|{\bm x}-{\bm x}'|}\;,
\ea\esu
where the scale factor $a$ in the left side of all equations is $a\equiv a\le[\eta(s,{\bm x})\ri]$, and the argument $s$ of the functions appearing in the integrands, is the retarded time
\be\la{pnc14}
s=t-|{\bm x}-{\bm x}'|\;.
\ee
The retarded time $s$ is a characteristic of the null cone in the conformal Minkowski spacetime that determines the causal nature of the gravitational field of the localized astronomical system in the expanding universe with $k=0$ \citep{2011rcms.book.....K}. Solutions (\ref{pnc10})--(\ref{pnc13}) are Lorentz-invariant as shown by calculations in Appendix \ref{lipr}.

Integration in (\ref{pnc10})--(\ref{pnc13}) is performed over the volume, $\cal V$, occupied by the matter of the localized astronomical system. In case of the system comprised of $N$ massive bodies that are separated by distances being much larger than their characteristic size, the matter occupies the volumes of the bodies. In this case the integration in equations (\ref{pnc10})--(\ref{pnc13}) is practically performed over the volumes of the bodies. It means that each post-Newtonian potential $q,p,p_\a,p^{\sst\intercal}_{\a\b}$ is split in the algebraic sum of $N$ pieces
\be\la{er1}
q=\sum_{{\sst A}=1}^N q_{\sst A}\;,\qquad  p=\sum_{{\sst A}=1}^N p_{\sst A}\;,\qquad p_\a=\sum_{{\sst A}=1}^N p_{{\sst A}\a}\;,\qquad p^{\sst\intercal}_{\a\b}=\sum_{{\sst A}=1}^N p^{\sst\intercal}_{{\sst A}\a\b}\;,
\ee
where each function with sub-index ${\st A}$ has the same form as one of the corresponding equations (\ref{pnc10})--(\ref{pnc13}) with the integration performed over the volume, ${\cal V}_{\sst A}$, of the body ${\st A}$.

\subsubsection{The gauge functions}

The residual gauge freedom describe the arbitrariness in adding solution of homogeneous equations (\ref{pnc10})--(\ref{pnc13}) with the right side being equal to zero. It is described by two functions, $\zeta\equiv\xi/a$ and $\zeta^\a$. Since we neglected the terms being quadratic with respect to the Hubble parameter, the gauge functions satisfy the following equations
\bsu\ba\la{pnc15}
\Box\zeta-2{\cal H}\bar{\mathtt v}^\b\zeta_{,\b}&=&0\;,\\
\la{jk4aa}
\Box\zeta^\a-2{\cal H}\bar{\mathtt v}^\b\zeta^\a{}_{;\b}&=&0\;.
\ea\esu
They are equivalent to the homogeneous wave equations in the conformal flat spacetime
\be\la{pnc16}
\Box\le(a\zeta\ri)=0\qquad,\qquad \Box\le(a\zeta^\a\ri)=0\;,
\ee
which point out that (in the approximation under consideration) the products, $a\zeta$ and $a\zeta^\a$, are the harmonic functions.

Potentials $q,p,p_\a,p^{\sst\intercal}_{\a\b}$ must satisfy the gauge conditions (\ref{gh1}), (\ref{gh2}).
Neglecting terms being quadratic with respect to the Hubble parameter, the gauge conditions (\ref{gh1}), (\ref{gh2}) can be written down as follows
\bsu\ba\la{qx1}
\le(ap^\a\ri)_{,\a}+\bar{\mathtt v}^\a\le(2aq-ap\ri)_{,\a}+{\cal H}ap&=&0\;,\\
\la{qx2}
\le(ap^{{\sst\intercal}\a\b}\ri){}_{,\b}+\bar{\mathtt{v}}^\b\le(ap^\a\ri){}_{,\b}+\frac13\bar \pi^{\a\b}(ap)_{,\b}+{\cal H}ap^\a&=&0\;,
\ea\esu
where the potentials $p^\a$ and $p^{{\sst\intercal}\a\b}$ are obtained from $p_\a$ and $p^{\sst\intercal}_{\a\b}$ by rising the indices with the Minkowski metric and taking into account that the indices of $\t_\a$ and $\t^{\sst\intercal}_{\a\b}$ in the integrands of (\ref{pnc12}) and (\ref{pnc13}) should be raised with the full background metric $\bar g^{\a\b}=a^{-2}\eta^{\a\b}$ taken at the point of integration. It yields
\bsu\ba\la{qx3}
ap^\a(t,{\bm x})&=&-4\int_{\cal V}\frac{a^4\le[\eta(s,{\bm x}')\ri]\t^\a\le(s,{\bm x}'\ri)d^3x'}{|{\bm x}-{\bm x}'|}\;,\\
\la{qx4}
ap^{{\sst\intercal}\a\b}(t,{\bm x})&=&-4\int_{\cal V}\frac{a^5\le[\eta(s,{\bm x}')\ri]\t^{{\sst\intercal}\a\b}\le(s,{\bm x}'\ri)d^3x'}{|{\bm x}-{\bm x}'|}\;.
\ea\esu
It is instrumental to write down solutions for the products of the potentials $p$ and $p^\a=\eta^{\a\b}p_\b$ with the Hubble parameter. Multiplying both sides of equations (\ref{pnc6}), (\ref{pnc7}) with the Hubble parameter ${\cal H}$, and neglecting the quadratic with respect to ${\cal H}$ terms, we obtain
\be
\la{qx22}
\Box{\le({\cal H}ap\ri)}=16\pi a^3{\cal H}\t\qquad,\qquad
\Box{\le({\cal H}ap^\a\ri)}=16\pi a^4{\cal H}\t^\a\;,
\ee
which solutions are the retarded potential
\bsu\ba\la{qx23}
{\cal H}ap(t,{\bm x})&=&-4\int_{\cal V}\frac{a^3\le[\eta(s,{\bm x}')\ri]{\cal H}\le[\eta(s,{\bm x}')\ri]\t\le(s,{\bm x}'\ri)d^3x'}{|{\bm x}-{\bm x}'|}\;,\\
\la{qx24}
{\cal H}ap^\a(t,{\bm x})&=&-4\int_{\cal V}\frac{a^4\le[\eta(s,{\bm x}')\ri]{\cal H}\le[\eta(s,{\bm x}')\ri]\t\le(s,{\bm x}'\ri)d^3x'}{|{\bm x}-{\bm x}'|}\;.
\ea\esu

Substituting functions $q,p,p^\a,p^{{\sst\intercal}\a\b}$ and ${\cal H}ap$, ${\cal H}ap^\a$ to the gauge equations (\ref{qx1}), (\ref{qx2}), bring about the following integral equations
\bsu\ba\la{qx5}
\int_{\cal V}\le[\le(a^4\t^\a+\bar{\mathtt v}^\a a^3\s\ri)_{,\a}+a^3{\cal H}\t\ri]\frac{d^3x'}{|{\bm x}-{\bm x}'|}&=&0\;,\\
\la{qx6}
\int_{\cal V}\le[\le(a^5\t^{{\sst\intercal}\a\b}+a^4\bar{\mathtt v}^\b\t^\a+\frac13\bar\pi^{\a\b}a^3\t \ri)_{,\b}+a^4{\cal H}\t^\a\ri]\frac{d^3x'}{|{\bm x}-{\bm x}'|}&=&0\;,
\ea\esu
where all functions in the integrands are taken at the retarded time $s$ and at the point ${\bm x}'$, for example, $a=a[\eta(s,{\bm x}')]$, ${\cal H}={\cal H}[\eta(s,{\bm x}')]$, $\sigma=\s[(s,{\bm x}')]$, and so on. These equations are satisfied by the equations of motion (\ref{qq2}), (\ref{qq3}) of the localized matter distribution. Indeed, divergences of any vector $F^\a$ and a symmetric tensor $F^{\a\b}$ obey the following equalities
\ba\la{qx7}
F^\a{}_{|\a}&=&\frac1{\sqrt{-\bar g}}\le(\sqrt{-\bar g}F^\a\ri)_{,\a}\;,\\
\la{qx8}
F^{\a\b}{}_{|\b}&=&\frac1{\sqrt{-\bar g}}\le(\sqrt{-\bar g}F^{\a\b}\ri)_{,\b}+\bar\Gamma^\a_{\b\g}F^{\b\g}\;.
\ea
Moreover, the root square of the determinant of the background metric tensor is expressed in terms of the scale factor, $\sqrt{-\bar g}=a^4$, while the four-velocity $\bar u^\a=\bar{\mathtt v}^\a/a$.
Applying these expressions along with equations (\ref{qx7}), (\ref{qx8}) in equations of motion (\ref{qq2}), (\ref{qq3}), transforms them to
\bsu\ba\la{qx9}
\le(a^4\t^\a+\bar{\mathtt v}^\a a^3\s\ri)_{,\a}+a^3{\cal H}\t&=&0\;,
\\
\la{qx10}
\le(a^4\t^{\a\b}+a^3\bar{\mathtt v}^\b\t^\a \ri)_{,\b}+2a^3{\cal H}\t^\a&=&0\;.
\ea\esu
Equation (\ref{qx9}) proves that the integral equation (\ref{qx5}) and, hence, the gauge condition (\ref{qx1}) are valid. In order to prove the second integral equation (\ref{qx6}), we multiply equation (\ref{qx10}) with the scale factor $a$, and reshuffle its terms. It brings (\ref{qx10}) to the following form
\be\la{qx11}
\le(a^5\t^{\a\b}+a^4\bar{\mathtt v}^\b\t^\a \ri)_{,\b}+a^4{\cal H}\t^\a=0\;.
\ee
Substituting, $\t^{\a\b}=\t^{{\sst\intercal}\a\b}+(1/3a^2)\bar\pi^{\a\b}\t$, to (\ref{qx11}) and comparing with the integrand of (\ref{qx7}) makes it clear that (\ref{qx6}) is valid. It proves the second gauge condition (\ref{qx2}). We conclude that the retarded integrals (\ref{pnc10})--(\ref{pnc13}) yield the complete solution of the linearised wave equations (\ref{pnc5})--(\ref{pnc8}). Thus, we can chose the gauge functions $\zeta=\zeta^\a=0.$

\subsubsection{The matter field perturbations}

What remains is to find out solutions for the scalar functions $V_{\rm m}$ and $V_{\rm q}$ and $\chi_{\rm m}$ and $\chi_{\rm q}$. In the linearized Hubble approximation equation for $V_{\rm m}$ is obtained from (\ref{pn10}) by discarding all terms of the order of ${\cal H}^2$. It yields
\be\la{qx12}
\Box V_{\rm m}+\le(1-\frac{c^2}{c_{\rm s}^2}\ri)\bar{\mathtt v}^\a\bar{\mathtt v}^\b V_{{\rm m},\a\b}+
\le(3-\frac{c^2}{c_{\rm s}^2}\ri){\cal H}\bar{\mathtt v}^\a V_{{\rm m},\a}=-4\pi a^2\le(\sigma+\t\ri)\;.
\ee
Applying relationships (\ref{wer10}), (\ref{wer11}) in equation (\ref{qx12}) allows us to recast it to
\be\la{qx13}
\frac1{a^n}\le[\Box \le(a^nV_{\rm m}\ri)+\le(1-\frac{c^2}{c_{\rm s}^2}\ri)\bar{\mathtt v}^\a\bar{\mathtt v}^\b \le(a^nV_{\rm m}\ri)_{,\a\b}\ri]+
\le[3+(2n-1)\frac{c^2}{c_{\rm s}^2}\ri]{\cal H}\bar{\mathtt v}^\a V_{{\rm m},\a}=-4\pi a^2\le(\sigma+\t\ri)\;,
\ee
where $n$ is yet undetermined real number.
Choosing, $n=n_{\rm s}$, with
\be\la{qx14}
n_{\rm s}=\frac12\le(1-3\frac{c^2_{\rm s}}{c^2}\ri)\;,
\ee
annihilates the term being proportional to ${\cal H}$ in the left side of (\ref{qx13}) and reduces it to
\be\la{qx15}
\Box \le(a^{n_{\rm s}}V_{\rm m}\ri)+\le(1-\frac{c^2}{c_{\rm s}^2}\ri)\bar{\mathtt v}^\a\bar{\mathtt v}^\b \le(a^{n_{\rm s}}V_{\rm m}\ri)_{,\a\b}
=-4\pi a^{2+n_{\rm s}}\le(\sigma+\t\ri)\;.
\ee
This equation describes propagation of perturbation $V_{\rm m}$ with the speed of sound $c_{\rm s}$. Indeed, let us introduce the sound-wave Laplace-Beltrami operator
\be\la{qx16}
\Box_{\rm s}\equiv\le(-\frac{c^2}{c^2_{\rm s}}\bar{\mathtt v}^\a\bar{\mathtt v}^\b+\bar\pi^{\a\b}\ri)\pd_{\a\b}\;.
\ee
Then, equation (\ref{qx15}) reads
\be\la{qx17}
\Box_{\rm s}\le(a^{n_{\rm s}}V_{\rm m}\ri)=-4\pi a^{2+n_{\rm s}}\le(\sigma+\t\ri)\;.
\ee
This equation has a well-defined Green function with characteristics propagating with the speed of sound $c_{\rm s}$. We discard the advanced Green function because we assume that at infinity the function $V_{\rm m}$ and its first derivatives vanish. Solution of (\ref{qx17}) is explained in Appendix \ref{lissw}, and has the following form
\be\la{qx18}
a^{n_{\rm s}}V_{\rm m}(t,{\bm x})=\int_{\cal V}\frac{a^{2+n_{\rm s}}\le(\varsigma,{\bm x}'\ri)\le[\s(\varsigma,{\bm x}')+\t(\varsigma,{\bm x}')\ri]}{\sqrt{1+\le(1-\frac{c^2}{c_{\rm s}^2}\ri)\g^2_\up({\bm\b}\times{\bm n})^2}}\frac{d^3x'}{|{\bm x}-{\bm x}'|}\;,
\ee
where the retarded time $\varsigma$ is given by equation (\ref{bp15}), ${\bm\b}={\b^i}=\bar{\mathtt v}^i/c$, $\g=1/\sqrt{1-{\bm\b}^2}$ is the Lorentz factor, and the unit vector ${\bm n}=({\bm x}-{\bm x}')/|{\bm x}-{\bm x}'|$.

Linearized equation for $V_{\rm q}$ is obtained from (\ref{vq1}) after discarding all terms being proportional to ${\cal H}^2$. It yields
\be\la{qx21}
\Box V_{\rm q}+2\le(1-\sqrt{\frac{3}{2\mathrm{x}_1}}\lambda \mathrm{x}_2\ri){\cal H}\bar{\mathtt v}^\m V_{{\rm q},\m}=-4\pi a^2\le(\sigma+\t\ri)\;.
\ee
Applying relationship (\ref{wer10}) in (\ref{qx21}) allows us to recast it to
\be\la{qx22z}
\frac1{a^n}\Box \le(a^nV_{\rm q}\ri)+
2\le(n+1-\sqrt{\frac{3}{2\mathrm{x}_1}}\lambda \mathrm{x}_2\ri){\cal H}\bar{\mathtt v}^\a V_{{\rm m},\a}=-4\pi a^2\le(\sigma+\t\ri)\;.
\ee
Choosing, $n=n_{\rm q}=-1+\sqrt{3/(2\mathrm{x}_1)}\lambda \mathrm{x}_2$,
eradicates the second term in the left side of (\ref{qx22z}) that results in
\be\la{qx24z}
\Box \le(a^{n_{\rm q}}V_{\rm q}\ri)
=-4\pi a^{2+n_{\rm q}}\le(\sigma+\t\ri)\;.
\ee
This is the wave equation in flat spacetime. We pick up the retarded solution as the most physical one. It reads
\be\la{qx25}
a^{n_{\rm q}}V_{\rm q}=\int_{\cal V}\frac{a^{2+n_{\rm q}}\le(s,{\bm x}'\ri)\le[\s\le(s,{\bm x}'\ri)+\t\le(s,{\bm x}'\ri)\ri]d^3x'}{|{\bm x}-{\bm x}'|}\;,
\ee
where the retarded time $s$ has been defined in (\ref{pnc14}).

Perturbations $\chi_{\rm m}$ and $\chi_{\rm q}$ can be found by integrating equations (\ref{sp7}) and (\ref{sp8}) that can be written as
\be\la{sp7m}
\bar{\mathtt v}^\a \chi_{{\rm m},\a}=a\le(V_{\rm m}+\frac{q}2\ri)\qquad,\qquad
\bar{\mathtt v}^\a \chi_{{\rm q},\a}=a\le(V_{\rm q}+\frac{q}2\ri)\;.
\ee
These are the ordinary differential equations of the first order. Their solutions are
\bsu\ba\la{qx26}
\chi_{\rm m}&=&\int_{t_0}^t a[t,{\bm x}(t)]\{V_{\rm m}[t,{\bm x}(t)]+\frac12 q[t,{\bm x}(t)]\}dt\;,\\
\la{qx27}
\chi_{\rm q}&=&\int_{t_0}^t a[t,{\bm x}(t)]\{V_{\rm q}[t,{\bm x}(t)]+\frac12 q[t,{\bm x}(t)]\}dt\;,\;.
\ea\esu
where $t_0$ is an initial epoch of integration, and the integration is performed along the characteristics of the unperturbed equations of motion of matter of the background universe
\be\la{qx28}
\frac{dx^i}{dt}=\bar{\mathtt v}^i(t,{\bm x})\;.
\ee
These characteristics make up the Hubble flow. Therefore, the most simple way to integrate equations (\ref{sp7m}) would be to work in the preferred coordinate frame $X^\a=(\eta,X^i)$ where the velocity $\bar{\mathtt v}^i=0$, and the coordinates $X^i={\rm const}$. After the calculations in the rest frame of the Hubble flow are finished, the transformation to a moving frame can be done with the Lorentz boost.

\acknowledgments
{This paper was made possible through the support of a grant from the John Templeton Foundation. The opinions expressed in this publication are those of the authors and do not necessarily reflect the views of the John Templeton Foundation. The funds from John Templeton Foundation were awarded in a grant to the University of Chicago which also managed the program in conjunction with National Astronomical Observatories, Chinese Academy of Sciences.

The work of A. Petrov was supported by the internal grant of the University of Missouri Research Board. S. Kopeikin was partially supported by the faculty incentive grant of the College of Arts and Sciences Alumni Organization of the University of Missouri as well as by the Chinese Academy of Sciences Visiting Professorship for International Senior Scientists. 

S. Kopeikin thanks Dr. Tang Zhenghong, Dr. Tao Jinhe, and the stuff of the Shanghai Astronomical Observatory of the People Republic of China for hospitality, and Dr. Xie Yi for valuable discussions. A. Petrov is grateful to the Department of Physics \& Astronomy of the University of Missouri for hospitality during his visiting Columbia, Missouri in 2007-08. 

The authors thank the anonymous referees for a number of fruitful comments and constructive suggestions that helped us to improve the manuscript. 
}

\bibliographystyle{plainnat}
\bibliography{NF-cosmology}

\appendix
\section{Lorentz Invariance of the Retarded Potential}\la{lipr}

We use a prime in the appendices exclusively as a notation for time and spatial coordinates which are used as variables of integration in volume integrals (see, for example, equations (\ref{app2}), (\ref{app3}), and so on). It should not be confused with the time derivative with respect to the conformal time used in the main text of the present paper.

Let us consider an inhomogeneous wave equation for a scalar field, $V=V(\eta,{\bm X})$, written down in a coordinate chart $X^\a=(X^0,X^i)=(\eta,{\bm X})$,
\be\la{app1}
\Box V=-4\pi \s_{\sst X} \;,
\ee
where $\Box\equiv\eta^{\a\b}\pd_{\a\b}$, $\pd_\a=\pd/\pd X^\a$, and $\s_{\sst X}=\s_{\sst X}(\eta,{\bm X})$ is the source (a scalar function) of the field $V$ with a compact support (bounded by a finite volume in space). Equation (\ref{app1}) has a solution given as a linear combination of advanced and retarded potentials. Let us focus only on the retarded potential which is more common in physical applications. Advanced potential can be treated similarly.

We assume the field, $V$, and its first derivatives vanish at past null infinity. Then, the retarded solution (retarded potential) of (\ref{app1}) is given by an integral,
\be\la{app2}
V(\eta,{\bm X})=\int_{\cal V}\frac{\s_{\sst X}(\zeta,{\bm X}')\;d^3X'}{|{\bm X}-{\bm X}'|}\;,
\ee
where
\be\la{app3}
\zeta=\eta-|{\bm X}-{\bm X}'|\;,
\ee
is the retarded time, and we assume the fundamental speed $c=1$. Physical meaning of the retardation is that the field $V$ propagates in spacetime with the fundamental speed $c$ from the source $\s_{\sst X}$, to the point with coordinates $X^\a=(\eta,{\bm X})$ where the field $V$ is measured in correspondence with equation (\ref{app2}).
Left side of equation (\ref{app1}) is Lorentz-invariant. Hence, we expect that solution (\ref{app3}) must be Lorentz-invariant as well. As a rule, textbooks prove this statement for a particular case of the retarded (Li\'enard-Wiechert) potential of a moving point-like source but not for the retarded potential given in the form of the integral (\ref{app2}). This appendix fulfils this gap.

Lorentz transformation to coordinates, $x^\a=(t,{\bm x})$ linearly transforms the isotropic coordinates $X^\a=(\eta,{\bm X})$ of the FLRW universe as follows
\be\la{app4}
x^\a=\Lambda^\a{}_\b X^\b\;,
\ee
where the matrix of the Lorentz boost \citep{mtw}
\be\la{app5}
\Lambda^0{}_0={\gamma}\;,\qquad \Lambda^i{}_0=\Lambda^0{}_i=-{\gamma}{\b}^i\;,\qquad\Lambda^i{}_j=\d^{ij}+\frac{\g-1}{{\b}^2}{\b}^i{\b}^j\;,
\ee
the boost four-velocity $u^\a=\{u^0, u^i\}=u^0\{1,\b^i\}$ is constant, and
\be\la{app6}
{\gamma}=u^0=\frac1{\sqrt{1-{\b}^2}}\;,
\ee
is the constant Lorentz-factor.

The inverse Lorentz transformation is given explicitly as follows
\ba\la{app7}
\eta&=&{\gamma}(t+{\bm \b}\cdot{\bm x})\;,\\\la{qx20b}
{\bm X}&=&{\bm r}+\frac{\g^2}{1+{\gamma}}({\bm \b}\cdot{\bm r}){\bm b}\;,
\ea
where
\be\la{app8}
{\bm r}={\bm x}+{\bm \b}t\;,
\ee
and the boost three-velocity, ${\bm \b}=\{\b^i\}=\{u^i/u^0\}$.

Let us reiterate (\ref{app2}) by introducing a one-dimensional Dirac's delta function and integration with respect to time $\eta$,
\be\la{app11}
V(\eta,{\bm X})=\int_{-\infty}^{\infty}\int_{\cal V}\frac{\s_{\sst X}(\eta',{\bm X}')\delta(\eta'-\zeta)\;d\eta'd^3X'}{|{\bm X}-{\bm X}'|}\;,
\ee
where $\zeta$ is the retarded time given by (\ref{app3}).
Then, we transform coordinates $X'^\a=(\eta',{\bm X}')$ to $x'^\a=(t',{\bm x}')$ with the Lorentz boost (\ref{app4}). The
Lorentz transformation changes functions entering the integrand of (\ref{app11}) as follows,
\ba\la{app12}
\s_{\sst}(\eta',{\bm X}')&=&\s_x(t',{\bm x}')\;,\\
\la{app9}
|{\bm X}-{\bm X}'|&=&\sqrt{|{\bm r}-{\bm r}'|^2+\g^2[{\bm \b}\cdot({\bm r}-{\bm r}')]^2}\;,
\ea
where the coordinate difference
\be\la{app10}
{\bm r}-{\bm r}'={\bm x}-{\bm x}'+{\bm \b}(t-t')\;.
\ee
The coordinate volume of integration remains Lorentz-invariant
\be\la{app13}
d\eta'd^3X'=dt'd^3x'\;.
\ee
Let us denote $F_\eta(\eta')\equiv \eta'-\zeta$ where $\zeta$ is given by (\ref{app3}). After making the Lorentz transformation this function changes to
\be\la{app14}
F_\eta(\eta')=F_t(t')={\gamma}\le[t'-t-{\bm \b}\cdot\le({\bm x}-{\bm x}'\ri)\ri]+\sqrt{|{\bm x}-{\bm x}'|^2-(t'-t)^2+\g^2[{\bm \b}\cdot({\bm x}-{\bm x}')-(t'-t)]^2}\;,
\ee
where we have used equations (\ref{app7}), (\ref{qx20b}) and (\ref{app9}) and relationship $\g^2\b^2=\g^2-1$, to perform the transformation.
Integral (\ref{app11}) in coordinates $x^\a$ becomes
\be\la{newin5}
V(t,{\bm x})=\int_{-\infty}^{\infty}\int_{\cal V}\frac{\s_x(t',{\bm x}')\delta(F_t(t'))\;dt'd^3x'}{\sqrt{|{\bm r}-{\bm r}'|^2+\g^2[{\bm \b}\cdot({\bm r}-{\bm r}')]^2}}\;,
\ee

The delta function has a complicated argument $F_t(t')$ in coordinates $x^\a$. It can be simplified with a well-known formula
\be\la{pp14a}
\delta\le[F_t(t')\ri]=\frac{\delta(t'-s)}{\dot F_t(s)}\;,
\ee
where $\dot F_t(s)\equiv \le[dF_t(t')/dt'\ri]_{t'=s}$, and $s$ is one of the roots of equation $F_t(t')=0$ that is associated with the retarded interaction. It is straightforward to confirm by inspection that the root is given by formula,
\be\la{app15}
s=t-|{\bm x}-{\bm x}'|\;.
\ee

The time derivative of function $F_t(t')$ is
\be\la{app16}
\dot F_t(t')={\gamma}+\g^2\frac{{\b}^2(t'-t)-{\bm \b}\cdot({\bm x}-{\bm x}')}{\sqrt{|{\bm x}-{\bm x}'|^2-(t'-t)^2+\g^2[{\bm \b}\cdot({\bm x}-{\bm x}')-(t'-t)]^2}}\;.
\ee
After substituting $t'=s$, with $s$ taken from equation (\ref{app15}), we obtain,
\be\la{app17}
\dot F_t(s)=\frac1{\gamma}\frac{|{\bm x}-{\bm x}'|}{|{\bm x}-{\bm x}'|+{\bm \b}\cdot({\bm x}-{\bm x}')}\;.
\ee
Performing now integration with respect to $t'$ in equation (\ref{newin5}) with the help of the delta-function, we arrive to
\be\la{app18}
V(t,{\bm x})=\int_{\cal V}\frac{\s_x(s,{\bm x}')\;d^3x'}{\dot F_t(s)|{\bm X}-{\bm X}'|_{t'=s}}\;,
\ee
where $|{\bm X}-{\bm X}'|_{t'=s}$ must be calculated from (\ref{app9}) with $t'=s$ where $s$ is taken from (\ref{app15}). It yields
\be\la{app19}
\dot F_t(s)|{\bm X}-{\bm X}'|_{t'=s}=|{\bm x}-{\bm x}'|\;,
\ee
and proves that the retarded potential (\ref{app2}) is Lorentz-invariant
\be\la{app20}
\int_{\cal V}\frac{\s_{\sst X}(\zeta,{\bm X}')\;d^3X'}{|{\bm X}-{\bm X}'|}=\int_{\cal V}\frac{\s_x(s,{\bm x}')\;d^3x'}{|{\bm x}-{\bm x}'|}\;.
\ee
We have verified the Lorentz invariance for the {\it scalar} retarded potential. However, it is not difficult to check that it is valid in case of a source $\sigma_{\a_1\a_2...\a_l}$ that is a tensor field of rank $l$. Indeed, the Lorentz transformation of the source leads to $\Lambda^{\b_1}{}_{\a_1}\Lambda^{\b_2}{}_{\a_2}...\Lambda^{\b_l}{}_{\a_l}\sigma_{\b_1\b_2...\b_l}$ but the matrix $\Lambda^\a{}_\b$ is constant, and can be taken out of the sign of the retarded integral. Because of this property, all mathematical operations given in this appendix for a scalar retarded potential, remain the same for the tensor of any rank. Hence, the Lorentz invariance of the retarded integral is a general property of the wave operator in Minkowski spacetime.

\section{Retarded Solution of the Sound-Wave Equation}\la{lissw}

Let us consider an inhomogeneous sound-wave equation for a scalar field $U=U(\eta,{\bm X})$ written down in the isotropic coordinates $X^\a=(\eta,{\bm X})$,
\be\la{bp1}
\Box_{\rm s}U=-4\pi \t_{\sst X} \;,
\ee
where $\t_{\sst X}=\t_{\sst X}(\eta,{\bm X})$ is the source of $U$ having a compact support, and the sound-wave differential operator $\Box_{\rm s}$ was defined in (\ref{qx16}). It is Lorentz-invariant and reads
\be\la{bp2}
\Box_{\rm s}=\Box+\le(1-\frac{c^2}{c^2_s}\ri)\bar{\mathtt v}^\a\bar{\mathtt v}^\b\pd_{\a\b}\;,
\ee
where $\bar{\mathtt v}^\a$ is four-velocity of motion of the medium with respect to the coordinate chart, $c_{\rm s}$ is the constant speed of sound in the medium, and we keep the fundamental speed $c$ in the definition of the operator for dimensional purposes. We assume that $c_{\rm s}<c$. The case of $c_{\rm s}=c$ is treated in section \ref{lipr}, and the case of $c_{\rm s}\ge c$ makes a formal mathematical sense in discussion of the speed of propagation of gravity in alternative theories of gravity since the equation describing propagation of gravitational potential $U$ has the same structure as (\ref{bp1}) after formal replacement of $c_{\rm s}$ with the speed of gravity $c_{\rm g}$ \citep{2004CQGra..21.3251K,2006LRR.....9....3W}. In particular, in the Newtonian theory the speed of gravity $c_{\rm g}=\infty$, and the operator (\ref{bp2}) is reduced to the Laplace operator
\be\la{bp3}
\Delta=\Box+{\mathtt v}^\a{\mathtt v}^\b\pd_\a\pd_\b=\bar\pi^{\a\b}\pd_{\a\b}\;,
\ee
where the constant projection operator, $\bar\pi^{\a\b}$, has been defined in (\ref{opp3}).

We are looking for the solution of (\ref{bp1}) in the Cartesian coordinates $x^\a=(t,{\bm x})$ moving with respect to the isotropic coordinates $X^\a$ with constant velocity ${\b}^i$. Transformation from $X^\a$ to $x^\a$ is given by the Lorentz transformation (\ref{app4}).
In coordinates $X^\a$ the four-velocity $\bar{\mathtt v}^\a=(1,0,0,0)$. Therefore, in these coordinates, equation (\ref{bp1}) is just a wave equation for the field $U$ propagating with speed $c_{\rm s}$. It has a well-known retarded solution,
\be\la{bp4}
U(\eta,{\bm X})=\int_{\cal V}\frac{\t_{\sst X}(\eta_s,{\bm X}')\;d^3X'}{|{\bm X}-{\bm X}'|}\;,
\ee
where
\be\la{bp5}
\eta_s=\eta-\frac{c}{c_{\rm s}}|{\bm X}-{\bm X}'|\;,
\ee
is the retarded time.

Equation (\ref{bp1}) is Lorentz-invariant. Hence, its solution must be Lorentz-invariant as well. Our goal is to prove this statement. To this end, we take solution (\ref{bp4}) and perform the Lorentz transformation (\ref{app7}), (\ref{qx20b}). We recast the retarded integral (\ref{bp4}) to another form with the help of one-dimensional delta-function
\be\la{bp6}
U(\eta,{\bm X})=\int_\infty^\infty\int_{\cal V}\frac{\t_{\sst X}(\eta',{\bm X}')\delta(\eta'-\eta_s)\;d\eta'd^3X'}{|{\bm X}-{\bm X}'|}\;.
\ee
It looks similar to (\ref{app2}) but one has to remember that the retarded time $\eta_s$ differs from $\zeta$ that was defined in (\ref{app3}) on the characteristics of the null cone defined by the fundamental speed $c$. Transformation of functions entering integrand in (\ref{bp6}) is similar to what we did in section \ref{lipr} but, because $c_{\rm s}\not= c$, calculations become more involved. It turns out more preferable to handle the calculations in tensor notations, making transition to the coordinate language only at the end of the transformation procedure.

Let us consider two events with the isotropic coordinates $X^\a=(\eta,{\bm X})$ and $X'^\a=(\eta',{\bm X}')$. We postulate that in the coordinate chart, $x^\a$, these two events have coordinates, $x^\a=(t,{\bm x})$, and, $x'^\a=(t',{\bm x}')$, respectively.
We define the components of a four-vector, $r^\a=(t'-t,{\bm x}-{\bm x}')$ which is convenient for doing mathematical manipulations with the Lorentz transformations. For instance, the Lorentz transformation of the Euclidean distance between the spatial coordinates of the two events, is given by
a\be\la{bp7}
|{\bm X}-{\bm X}'|=\sqrt{\bar\pi_{\a\b}r^\a r^\b}\;,
\ee
where $\bar\pi^{\a\b}$ is the operator of projection on the hyperplane being orthogonal to $\bar{\mathtt v}^\a$ (the same operator as in (\ref{bp3})). Equation (\ref{bp7}) is a Lorentz-invariant analogue of expression (\ref{app9}) and matches it exactly. Transformation of the source, $\t_{\sst X}(X^\a)=\t_x(x^\a)$ is fully equivalent to that of $\s_{\sst X}$ as given by equation (\ref{app12}). Coordinate volume of integration transforms in accordance with (\ref{app13}). We need to transform the argument, $\eta'-\eta_{\rm s}$, of delta-function which we shall denote in coordinates $X^\a$ as $f_\eta(\eta')\equiv\eta'-\eta_s$. The argument is a scalar function which is transformed as $f_\eta(\eta')=f_t(t')$ where,
\be\la{bp8}
f_t(t')=-\bar{\mathtt v}_\a r^\a+\frac{c}{c_{\rm s}}\sqrt{\bar\pi_{\a\b}r^\a r^\b}\;.
\ee
Transformation of the delta-function in the integrand of integral (\ref{bp6}) is
\be\la{bp9}
\delta\le[f_t(t')\ri]=\frac{\delta(t'-\varsigma)}{\dot f_t(\varsigma)}\;,
\ee
where $\dot f_t(\varsigma)\equiv\left[df_t(t')/dt'\right]_{t'=\varsigma}$, and $\varsigma$ is one of the roots of equation $f_t(t')=0$ that is associated with the retarded interaction. Eventually, after accounting for transformation of all functions and performing integration with respect to time, integral (\ref{bp6}) assumes the following form
\be\la{bp9a}
U(t,{\bm x})=\int_{\cal V}\frac{\t_x(\varsigma,{\bm x}')d^3x'}{\dot f_t(\varsigma)|{\bm X}-{\bm X}'|_{t'=\varsigma}}\;,
\ee
where $|{\bm X}-{\bm X}'|_{t'=\varsigma}$ denotes the expression (\ref{bp7}) taken at the value of $t'=\varsigma$.
What remains is to calculate the instant of time, $\varsigma$, and the value of functions entering denominator of the integrand in (\ref{bp9a}).

Calculation of $\varsigma$ is performed by solving equation $f_t(\varsigma)=0$, that defines the characteristic cone of the sound waves, and has the following explicit form,
\be\la{bp10}
\le[\eta_{\a\b}+\le(1-\frac{c_{\rm s}^2}{c^2}\ri)\bar{\mathtt v}_\a \bar{\mathtt v}_\b\ri]r^\a r^\b=0\;,
\ee
which is derived from (\ref{bp8}).
This is a quadratic algebraic equation with respect to the time variable $r^0=\varsigma-t$. It reads
\be\la{bp11}
A(\varsigma-t)^2+2B(\varsigma-t)+C=0\;,
\ee
where the coefficients $A,B,C$ of the quadratic form are,
\ba\la{bp12}
A&=&-1+\le(1-\frac{c_{\rm s}^2}{c^2}\ri)\gamma^2\;,\\
B&=&-\le(1-\frac{c_{\rm s}^2}{c^2}\ri)\gamma^2{\bm \b}\cdot({\bm x}-{\bm x}')\;,\\
C&=&|{\bm x}-{\bm x}'|^2+\le(1-\frac{c_{\rm s}^2}{c^2}\ri)\gamma^2\le[{\bm \b}\cdot({\bm x}-{\bm x}')\ri]^2\;.
\ea
Equation (\ref{bp11}) has two roots corresponding to the advanced and retarded times. The root corresponding to the retarded-time solution of (\ref{bp11}) is
\be\la{bp13}
\varsigma=t-\frac1{A}\le(B-\sqrt{B^2-AC}\ri)\;,
\ee
or, more explicitly,
\be\la{bp14}
\varsigma=t-|{\bm x}-{\bm x}'|
\frac{\le(1-\cfrac{c_{\rm s}^2}{c^2}\ri)\gamma^2\le({\bm \b}\cdot{\bm n}\ri)+\sqrt{1-\le(1-\cfrac{c_{\rm s}^2}{c^2}\ri)\gamma^2\le[1-({\bm \b}\cdot{\bm n})^2\ri]}}{1-\le(1-\cfrac{c_{\rm s}^2}{c^2}\ri)\gamma^2}\;,
\ee
where the unit vector ${\bm n}=({\bm x}-{\bm x}')/|{\bm x}-{\bm x}'|$. After some algebra equation (\ref{bp14}) can be simplified to
\be\la{bp15}
\varsigma=t-\frac{\a_{\rm s}}{c_{\rm s}}|{\bm x}-{\bm x}'|\;,
\ee
where
\be\la{bp16}
\alpha_{\rm s}=\frac{1-\b^2}{1-\displaystyle\frac{{\b}^2}{c_{\rm s}^2}}
\le[\sqrt{1+\le(1-\frac{c^2}{c_{\rm s}^2}\ri)\gamma^2({\bm \b}\times{\bm n})^2}-
\le(1-\frac{c^2}{c_{\rm s}^2}\ri)\g^2({\bm \b}\cdot{\bm n})\ri]\;.
\ee
Coefficient $\a_{\rm s}$ defines the speed of propagation of the sound waves, $v_{\rm s}\equiv c_{\rm s}/\a_{\rm s}$, as measured by observer moving with speed $\b^i$ with respect to the Hubble flow. Thus, the value of the speed of sound, $v_{\rm s}$, depends crucially on the motion of observer.

Derivative of the function, $\dot f_t(\varsigma)$, is given by
\be\la{bp17}
\dot f_t(\varsigma)=\frac{\pd f_t}{\pd r^\a}\frac{\pd r^\a}{\pd\varsigma}\;,
\ee
where the partial derivative $\pd r^\a/\pd\varsigma=\delta^\a_0=(1,0,0,0)$. Making use of (\ref{bp8}), the partial derivative
\be\la{bp18}
\frac{\pd f_x}{\pd r^\a}=-\bar{\mathtt v}_\a+\frac{c}{c_{\rm s}}\frac{\bar\pi_{\a\b}r^\b}{\sqrt{\bar\pi_{\a\b}r^\a r^\b}}\;,
\ee
which has to be calculated at the instant of time, $t'=\varsigma$, where $\varsigma$ is given by (\ref{bp15}).

In order to calculate the denominator in the integrand in (\ref{bp9a}), we account for (\ref{bp7}), (\ref{bp10}) and combine (\ref{bp17}), (\ref{bp18})  together. We get
\be\la{bp19}
|{\bm X}-{\bm X}'|\dot f_x(\varsigma)=\frac{c}{c_{\rm s}}\le[r_\a+\le(1-\frac{c_{\rm s}^2}{c^2}\ri)\bar{\mathtt v}_\a\bar{\mathtt v}_\b r^\b\ri]\delta^\a_0\;.
\ee
It is straightforward to check that after using (\ref{bp13}) the above equation is  reduced to $|{\bm X}-{\bm X}'|\dot f_x(\varsigma)=(c/c_{\rm s})\sqrt{B^2-AC}$, or more explicitly,
\be\la{bp20}
|{\bm X}-{\bm X}'|\dot f_x(\varsigma)=|{\bm x}-{\bm x}'|\sqrt{1+\le(1-\frac{c^2}{c_{\rm s}^2}\ri)\g^2({\bm \b}\times{\bm n})^2}\;,
\ee
Finally, the retarded Lorentz-invariant solution of (\ref{bp1}) is
\be\la{bp21}
U(t,{\bm x})=\int_{\cal V}\frac{\t_x(\varsigma,{\bm x}')}{\sqrt{1+\le(1-\cfrac{c^2}{c_{\rm s}^2}\ri)\g^2({\bm \b}\times{\bm n})^2}}\frac{d^3x'}{|{\bm x}-{\bm x}'|}\;,
\ee
with the retarded time $\varsigma$ calculated in accordance with (\ref{bp15}). This solution reduces to the retarded potential (\ref{app20}) in the limit of $c_{\rm s}\rightarrow c$.
\end{document}